\documentclass[12pt,a4paper]{article}
\usepackage{geometry}
\usepackage{amsmath}
\usepackage{amsmath}
\usepackage{amssymb}
\usepackage{graphicx}
\usepackage{cite}
\usepackage{pstricks}
\usepackage{bm}
\usepackage{pbox}
\usepackage{placeins}
\usepackage{graphicx}
\usepackage{caption}
\usepackage{comment}
\usepackage{subcaption}
\usepackage[T1]{fontenc}
\usepackage{footnote}
\usepackage{pdfpages}
\usepackage{hhline}
\usepackage{multirow}
\usepackage{boldline}
\usepackage{feynmf}
\usepackage{multicol}
\usepackage{enumitem}
\usepackage{multirow}
\usepackage{amsmath}
\usepackage{relsize}
\usepackage{slashed}
\usepackage{graphicx}
\usepackage[citecolor=green]{hyperref}
\usepackage[toc,page]{appendix}
\usepackage[english]{babel}
\usepackage[compat=1.1.0]{tikz-feynman}
\allowdisplaybreaks

\usepackage{slashed}
\usepackage{cases}

\usepackage{empheq}

\definecolor{MyDarkBlue}{rgb}{0.1, 0.1, 0.8} %defining the color 'MyDarkBlue'
\definecolor{SBlue}{rgb}{0.2, 0.4, 0.7} %defining the color 'MyDarkBlue'
\definecolor{MyLightBlue}{rgb}{0.22,0.51,0.9}
\definecolor{MyGreen}{rgb}{0.0, 0.5, 0.0}
\definecolor{BrickRed}{rgb}{0.8, 0.25, 0.33}
\RequirePackage{hyperref}
\hypersetup{colorlinks, citecolor=SBlue,linkcolor=MyDarkBlue, urlcolor=BrickRed}

\begin{document}
\vspace*{-0.2in}
\begin{flushright}
\end{flushright}
\vspace{0.1cm}
\begin{center}
{\Large \bf
Discovery prospects of a singly-charged scalar at $\mu$TRISTAN
}
\end{center}
\renewcommand{\thefootnote}{\fnsymbol{footnote}}
\begin{center}
{
{}~\textbf{Joseph George$^{1,2}$}\footnote{ E-mail: \textcolor{MyDarkBlue}{ joseph.george@ncbj.gov.pl}},
{}~\textbf{Nobuchika Okada$^3$}\footnote{ E-mail: \textcolor{MyDarkBlue}{okadan@ua.edu}},
{}~\textbf{Dibyashree Sengupta$^{4,5}$}\footnote{ E-mail: \textcolor{MyDarkBlue}{sengupta.dibyashree@ucy.ac.cy}},
{}~\textbf{Sudhir K. Vempati$^6$}\footnote{ E-mail: \textcolor{MyDarkBlue}{vempati@iisc.ac.in}}
}
\vspace{0.5cm}
{
\\
\em $^1$Department of Physics, Indian Institute of Science Education and Research Pune, Pune 411008, India\\
$^2$ National Centre for Nuclear Research, Warsaw 02-093, Poland
\\
$^3$Department of Physics and Astronomy, University of Alabama, Tuscaloosa, Alabama 35487, USA
\\
$^4$INFN, Sezione di Roma,  
c/o Dip. di Fisica, Sapienza Università di Roma, \\
Piazzale Aldo Moro 2, I-00185 Rome, Italy
\\
$^5$Department of Physics, University of Cyprus, P.O. Box 20537, 1678 Nicosia, Cyprus
\\
$^6$Centre for High Energy Physics, Indian Institute of Science, Bangalore 560012, India
} 
\end{center}

%\vspace{0.6cm}
\renewcommand{\thefootnote}{\arabic{footnote}}
\setcounter{footnote}{0}
\thispagestyle{empty}

%%%%%%%%%%%%%%%%%%%%%%%%%%%%%%%%%%%%%%%%%%%%%%%
%%%%%%%%%%%%%%%%%%%%%%%%%%%%%%%%%%%%%%%%%%%%%%%
\begin{abstract}
In this article, we study the associated production of a singly-charged ($\Delta^+$) scalar along with a $W^+$ boson in the newly proposed $\mu^+\mu^+$ collider (also known as $\mu$TRISTAN) at $\sqrt{s} = 2~$ TeV. Such a singly-charged scalar is naturally accommodated in an extremely well-motivated neutrino mass model, namely, the Type-II seesaw model. This model, beside providing a viable explanation of neutrino mass generation, also allows for lepton flavor violating (LFV) processes. Since LFV processes are not allowed in the Standard Model (SM), we focus on the discovery prospect of the singly-charged scalar in the Type-II seesaw model at $\mu$TRISTAN through a LFV process, owing to the advantage of this process being free of any SM background. Additionally, this article also proposes a method to indicate if the underlying theory follows a Normal or an Inverted hierarchy depending on the distribution of lepton flavors in the final state.
\end{abstract}
%%%%%%%%%%%%%%%%%%%%%%%%%%%%%%%%%%%%%%%%%%%%%%%
%%%%%%%%%%%%%%%%%%%%%%%%%%%%%%%%%%%%%%%%%%%%%%%

%\newpage
\setcounter{footnote}{0}

%{
%  \hypersetup{linkcolor=black}
%  \tableofcontents
%}
\newpage

%%%%%%%%%%%%%%%%%%%%%%%%%%%%%%%%%%%%%%%%%%%%%%%
%%%%%%%%%%%%%%%%%%%%%%%%%%%%%%%%%%%%%%%%%%%%%%%

%%%%%%%%%%%%%%%%%%%%%%%%%%%%%%%%%%%%%%%%%%%%%%%
\section{Introduction}
\label{sec:intro}
%%%%%%%%%%%%%%%%%%%%%%%%%%%%%%%%%%%%%%%%%%%%%%%
The Standard Model (SM), till date, is the most successful theory of nature which attained its peak with the discovery of the Higgs boson in 2012 at ATLAS and CMS~\cite{Aad:2012tfa, Chatrchyan:2012ufa}. Despite such massive success of the SM, it cannot explain various natural phenomena such as : 1. Neutrino mass generation mechanism~\cite{Mohapatra:2006gs, deGouvea:2016qpx}, 2. Higgs mass instability in the electroweak sector~\cite{Branchina:2013jra, Susskind:1978ms, Veltman:1980mj}, 3. Matter-Antimatter assymmetry~\cite{Sakharov:1967dj}, 4. the Strong CP problem~\cite{Peccei:1977np, Peccei:1977hh, Peccei:1977ur} and several others. Such drawbacks of the SM motivate the existence of Beyond Standard Model (BSM) scenarios. 

An extremely well-motivated BSM scenario is the Type-II seesaw scenario~\cite{Magg:1980ut, Schechter:1980gr, Mohapatra:1979ia, Lazarides:1980nt}  which not only provides a valid explanation for neutrino mass generation mechanism but also allows for LFV processes which the SM does not accommodate. Therefore, it is highly lucrative to look for the exotic BSM particles which can vouch for the existence of the Type-II seesaw model through a LFV process since such a process will have no relevant SM background and hence will be extremely clean. All existing collider constraints on the Type-II seesaw model can be found in Ref.~\cite{Deppisch:2015qwa, Cai:2017mow}. Ref.~\cite{Calibbi:2017uvl, Davidson:2022jai} summarizes the LFV constraints. 

Lepton colliders are extremely advantageous for looking into LFV processes as they provide a much cleaner environment than hadron colliders. Additionally, since leptons are fundamental particles, the entire beam energy is available for the hard collision. Whereas, in a hadron collider such as the LHC, only a fraction of the proton-beam energy that is carried by the colliding partons is available for collision. Therefore, a higher physics reach is obtained in lepton colliders. Lepton colliders are constructed by colliding Electrons or Muons. However, since Muons are heavier than Electrons, colliding Muons leads to much smaller loss of energy due to synchrotron radiations as compared to colliding Electrons. Thanks to the huge development in technology we can now solve several technical difficulties that had been hindering the operation of a $\mu^+ \mu^-$ collider, thereby, making the $\mu^+ \mu^-$ collider gain huge interest in the community in recent years~\cite{Delahaye:2019omf, Long:2020wfp, AlAli:2021let, MuonCollider:2022xlm, Accettura:2023ked}. Numerous studies have been performed showing the huge prospect of the $\mu^+ \mu^-$ collider in search of new physics~\cite{Aime:2022flm}.

However, in this article we concentrate on a slightly different setup, namely the $\mu$TRISTAN~\cite{Hamada:2022mua} which plans on operating as a $\mu^+ e^-$ collider at $\sqrt{s} = 346~$GeV and a $\mu^+ \mu^+$ collider at $\sqrt{s} = 2~$TeV and higher~\cite{Heusch:1995yw}. We study our signature of choice in the  $\mu^+ \mu^+$ mode of the $\mu$TRISTAN. Apparently the $\mu^+ \mu^+$ collider is quite convenient to build, thanks to the cooling technology developed at J-PARC~\cite{Abe:2019thb}. Several studies have been performed showing the discovery prospects of new physics in $\mu$TRISTAN~\cite{Bossi:2020yne, Lu:2020dkx, Lichtenstein:2023iut, Das:2022mmh, Yang:2023ojm, Fridell:2023gjx, Dev:2023nha, Jia:2024wqi, Kitano:2025xaj, Das:2024gfg, Das:2024kyk}. In this article we intend to look for the singly-charged scalar that can be naturally accommodated in the Type-II seesaw model via its associated production along with a $W^+$ boson in the $\mu^+ \mu^+$ collider at $\sqrt{s} = 2~$TeV, followed by its decay in channels such that our signature of interest is a LFV process. Following the discovery prospects of the singly-charged scalar, we also propose a technique to identify whether the underlying theory respects Normal hierarchy or Inverted hierarchy depending on the flavors of the lepton in the final state.

The rest of the article is organized as follows. In Sec.~\ref{sec:model}, we discuss in detail the Type-II seesaw model. In Sec.~\ref{sec:bm}, we present our signature of interest and motivation behind choosing this signal along with the benchmark points. Sec.~\ref{sec:analysis} discusses the signal analysis and mass reach of the singly charged scalar, while Sec.~\ref{sec:diff} presents the technique to distinguish between the Normal and Inverted hierarchy. Finally in Sec.~\ref{sec:con} we provide the summary and outlook.

%%%%%%%%%%%%%%%%%%%%%%%%%%%%%%%%%%%%%%%%%%%%%%%
\section{Type-II seesaw model}
\label{sec:model}
%%%%%%%%%%%%%%%%%%%%%%%%%%%%%%%%%%%%%%%%%%%%%%%
In Type-II seesaw scenario~\cite{Magg:1980ut, Schechter:1980gr, Mohapatra:1979ia, Lazarides:1980nt} the SM particle spectrum is augmented with a $SU(2)_L$ triplet scalar $\Delta$ = ($\Delta^{++}$, $\Delta^{+}$, $\Delta^{0}$) with hypercharge $Y_{\Delta}$= 1. Therefore, the scalar sector of this BSM scenario consists of the following: 
\begin{equation}
    \Delta = \begin{pmatrix}
\Delta^+/\sqrt{2} & \Delta^{++} \\
\Delta^0 &  -\Delta^+/\sqrt{2}
\end{pmatrix}, \hspace{1in}
\Phi =  \begin{pmatrix}
\phi^+ \\
\phi^0
\end{pmatrix},
\end{equation}
where $\Phi$ is the SM Higgs doublet.

This additional scalar $\Delta$, being charged under $SU(2)_L$, interacts with the SM gauge sector and also has Yukawa interaction terms with the SM lepton doublets. Therefore, the Lagrangian terms that arise due to this additional scalar beside the SM Lagrangian are:
\begin{equation}
    \mathcal{L}_{BSM} = Tr[(D_{\mu} \Delta)^{\dagger}D^{\mu} \Delta] 
    + (Y_{\alpha \beta}\overline{L^{\alpha \, c}}(i\sigma_2)\Delta L^{\beta} + h.c.) - V(\Delta, \Phi),
    \label{eq:l}
\end{equation}
where the first term with covariant derivative $D_{\mu}$ denotes the interaction between $\Delta$ and the SM gauge sector where $D_{\mu}$ is given by
\begin{equation}
    D_{\mu} \Delta = \partial_{\mu} \Delta - \frac{i}{2} g~(\sum\limits_{k=1,2,3} W^k_{\mu} [\sigma_k \Delta - \Delta \sigma_k] )  - i g^{\prime} y_{\Delta} B_{\mu} \Delta,
\end{equation}
where $\sigma_k$ are the Pauli spin matrices, $g$ and $g^{\prime}$ are the gauge couplings of $\Delta$ with $SU(2)_L$ and $U(1)_Y$ SM gauge bosons, respectively, and $y_{\Delta}$ is the hypercharge of $\Delta$ which is postulated to be 1.

The second term in Eqn.~(\ref{eq:l}) denotes the Yukawa interaction between $\Delta$ and the SM  left-handed weak lepton doublets denoted by $L^{\alpha}$ and $L^{\beta}$ where $\alpha$ and $\beta$ are the flavor indices. The generic form of the SM left-handed weak lepton doublet is given by
\begin{equation}
    L = \begin{pmatrix}
        \nu_e \\
        e^{-}
    \end{pmatrix}_L
\end{equation}

The term $V(\Delta, \Phi)$ in Eqn.~(\ref{eq:l}) is the scalar potential which can be written as:
\begin{equation}
\begin{split}
    V(\Delta, \Phi) = -m^2_{\Phi}\Phi^{\dagger}\Phi + \frac{\lambda}{4}(\Phi^{\dagger}\Phi)^2+ M_{\Delta}^2 Tr [\Delta^{\dagger} \Delta] + \lambda_1 [Tr \Delta^{\dagger} \Delta]^2 + \lambda_2 Tr [\Delta^{\dagger} \Delta]^2 \\
    + [\mu \Phi^T i \sigma_2 \Delta^{\dagger}\Phi + h.c.] + \lambda_3 (\Phi^{\dagger}\Phi)  Tr [\Delta^{\dagger} \Delta] + \lambda_4 \Phi^{\dagger} \Delta \Delta^{\dagger} \Phi,
    \end{split}
    \label{eq:scp}
\end{equation}
where $\mu$ and $\lambda_i$ ($i$ = 1, 2, 3, 4) are the relevant couplings among the SM and BSM scalars. 

On minimizing the scalar potential in Eqn.~(\ref{eq:scp}) one can obtain a VEV of $\Delta^0$ (denoted by $v_{\Delta}$) in Eqn.~(\ref{eq:l}) and with the SM Higgs VEV (denoted by $v$), one can obtain the neutrino mass as :
\begin{equation}
    m_{\nu} = \sqrt{2}Yv_{\Delta},
\end{equation}
where $v_{\Delta} \sim \mu v^2 /(\sqrt{2}M_{\Delta}^2)$.

The neutrino mass matrix can be diagonalized using the PMNS matrix denoted by $U_{PMNS}$ as follows: 

\begin{equation}
    m_{\nu}^{diag} = Diag [m_{\nu_1}, m_{\nu_2}, m_{\nu_3}] = U_{PMNS}^* m_{\nu} U^{\dagger}_{PMNS}.
    \label{eq:mnu}
\end{equation}
The neutrino mass hierarchy can either respect normal ordering (i.e., $\nu_1$ is the lightest) or inverted ordering (i.e., $\nu_3$ is the lightest). Here we study both cases and propose a method to distinguish between these two cases depending on the phenomenological manifestation at the $\mu^+ \mu^+$ Collider. 

After the electroweak symmetry breaking, this BSM scenario yields seven physical states of definite mass in the scalar sector. These states are denoted by $H^{\pm \pm}$, $H^{\pm}$, $H^0$,  $A^0$ and $h$ where $h$ is the SM Higgs boson, $A^0$ is a pseudoscalar, $H^0$ is the electrically-neutral heavy scalar, and $H^{\pm \pm}$ and $H^{\pm}$ are electrically-charged heavy scalars.

From Eqn.~(\ref{eq:mnu}), one can deduce that in order to obtain a small mass for the neutrino as supported by experiments, $v_{\Delta}$ must be quite small. In the simplest type-II seesaw model~\cite{Magg:1980ut, Schechter:1980gr, Mohapatra:1979ia, Lazarides:1980nt}, $v_{\Delta}$ is bounded by the electroweak $\rho$ parameter constraining $v_{\Delta}$ $\lesssim$ 3 GeV ~\cite{Zyla:2020zbs}. In the limit $v_{\Delta} \ll v$, the mass eigen states of the scalar triplet appear as :
\begin{equation}
\begin{split}
    m_{H^{++}} \sim m_{\Delta^{++}}, m_{H^{+}} \sim m_{\Delta^{+}}, m_{H^0} \sim Re [m_{\Delta^{0}}], m_{A^0} \sim Im [m_{\Delta^{0}}] \\
    m_{H^0}^2 \approx m_{A^0}^2 \approx  m_{H^{+}}^2 + \frac{\lambda_4}{4} v^2 \approx  m_{H^{++}}^2 + \frac{\lambda_4}{2} v^2.
\end{split}
\label{eq:mdel}
\end{equation}
Depending on the value and sign of $\lambda_4$, the mass ordering of the scalar triplets can be : 
\\
$i)$ $\lambda_4$ = 0 : $m_{H^0}^2 = m_{A^0}^2 =  m_{H^{+}}^2 = m_{H^{++}}^2$.
\\
$ii)$ $\lambda_4 <$  0 : $m_{H^0}^2 = m_{A^0}^2 <  m_{H^{+}}^2 < m_{H^{++}}^2$.
\\
$iii)$ $\lambda_4 >$  0 : $m_{H^0}^2 = m_{A^0}^2 >  m_{H^{+}}^2 > m_{H^{++}}^2$.
\\
In this article, we choose the second case with the mass hierarchy $m_{H^0}^2 = m_{A^0}^2 <  m_{H^{+}}^2 < m_{H^{++}}^2$ with the mass difference of $\sim$ 1-2 GeV. In the rest of the article, we denote the physical states $H^{\pm \pm}$, $H^{\pm}$ and $H^0$ by $\Delta^{\pm \pm}$, $\Delta^{\pm}$ and $\Delta^0$, respectively.

\section{Signature of interest and Benchmark points}
\label{sec:bm}

\begin{figure}[ht!]
\centering
\begin{subfigure}[h]{0.45\textwidth}
  \centering
  \includegraphics[width=1\linewidth]{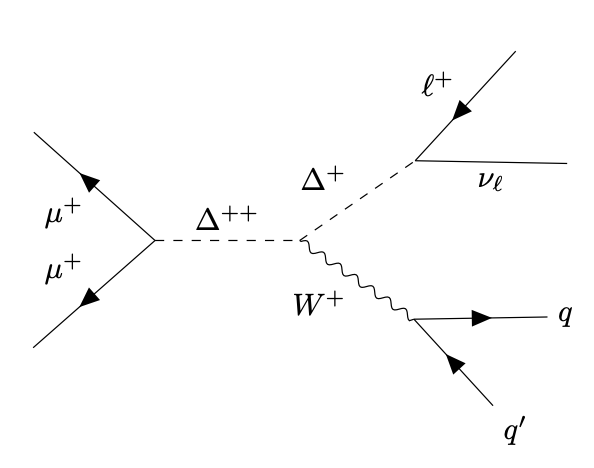}
  \caption{}
  \label{fig:gen1}
\end{subfigure}%
\begin{subfigure}[h]{0.45\textwidth}
  \centering
\includegraphics[width=1\linewidth]{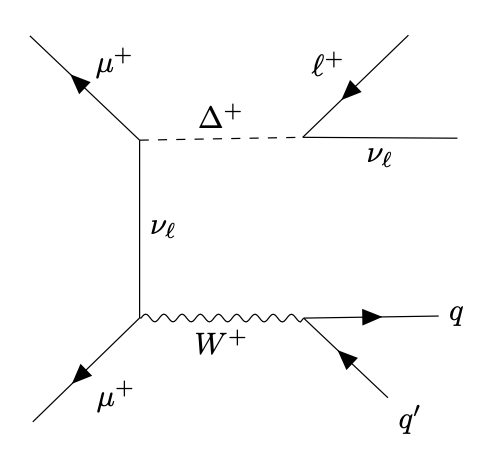}
\caption{}
  \label{fig:gen2}
\end{subfigure}
\vspace*{-0.1in}
\caption{Generic Feynman diagram for signal}
\label{fig:generic}
\end{figure}

Several studies have previously been performed to look for the doubly-charged scalar $\Delta^{++}$ and electrically-neutral scalar $\Delta^0$ in the Muon collider~\cite{Aime:2022flm, Ghosh:2025gdx} as well as the $\mu^+ \mu^+$ mode of the $\mu$TRISTAN~\cite{Yang:2023ojm, Fridell:2023gjx, Dev:2023nha, Jia:2024wqi}. A comparative study of the doubly-charged scalar $\Delta^{++}$ at the Muon collider and the $\mu^+ \mu^+$ mode of the $\mu$TRISTAN for the chosen benchmark points has been shown in Appendix~\ref{sec:dppmu} and Appendix~\ref{sec:dpp}. However, the discovery prospects of the singly-charged scalar $\Delta^{+}$ have not yet received much deserved attention. Therefore, here in this article we study the associated production of $\Delta^+$ along with the $W^+$ boson in the $\mu^+ \mu^+$ mode of the $\mu$TRISTAN, followed by hadronic decay of the $W^+$ boson and leptonic decay of $\Delta^+$. This associated production of $\Delta^+$ and $W^+$ involves two types of contribution: $(i)$ mediated by a virtual $\Delta^{++}$ in s-channel and $(ii)$ mediated by exchange of a $\nu_l$ in t-channel. The generic Feynman diagrams of these processes are shown in Fig.~\ref{fig:gen1} and Fig~\ref{fig:gen2}, respectively. 

\begin{figure}[ht!]
\centering
\begin{subfigure}[h]{0.45\textwidth}
  \centering
  \includegraphics[width=1\linewidth]{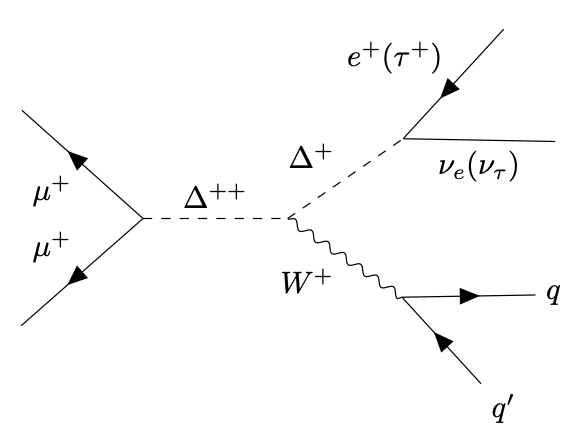}
  \caption{}
  \label{fig:sig1}
\end{subfigure}%
\begin{subfigure}[h]{0.45\textwidth}
  \centering
\includegraphics[width=1\linewidth]{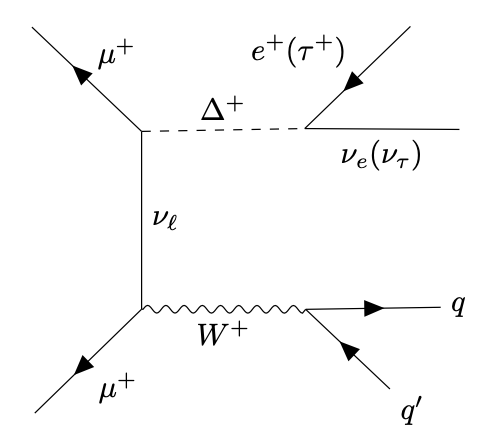}
\caption{}
  \label{fig:sig2}
\end{subfigure}
\vspace*{-0.1in}
\caption{Feynman diagram for lepton flavor violating (LFV) signal processes chosen to get rid of any SM background.}
\label{fig:signal}
\end{figure}

If the lepton arising from the decay of the $\Delta^+$ is a $\mu^+$, then the process in Fig.~\ref{fig:generic} has a sizeable SM background $\approx$ 0.92 pb. However, if one considers the final state lepton to be $e^+$ or $\tau^+$, then there is no relevant SM background as in order to generate such a process, lepton flavor violation is needed which the SM does not accommodate. Hence, finally our chosen signature of interest consists of only $e^+$ or $\tau^+$ as the final state charged leptons as shown in Fig.~\ref{fig:signal}. This choice helps to get rid of any SM background as discussed above.

\textbf{Benchmark points:}

\begin{figure}[ht!]
    \centering
    \includegraphics[width=1.0\linewidth]{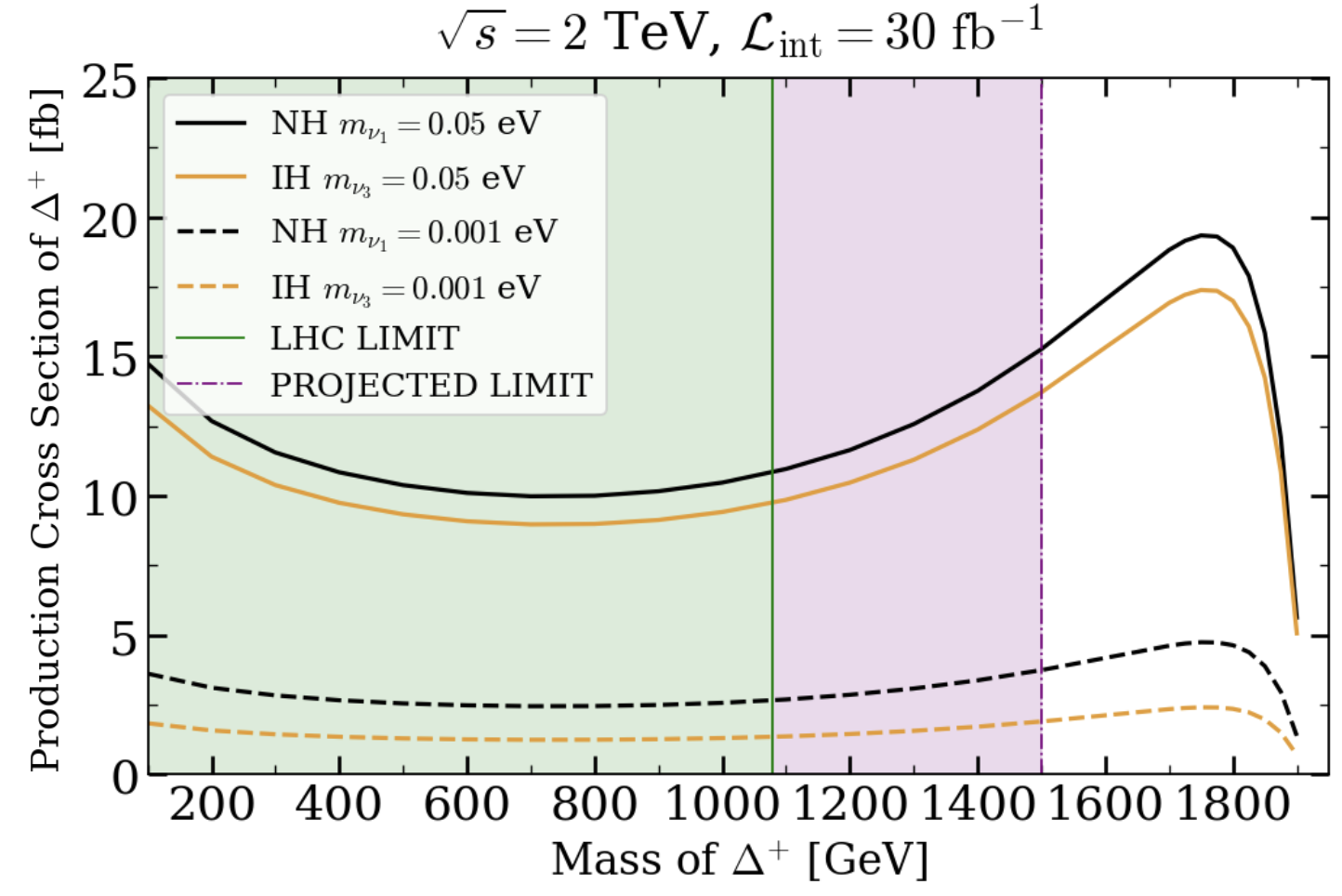}
    \caption{Cross-Section for the process $\mu^+ \mu^+ \longrightarrow \Delta^+ W^+$ for masses of $\Delta^+$ varying between 101 and 1901 GeV.}
    \label{fig:prodxsec}
\end{figure}

\begin{figure}[ht!]
\centering
\begin{subfigure}[h]{0.45\textwidth}
  \centering
  \includegraphics[width=1\linewidth]{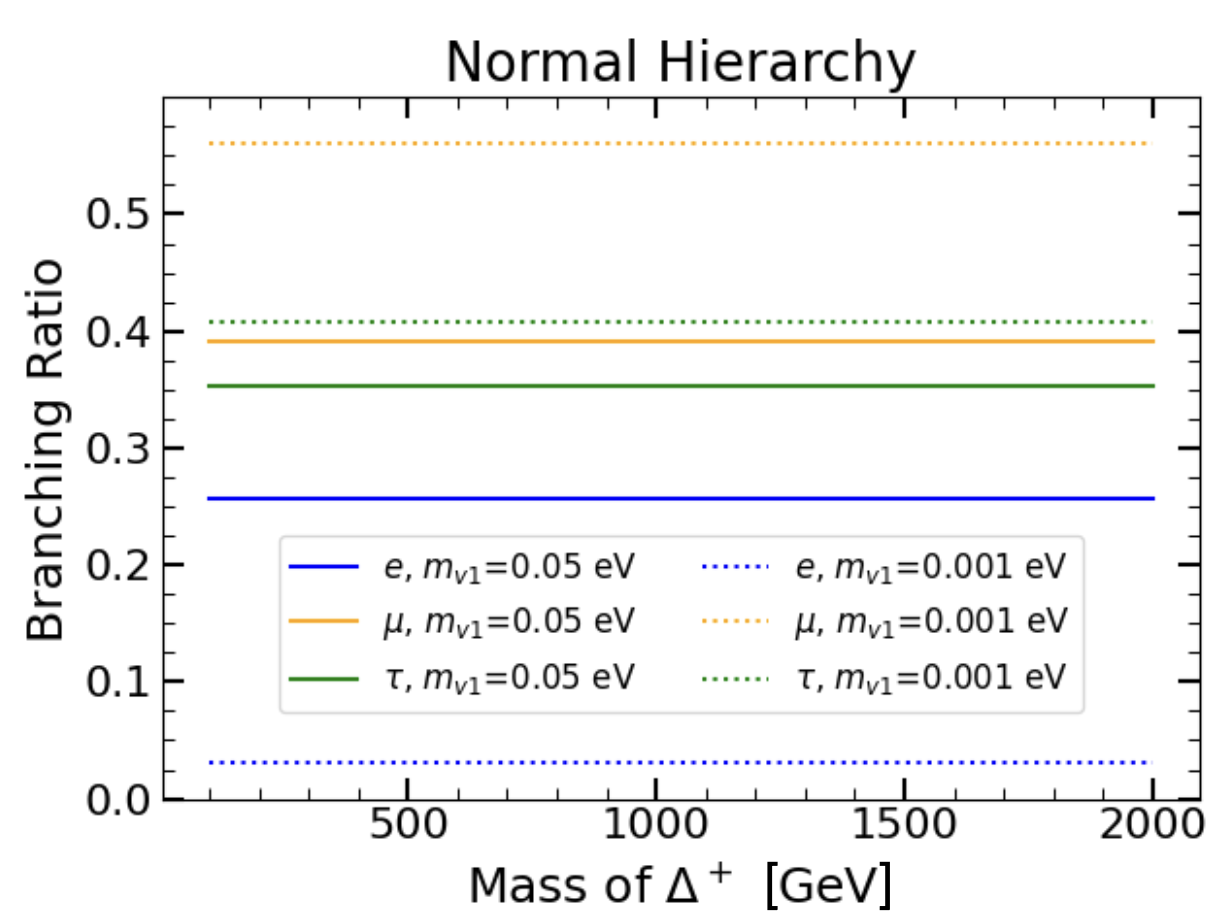}
  \caption{Normal Hierarchy}
  \label{fig:br1nh}
\end{subfigure}%
\begin{subfigure}[h]{0.45\textwidth}
  \centering
  \includegraphics[width=1\linewidth]{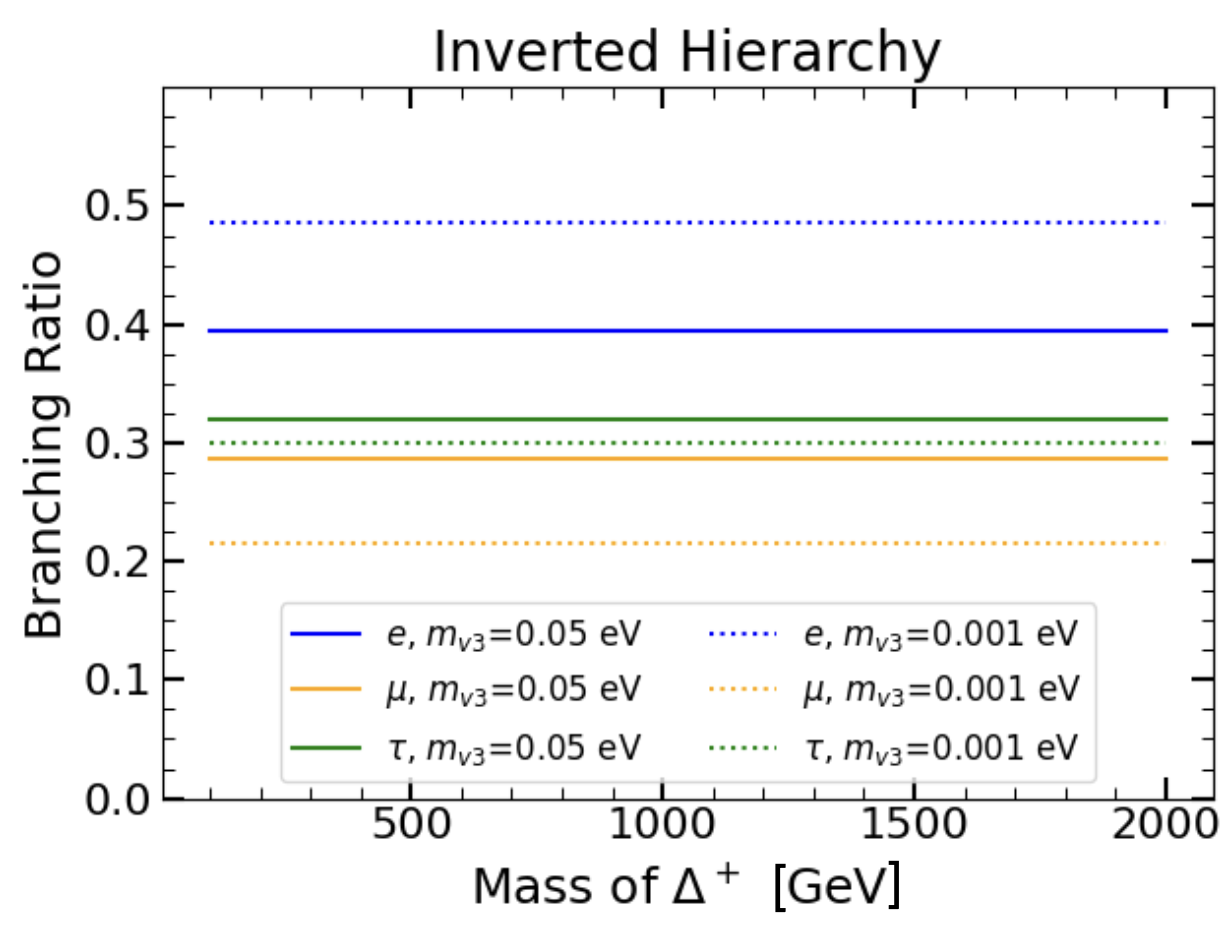}
  \caption{Inverted Hierarchy}
  \label{fig:br1ih}
\end{subfigure}
\vspace*{-0.1in}
\caption{Branching Ratio of $\Delta^+$ for masses of $\Delta^+$ varying between 101 and 1901 GeV.}
\label{fig:br1}
\end{figure}

\begin{figure}[ht!]
\centering
\begin{subfigure}[h]{0.45\textwidth}
  \centering
  \includegraphics[width=1\linewidth]{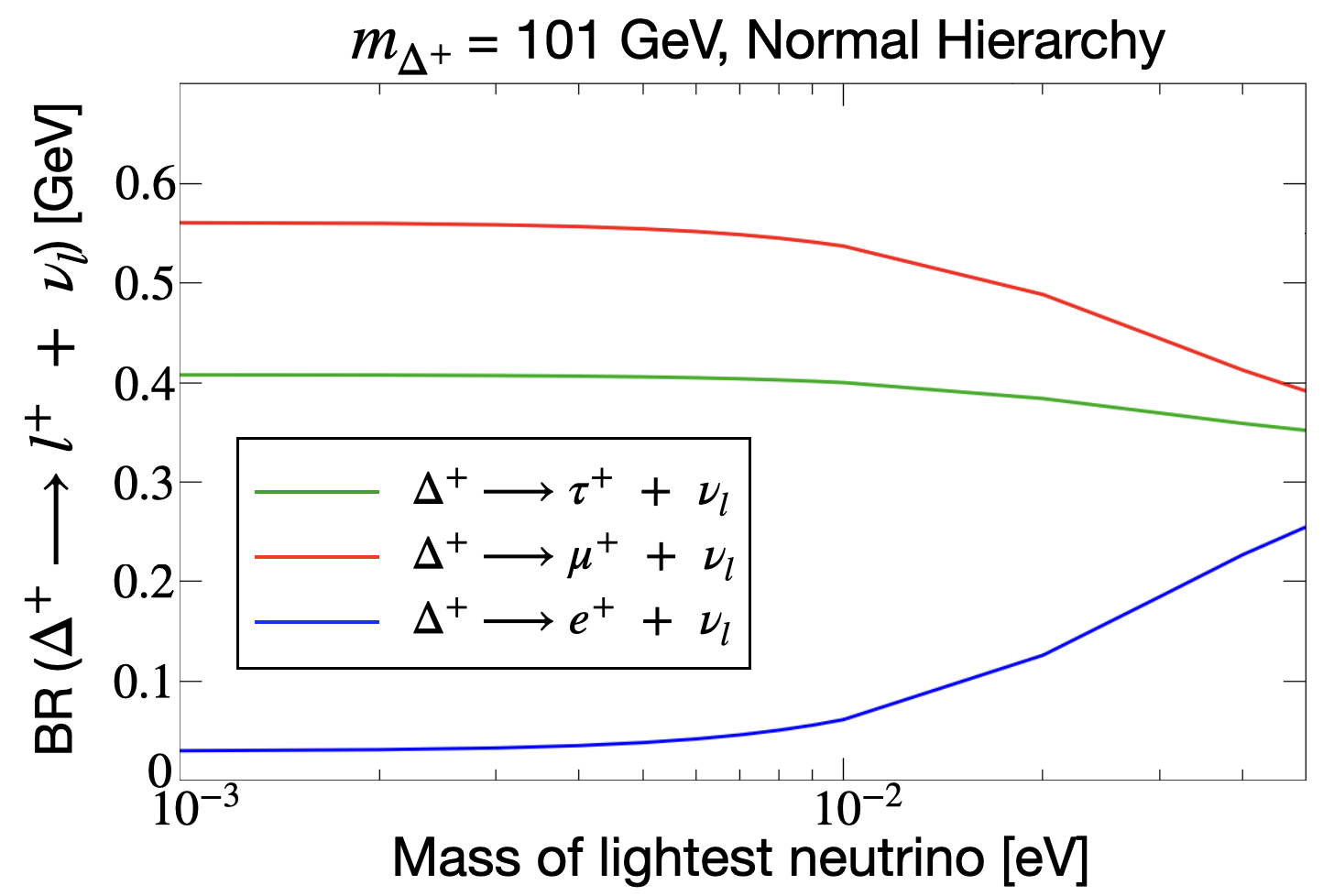}
  \caption{Normal Hierarchy}
  \label{fig:br2nh}
\end{subfigure}%
\begin{subfigure}[h]{0.45\textwidth}
  \centering
  \includegraphics[width=1\linewidth]{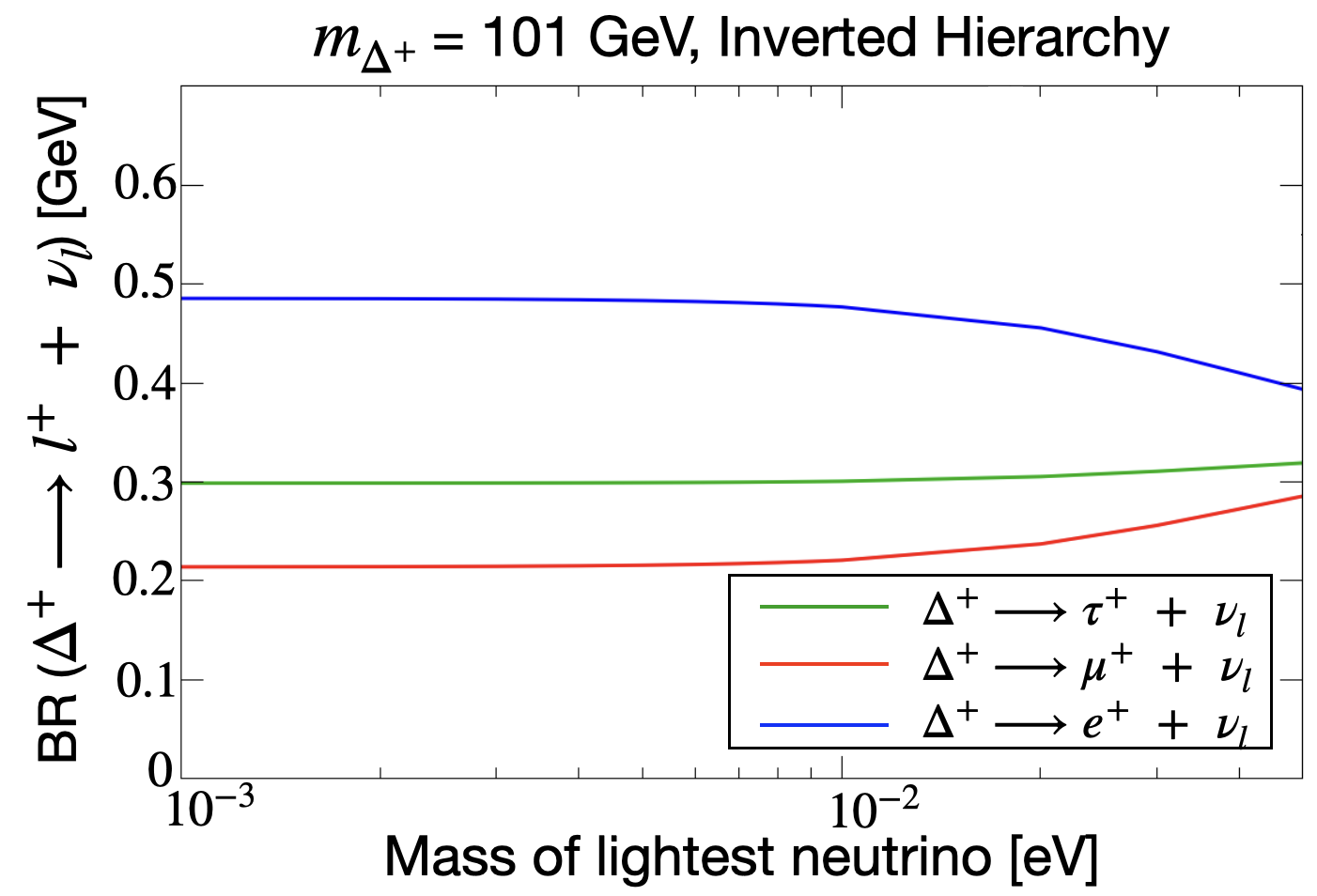}
  \caption{Inverted Hierarchy}
  \label{fig:br2ih}
\end{subfigure}
\vspace*{-0.1in}
\caption{Branching Ratio of $\Delta^+$ for masses of the lightest neutrino varying between 0.05 and 0.001 eV for a fixed $m_{\Delta^+} = 101~$ GeV.}
\label{fig:br2}
\end{figure}

As benchmark points, we consider two values of the lightest neutrino mass : 0.05 eV and 0.001 eV. The VEV of $\Delta^0$, denoted as $v_{\Delta}$, is fixed at $10^{-9}$ GeV as at such a small value of $v_{\Delta}$, $\Delta^+$ primarily decays to leptons. The analysis is performed by varying the mass of $\Delta^+$ in between 101 GeV and 1901 GeV at an interval of 100 GeV followed by a more detailed scan in the high mass region i.e., 1700-1900 GeV at an interval of 25 GeV for both Normal and Inverted Hierarchy at $\sqrt{s} = 2~$TeV \footnote{The mass limits on $\Delta^+$ and the couplings has been carefully chosen motivated by several experimental constraints~\cite{ATLAS:2012hi, ATLAS:2014kca, ATLAS:2017xqs, ATLAS:2018ceg, ATLAS:2021jol, CMS:2012dun, CMS:2014mra, CMS:2017fhs} and the reach of the $\mu$TRISTAN collider used in this analysis. Furthermore, the couplings can be changed by changing $v_{\Delta}$ and we have checked that the cross-section of the signature of our interest is directly proportional to $v_{\Delta}^2$. Therefore, one can easily scale the cross-section for different values of $v_{\Delta}$ and hence different couplings.}. The benchmark points for $m_{\Delta^+} = 101, 501, 1001, 1501, 1701, 1801, 1851$ and $1901$ GeV and the lightest neutrino mass : 0.05 eV and 0.001 eV are shown in Table.~\ref{tab: bmpoints1} and Table.~\ref{tab: bmpoints2}, respectively. We emphasize on benchmark points with $m_{\Delta^+} \geq 1.5$ TeV following the current and future LHC limit on $m_{\Delta^{++}}$ as found in Ref. ~\cite{ATLAS:2014kca, ATLAS:2017xqs, ATLAS:2022pbd} and Ref.~\cite{Ashanujjaman:2021txz}, respectively. The production cross-section for our signal of interest is shown in Fig.~\ref{fig:prodxsec} which exhibits a fall and rise behavior with increase in mass of $\Delta^+$. This typical behavior of the cross-section for our signal of interest is due to the contribution from Fig.~\ref{fig:sig2} as discussed in detail in Appendix~\ref{sec:analytical}. The solid green(dashed purple) vertical line in Fig.~\ref{fig:prodxsec} correspond to the current(future) LHC limit on $m_{\Delta^{++}}$ obtained from Ref.~\cite{ATLAS:2014kca, ATLAS:2017xqs, ATLAS:2022pbd, Ashanujjaman:2021txz}, while the green (purple) shaded region denotes the parameter space excluded by the current (projected) reach of the LHC. The branching ratios of $\Delta^+$ in different channels are shown for Normal and Inverted Hierarchy in Fig.~\ref{fig:br1nh} and Fig.~\ref{fig:br1ih}, respectively. The branching ratios of $\Delta^+$ do not change with change in its mass, as can be seen in Fig.~\ref{fig:br1}. 
Although for Normal hierarchy the branching ratio for $\Delta^+$ to decay into $\mu^+$ is higher than that for decaying into $e^+$ or $\tau^+$ for a particular value of lightest neutrino mass, as evident from Fig:~\ref{fig:br1nh}, we consider the decay of $\Delta^+$ to $e^+$ and $\tau^+$ only, as then the process is absolutely free of SM background, as mentioned earlier, although at a cost of losing a reasonable number of signal events. But since the surviving events don't have any SM background, those small number of events are sufficient to yield 5$\sigma$ significance up to a certain mass of $\Delta^+$ which will be discussed in detail in Sec.~\ref{sec:analysis}. For Inverted Hierarchy, however, the branching ratio for $\Delta^+$ to decay into $\mu^+$ is smaller than that for decaying into $e^+$ or $\tau^+$. Therefore, for Inverted Hierarchy, selecting the decay of $\Delta^+$ to $e^+$ and $\tau^+$ only makes the process free of SM background without much loss of signal events.
Although the branching ratios of $\Delta^+$ do not change with its mass, they do change with change in the mass of the lightest neutrino, as evident from Fig~\ref{fig:br2}.

\begin{table}[ht!]
\centering
\resizebox{\textwidth}{!}{\begin{tabular}{|c|c|c V{5}c |c|c|c|c|c V{5}c|c|c|c|c|c|c|}
\hline
\multicolumn{3}{|c V{5}}{Masses [GeV]}         & \multicolumn{6}{c V{5}}{Normal Hierarchy}    & \multicolumn{6}{c|}{Inverted Hierarchy}                                                                                 \\ \hline
  $m_{\Delta^0}$    &    $m_{\Delta^+}$                    &  $m_{\Delta^{++}}$    &      $\Gamma_{\Delta^0}$      &   $\Gamma_{\Delta^+}$   &  $\Gamma_{\Delta^{++}}$   &  $\sigma_{Production}$ & $\sigma$  &  $S/\sqrt{S}$ & $\Gamma_{\Delta^0}$      &   $\Gamma_{\Delta^+}$   &  $\Gamma_{\Delta^{++}}$       &   $\sigma_{Production}$ &  $\sigma$      &      $S/\sqrt{S}$    \\ \hline
    100   & 101   & 103   & 0.02004 & 0.02024 & 0.02063 & 14.74 & 5.500 & 12.85 & 0.02472 & 0.02496 & 0.02545 & 13.24 & 5.802 & 13.19 \\ \hline
    % 200   & 201   & 203   & 0.04008 & 0.04028 & 0.04068 & 12.68 & 4.735 & 11.92 & 0.04944 & 0.04969 & 0.05018 & 11.40 & 4.993 & 12.24 \\ \hline
    %300   & 301   & 303   & 0.06012 & 0.06032 & 0.06072 & 11.55 & 4.315 & 11.38 & 0.07416 & 0.07441 & 0.07490 & 10.38 & 4.550 & 11.68 \\ \hline
    % 400   & 401   & 403   & 0.08017 & 0.08036 & 0.08077 & 10.84 & 4.051 & 11.02 & 0.09888 & 0.09913 & 0.09962 & 9.745 & 4.271 & 11.32 \\ \hline
    500   & 501   & 503   & 0.1002  & 0.1004  & 0.1008  & 10.38 & 3.876 & 10.78 & 0.1236  & 0.1239  & 0.1243  & 9.332 & 4.087 & 11.07 \\ \hline
    % 600   & 601   & 603   & 0.1202  & 0.1204  & 0.1208  & 10.10 & 3.776 & 10.64 & 0.1483  & 0.1486  & 0.1491  & 9.082 & 3.978 & 10.93 \\ \hline
    % 700   & 701   & 703   & 0.1403  & 0.1405  & 0.1409  & 9.980 & 3.725 & 10.57 & 0.1730  & 0.1733  & 0.1738  & 8.970 & 3.927 & 10.85 \\ \hline
    % 800   & 801   & 803   & 0.1603  & 0.1605  & 0.1609  & 9.998 & 3.732 & 10.58 & 0.1978  & 0.1980  & 0.1985  & 8.986 & 3.935 & 10.87 \\ \hline
    % 900   & 901   & 903   & 0.1804  & 0.1806  & 0.1810  & 10.16 & 3.792 & 10.67 & 0.2225  & 0.2227  & 0.2232  & 9.129 & 3.999 & 10.95 \\ \hline
    1000  & 1001  & 1003  & 0.2004  & 0.2006  & 0.2010  & 10.47 & 3.911 & 10.83 & 0.2472  & 0.2475  & 0.2480  & 9.415 & 4.123 & 11.12 \\ \hline
    % 1100  & 1101  & 1103  & 0.2205  & 0.2207  & 0.2211  & 10.96 & 4.088 & 11.07 & 0.2719  & 0.2722  & 0.2727  & 9.852 & 4.312 & 11.37 \\ \hline
    % 1200  & 1201  & 1203  & 0.2405  & 0.2407  & 0.2411  & 11.64 & 4.347 & 11.42 & 0.2966  & 0.2969  & 0.2974  & 10.47 & 4.582 & 11.72 \\ \hline
    % 1300  & 1301  & 1303  & 0.2605  & 0.2607  & 0.2611  & 12.57 & 4.690 & 11.86 & 0.3214  & 0.3216  & 0.3221  & 11.29 & 4.943 & 12.18 \\ \hline
    % 1400  & 1401  & 1403  & 0.2806  & 0.2808  & 0.2812  & 13.76 & 5.132 & 12.41 & 0.3461  & 0.3463  & 0.3468  & 12.37 & 5.413 & 12.74 \\ \hline
    1500  & 1501  & 1503  & 0.3006  & 0.3008  & 0.3012  & 15.26 & 5.7   & 13.08 & 0.3708  & 0.3711  & 0.3716  & 13.71 & 6.0   & 13.42 \\ \hline
    % 1600  & 1601  & 1603  & 0.3207  & 0.3209  & 0.3213  & 17.04 & 6.354 & 13.81 & 0.3955  & 0.3958  & 0.3963  & 15.32 & 6.703 & 14.18 \\ \hline
    1700  & 1701  & 1703  & 0.3407  & 0.3409  & 0.3413  & 18.82 & 7.008 & 14.50 & 0.4203  & 0.4205  & 0.4210  & 16.92 & 7.394 & 14.89 \\ \hline
    %1725  & 1726  & 1728  & 0.3457  & 0.3459  & 0.3463  & 19.15 & 7.128 & 14.62 & 0.4264  & 0.4267  & 0.4272  & 17.21 & 7.518 & 15.02 \\ \hline
    %1750  & 1751  & 1753  & 0.3507  & 0.3509  & 0.3513  & 19.34 & 7.201 & 14.70 & 0.4326  & 0.4329  & 0.4334  & 17.38 & 7.591 & 15.09 \\ \hline
    %1775  & 1776  & 1778  & 0.3557  & 0.3559  & 0.3563  & 19.30 & 7.183 & 14.68 & 0.4388  & 0.4390  & 0.4395  & 17.35 & 7.578 & 15.08 \\ \hline
    1800  & 1801  & 1803  & 0.3607  & 0.3609  & 0.3613  & 18.90 & 7.030 & 14.52 & 0.4450  & 0.4452  & 0.4457  & 16.99 & 7.412 & 14.91 \\ \hline
    %1825  & 1826  & 1828  & 0.3658  & 0.3660  & 0.3664  & 17.88 & 6.642 & 14.12 & 0.4511  & 0.4514  & 0.4519  & 16.07 & 7.004 & 14.50 \\ \hline
    1850  & 1851  & 1853  & 0.3708  & 0.3710  & 0.3714  & 15.84 & 5.874 & 13.28 & 0.4573  & 0.4576  & 0.4581  & 14.23 & 6.195 & 13.63 \\ \hline
    %1875  & 1876  & 1878  & 0.3758  & 0.3760  & 0.3764  & 12.08 & 4.468 & 11.58 & 0.4635  & 0.4638  & 0.4643  & 10.86 & 4.712 & 11.89 \\ \hline
    1900  & 1901  & 1903  & 0.3808  & 0.3810  & 0.3814  & 5.63  & 2.1   & 7.94  & 0.4697  & 0.4700  & 0.4704  & 5.06  & 2.2   & 8.12  \\ \hline
\end{tabular}}
\caption{Benchmark Points for Normal and Inverted Hierarchy at $v_{\Delta} = 10^{-9}$ GeV and lightest neutrino mass = 0.05 eV. The cross-sections listed here are for the processes before and after decay of $\Delta^+$ and $W^+$ (as in Fig.~\ref{fig:signal}) at $\sqrt{s} = 2$~ TeV. The significance listed here is at Integrated Luminosity $\mathcal{L}$ = 30 fb$^{-1}$. All the masses and decay widths($\Gamma$) are in units of GeV and the cross-sections $\sigma_{Production}$ and $\sigma$ are in units of fb.}
    \label{tab: bmpoints1}
\end{table}

\begin{table}[ht!]
\centering
\resizebox{\textwidth}{!}{\begin{tabular}{|c|c|c V{5}c |c|c|c|c|c V{5}c|c|c|c|c|c|c|}
\hline
\multicolumn{3}{|c V{5}}{Masses [GeV]}         & \multicolumn{6}{c V{5}}{Normal Hierarchy}    & \multicolumn{6}{c|}{Inverted Hierarchy}                                                                                 \\ \hline
  $m_{\Delta^0}$    &    $m_{\Delta^+}$                    &  $m_{\Delta^{++}}$    &      $\Gamma_{\Delta^0}$      &   $\Gamma_{\Delta^+}$   &  $\Gamma_{\Delta^{++}}$   &  $\sigma_{Production}$ & $\sigma$  &  $S/\sqrt{S}$ & $\Gamma_{\Delta^0}$      &   $\Gamma_{\Delta^+}$   &  $\Gamma_{\Delta^{++}}$       &   $\sigma_{Production}$ &  $\sigma$      &      $S/\sqrt{S}$    \\ \hline
    100   & 101   & 103   & 0.005127 & 0.005176 & 0.005278 & 3.614 & 0.9744 & 5.407  & 0.009806 & 0.009902 & 0.01010  & 1.833 & 0.8835 & 5.149  \\ \hline
    % 200   & 201   & 203   & 0.01025  & 0.01030  & 0.01041  & 3.110 & 0.8386 & 5.015  & 0.01961  & 0.01971  & 0.01990  & 1.577 & 0.7602 & 4.776  \\ \hline
    %300   & 301   & 303   & 0.01538  & 0.01543  & 0.01553  & 2.834 & 0.7642 & 4.788  & 0.02942  & 0.02952  & 0.02971  & 1.437 & 0.6931 & 4.560  \\ \hline
    % 400   & 401   & 403   & 0.02051  & 0.02056  & 0.02066  & 2.660 & 0.7166 & 4.637  & 0.03922  & 0.03932  & 0.03952  & 1.349 & 0.6505 & 4.417  \\ \hline
    500   & 501   & 503   & 0.02563  & 0.02568  & 0.02579  & 2.546 & 0.6863 & 4.538  & 0.04903  & 0.04913  & 0.04932  & 1.291 & 0.6223 & 4.320  \\ \hline
    % 600   & 601   & 603   & 0.03076  & 0.03081  & 0.03091  & 2.478 & 0.6680 & 4.477  & 0.05884  & 0.05893  & 0.05913  & 1.257 & 0.6057 & 4.262  \\ \hline
    % 700   & 701   & 703   & 0.03589  & 0.03594  & 0.03604  & 2.447 & 0.6597 & 4.449  & 0.06864  & 0.06874  & 0.06894  & 1.241 & 0.5982 & 4.236  \\ \hline
    % 800   & 801   & 803   & 0.04101  & 0.04106  & 0.04117  & 2.452 & 0.6607 & 4.453  & 0.07845  & 0.07855  & 0.07874  & 1.244 & 0.5995 & 4.240  \\ \hline
    % 900   & 901   & 903   & 0.04614  & 0.04619  & 0.04629  & 2.491 & 0.6711 & 4.488  & 0.08825  & 0.08835  & 0.08855  & 1.263 & 0.6090 & 4.274  \\ \hline
    1000  & 1001  & 1003  & 0.05127  & 0.05132  & 0.05142  & 2.568 & 0.6922 & 4.559  & 0.09806  & 0.09816  & 0.09835  & 1.303 & 0.6279 & 4.339  \\ \hline
    % 1100  & 1101  & 1103  & 0.05639  & 0.05644  & 0.05655  & 2.688 & 0.7241 & 4.661  & 0.1079   & 0.1080   & 0.1082   & 1.363 & 0.6565 & 4.437  \\ \hline
    % 1200  & 1201  & 1203  & 0.06152  & 0.06157  & 0.06167  & 2.856 & 0.7694 & 4.805  & 0.1177   & 0.1178   & 0.1180   & 1.448 & 0.6979 & 4.576  \\ \hline
    % 1300  & 1301  & 1303  & 0.06665  & 0.06670  & 0.06680  & 3.081 & 0.8298 & 4.989  & 0.1275   & 0.1276   & 0.1278   & 1.563 & 0.7529 & 4.752  \\ \hline
    % 1400  & 1401  & 1403  & 0.07177  & 0.07182  & 0.07193  & 3.374 & 0.9084 & 5.220  & 0.1373   & 0.1374   & 0.1376   & 1.711 & 0.8237 & 4.971  \\ \hline
    1500  & 1501  & 1503  & 0.07690  & 0.07695  & 0.07705  & 3.742 & 1.007  & 5.495  & 0.1471   & 0.1472   & 0.1474   & 1.898 & 0.9135 & 5.234  \\ \hline
    % 1600  & 1601  & 1603  & 0.08203  & 0.08208  & 0.08218  & 4.178 & 1.124  & 5.807  & 0.1569   & 0.1570   & 0.1572   & 2.120 & 1.020  & 5.531  \\ \hline
    1700  & 1701  & 1703  & 0.08715  & 0.08720  & 0.08731  & 4.616 & 1.240  & 6.099  & 0.1667   & 0.1668   & 0.1670   & 2.342 & 1.125  & 5.809  \\ \hline
    %1725  & 1726  & 1728  & 0.08843  & 0.08848  & 0.08859  & 4.697 & 1.261  & 6.150  & 0.1692   & 0.1693   & 0.1694   & 2.382 & 1.144  & 5.858  \\ \hline
    %1750  & 1751  & 1753  & 0.08972  & 0.08977  & 0.08987  & 4.742 & 1.274  & 6.182  & 0.1716   & 0.1717   & 0.1719   & 2.406 & 1.156  & 5.889  \\ \hline
    %1775  & 1776  & 1778  & 0.09100  & 0.09105  & 0.09115  & 4.733 & 1.271  & 6.175  & 0.1741   & 0.1742   & 0.1744   & 2.401 & 1.153  & 5.881  \\ \hline
    1800  & 1801  & 1803  & 0.09228  & 0.09233  & 0.09243  & 4.634 & 1.244  & 6.109  & 0.1765   & 0.1766   & 0.1768   & 2.351 & 1.128  & 5.817  \\ \hline
    %1825  & 1826  & 1828  & 0.09356  & 0.09361  & 0.09371  & 4.386 & 1.176  & 5.940  & 0.1790   & 0.1791   & 0.1793   & 2.225 & 1.066  & 5.655  \\ \hline
    1850  & 1851  & 1853  & 0.09484  & 0.09489  & 0.09499  & 3.884 & 1.040  & 5.586  & 0.1814   & 0.1815   & 0.1817   & 1.970 & 0.9435 & 5.319  \\ \hline
    %1875  & 1876  & 1878  & 0.09612  & 0.09617  & 0.09628  & 2.962 & 0.7907 & 4.870  & 0.1839   & 0.1840   & 0.1842   & 1.503 & 0.7174 & 4.640  \\ \hline
    1900  & 1901  & 1903  & 0.09740  & 0.09746  & 0.09756  & 1.381 & 0.3692 & 3.328  & 0.1863   & 0.1864   & 0.1866   & 0.7004& 0.3350 & 3.171  \\ \hline
\end{tabular}}
\caption{Benchmark Points for Normal and Inverted Hierarchy at $v_{\Delta} = 10^{-9}~$ GeV and lightest neutrino mass = 0.001 eV. The cross-sections listed here are for the processes before and after decay of $\Delta^+$ and $W^+$ (as in Fig.~\ref{fig:signal}) at $\sqrt{s} = 2$~ TeV. The significance listed here is at Integrated Luminosity $\mathcal{L}$ = 30 fb$^{-1}$. All the masses and decay widths($\Gamma$) are in units of GeV and the cross-sections $\sigma_{Production}$ and $\sigma$ are in units of fb.}
    \label{tab: bmpoints2}
\end{table}

%\clearpage

%%%%%%%%%%%%%%%%%%%%%%%%%%%%%%%%%%%%%%%%%%%%%%%
\section{Signal analysis and Mass reach of \texorpdfstring{$\Delta^+$}{Delta+}}
\label{sec:analysis}
\begin{figure}[ht!]
\centering
\begin{subfigure}[h]{0.45\textwidth}
  \centering
  \includegraphics[width=1\linewidth]{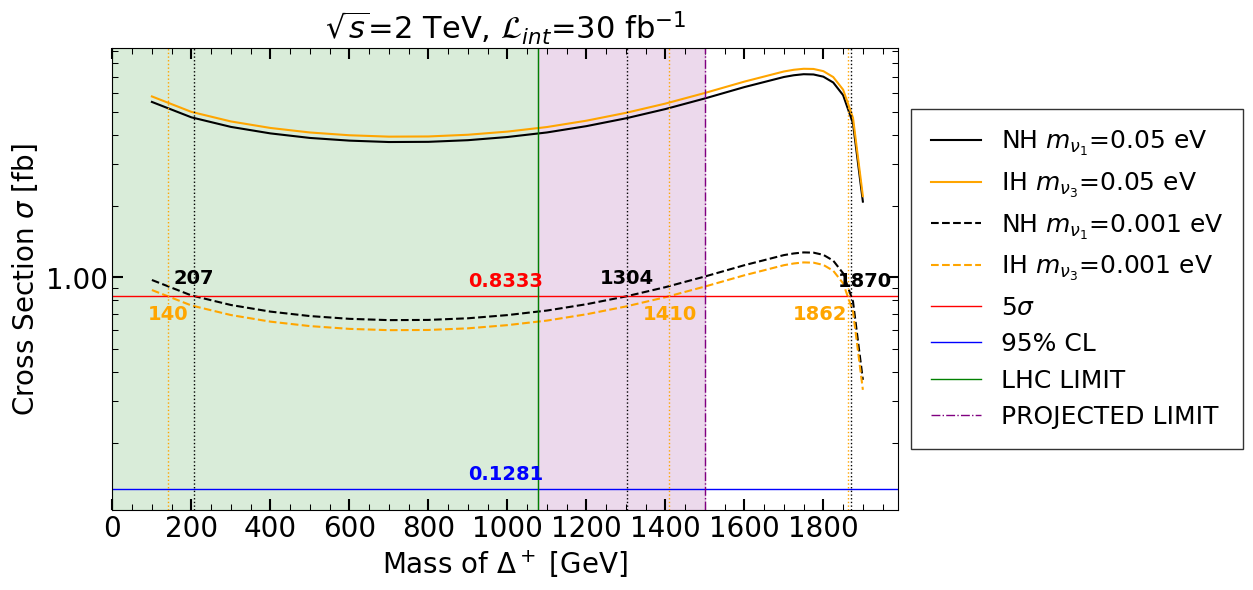}
  \caption{Cross-section}
  \label{fig:xsec}
\end{subfigure}%
\begin{subfigure}[h]{0.45\textwidth}
  \centering
  \includegraphics[width=1\linewidth]{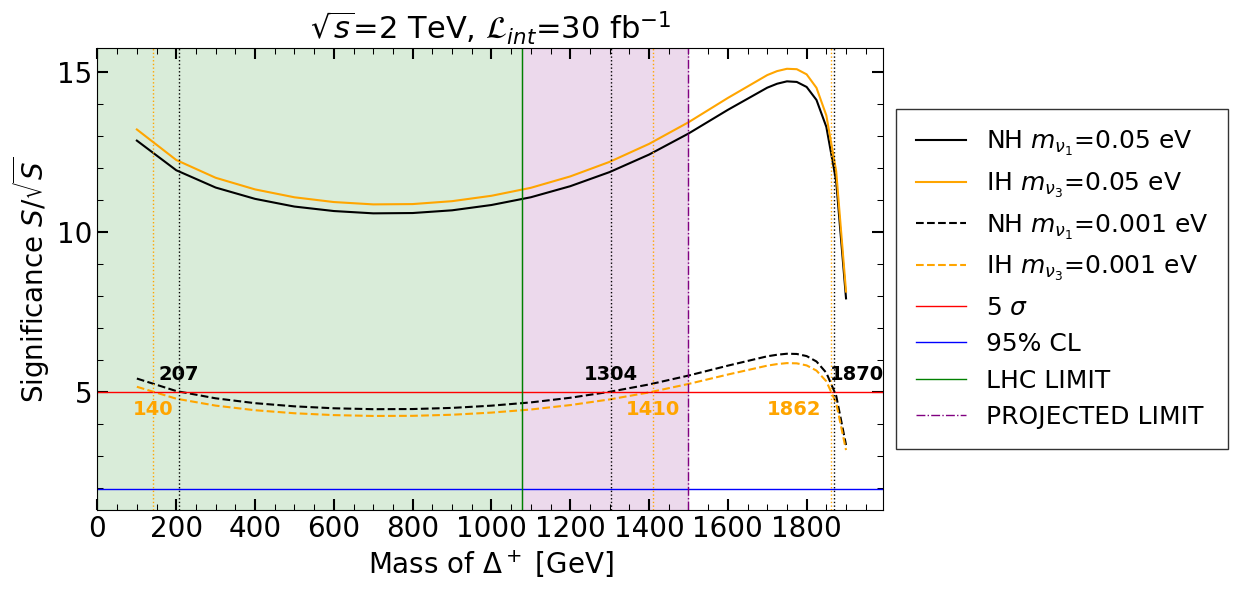}
  \caption{Significance}
  \label{fig:signif}
\end{subfigure}
\vspace*{-0.1in}
\caption{Plot of (a) Cross-section and (b) Significance of the signal process (shown in Fig.~\ref{fig:signal}) vs. $m_{\Delta^+}$ showing the 5$\sigma$ and 95\% CL level by the horizontal red and blue line, respectively.}
\label{fig:xsecsig}
\end{figure}

%%%%%%%%%%%%%%%%%%%%%%%%%%%%%%%%%%%%%%%%%%%%%%%
All benchmark points have been generated using the UFO model~\cite{Fuks:2019clu} and simulated using \textsc{Madgraph}~\cite{Alwall:2011uj} followed by showering in \textsc{Pythia}~\cite{Sjostrand:2014zea} and detector level analysis in \textsc{Delphes}~\cite{deFavereau:2013fsa} using the default Muon collider delphes card. Since the $\mu$TRISTAN collider has not yet been validated by full simulation, therefore, the default Muon collider delphes card is a hybrid between FCC-hh~\cite{Selvaggi:2717698} and the CLIC~\cite{Leogrande:2019qbe} colliders, as stated by the authors of the default Muon collider delphes card. Therefore, the default Muon collider delphes card has the provision of using several different algorithms to cluster and identify the final state particles. Out of these available algorithms, we have used the default Muon collider delphes card with the following specifications: 
\begin{itemize}
    \item The jets are clustered using the Valencia algorithm~\cite{Boronat:2014hva} with $R = 1.5$ and $p_{T}(jet) > 20$~GeV with the requirement that all the hadronic particles are clustered into exactly 2 jets as that is what is expected in the final state as stated by the Feynman diagrams in Fig. 2. We have used this particular setup for identifying and clustering the jets as this setup yielded the invariant mass of the two jets to be closest to mass of a $W^\pm$ boson.
    \item The b-jets are tagged based on the CLICdp-Note-2014-002~\cite{AlipourTehrani:1742993}. Although flavor of the jets do not play any significant role in the analysis since the focus is on the decay of $\Delta^+$ which is forced to decay leptonically. Therefore, using any other b-tagging algorithm would yield similar results.
    \item The $\tau$ lepton is tagged with 80$\%$ efficiency if it has $p_T > 10.0$ GeV, $\eta < 2.5$ and if it lies within $\Delta R = 0.5$ of the jet axis. 
    \item The $e$ lepton is isolated following the criterion that the ratio of the sum of transverse momenta greater than 0.5 GeV of all particles which lie within a cone of $\Delta R = 0.1$ around the $e$ lepton to the transverse momentum of that $e$ lepton is less than 0.2 and is tagged only if the $e$ lepton has $p_T > 3.0$~GeV and $\eta < 2.5$.
\end{itemize}
Since our signature of interest involves lepton number violation and lepton flavor violation, therefore, it is, in principle, free of any relevant SM background. Hence at the ROOT analysis level, no other cuts, beside the detector efficiency at the Delphes level mentioned above, have been employed to extract this rather clean signature. The total cross-section of the full process, as in Fig.~\ref{fig:signal}, is calculated at $\sqrt{s}=2~$TeV and shown in Fig.~\ref{fig:xsec}. Since the process involves lepton flavor violation, it is a noble signature with absolutely zero SM background. Therefore, the significance for Fig.~\ref{fig:signal} can be calculated as 
\begin{center}
    $\text{significance} = S/\sqrt{S}$,\footnote{For simplicity, Gaussian distribution has been used here to calculate the significance. In experimental analyses, Poisson statistics should be used to compute the statistical significance more accurately, particularly in regimes with limited background events. At the limit B $\longrightarrow$ 0, Poisson statistics will yield the significance $\sim \sqrt{2S\text{ln}(1/B)}$, where B is the number of background events and S is the number of signal events. Although theoretically, one would expect absolutely no background events for the signature considered here, in practice, a non-zero background will arise experimentally due to uncertainties. Such a background can then be incorporated into this expression to obtain a finite and physically meaningful significance. This approach generally yields a higher significance than that obtained here using Gaussian approximations. Consequently, our calculation is conservative: an experimental determination of the significance is expected to be equal to or greater than our result, but not smaller.}
\end{center}
where $S$ is the number of signal events which is calculated as 
\begin{equation}
    S= \sigma \times \mathcal{L}_{int},
\end{equation}
where $\sigma$ is the cross-section, which is shown in Fig.~\ref{fig:xsec} and $\mathcal{L}_{int}$ is the integrated Luminosity which has been considered to be 30 fb$^{-1}$. 

The significance for all the benchmark points have been shown in Fig.~\ref{fig:signif}. In Fig.~\ref{fig:xsec} and Fig.~\ref{fig:signif}, the solid curves are for the points with mass of lightest neutrino = 0.05 eV and the dashed curves are for the points with mass of lightest neutrino = 0.001 eV. The black curves denote Normal Hierarchy and the orange curves denote Inverted Hierarchy. The solid green (dashed purple) vertical line and shaded region hold the same meaning as in Fig.~\ref{fig:prodxsec}. The horizontal red and blue lines in Fig.~\ref{fig:xsec} and Fig.~\ref{fig:signif} denote significance of 5$\sigma$ and 95$\%$ CL, respectively. As can be noted from Fig.~\ref{fig:xsecsig}, all the points for $m_{\nu_{lightest}}$ = 0.05 eV and for both Normal and Inverted Hierarchy have significance above 5$\sigma$ while for $m_{\nu_{lightest}}$ = 0.001 eV, all the points have significance above 95$\%$ CL but not all of then have significance above 5$\sigma$. For Normal hierarchy points with $m_{\nu_{lightest}}$ = 0.001 eV and $m_{\Delta^+} \lesssim 207~$ GeV and  $1304 \lesssim m_{\Delta^+} \lesssim 1870~$ GeV have significance above 5$\sigma$ and for Inverted hierarchy points with $m_{\nu_{lightest}}$ = 0.001 eV and $m_{\Delta^+} \lesssim 140~$ GeV and  $1410 \lesssim m_{\Delta^+} \lesssim 1862~$ GeV have significance above 5$\sigma$, as evident from Fig.~\ref{fig:xsecsig}. However, owing to the current(future) LHC limits on $m_{\Delta^{++}}$, one must concentrate on the signal significance in the parameter space only to the right of the solid green(dashed purple) vertical line. One should note that apart from slight difference in the values of cross-section and significance in between Normal and Inverted Hierarchy, there is not much difference between these two cases. The shapes of the curves, as shown in Fig.~\ref{fig:xsecsig}, are same for both Normal and Inverted Hierarchy. Therefore, using these curves one cannot identify whether the underlying mechanism is Normal or Inverted Hierarchy. However, the flavors in the final-state leptons will vary a lot for these two cases and hence can be used to differentiate between Normal and Inverted Hierarchy, as discussed in detail in the next Section.

%%%%%%%%%%%%%%%%%%%%%%%%%%%%%%%%%%%%%%%%%%%%%%%
\section{Differentiating Normal Hierarchy from Inverted Hierarchy}
\label{sec:diff}
\begin{figure}[ht!]
\centering
\begin{subfigure}[h]{0.45\textwidth}
  \centering
  \includegraphics[width=1\linewidth]{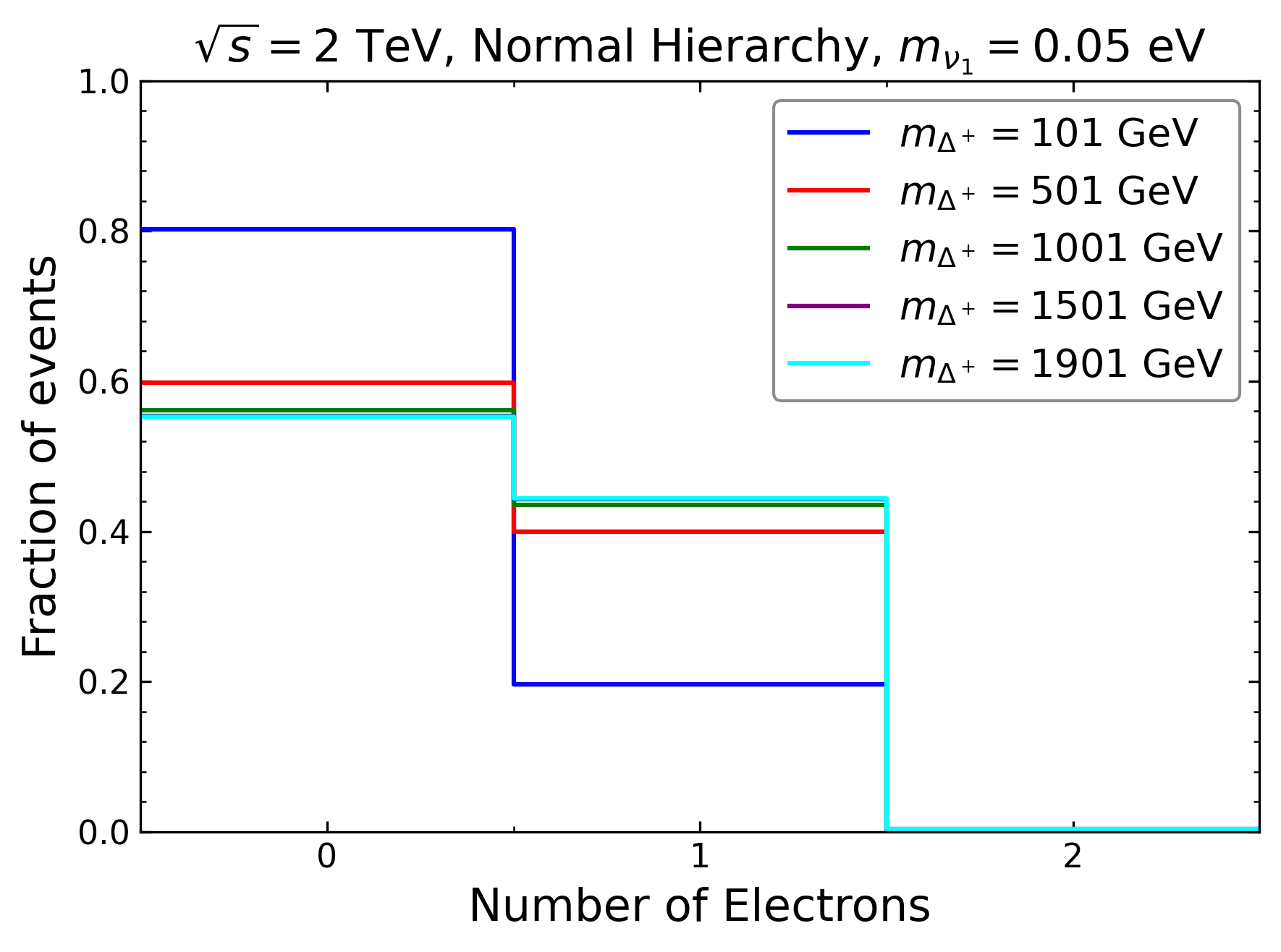}
  \caption{}
  \label{fig:bm1norme}
\end{subfigure}%
\begin{subfigure}[h]{0.45\textwidth}
  \centering
  \includegraphics[width=1\linewidth]{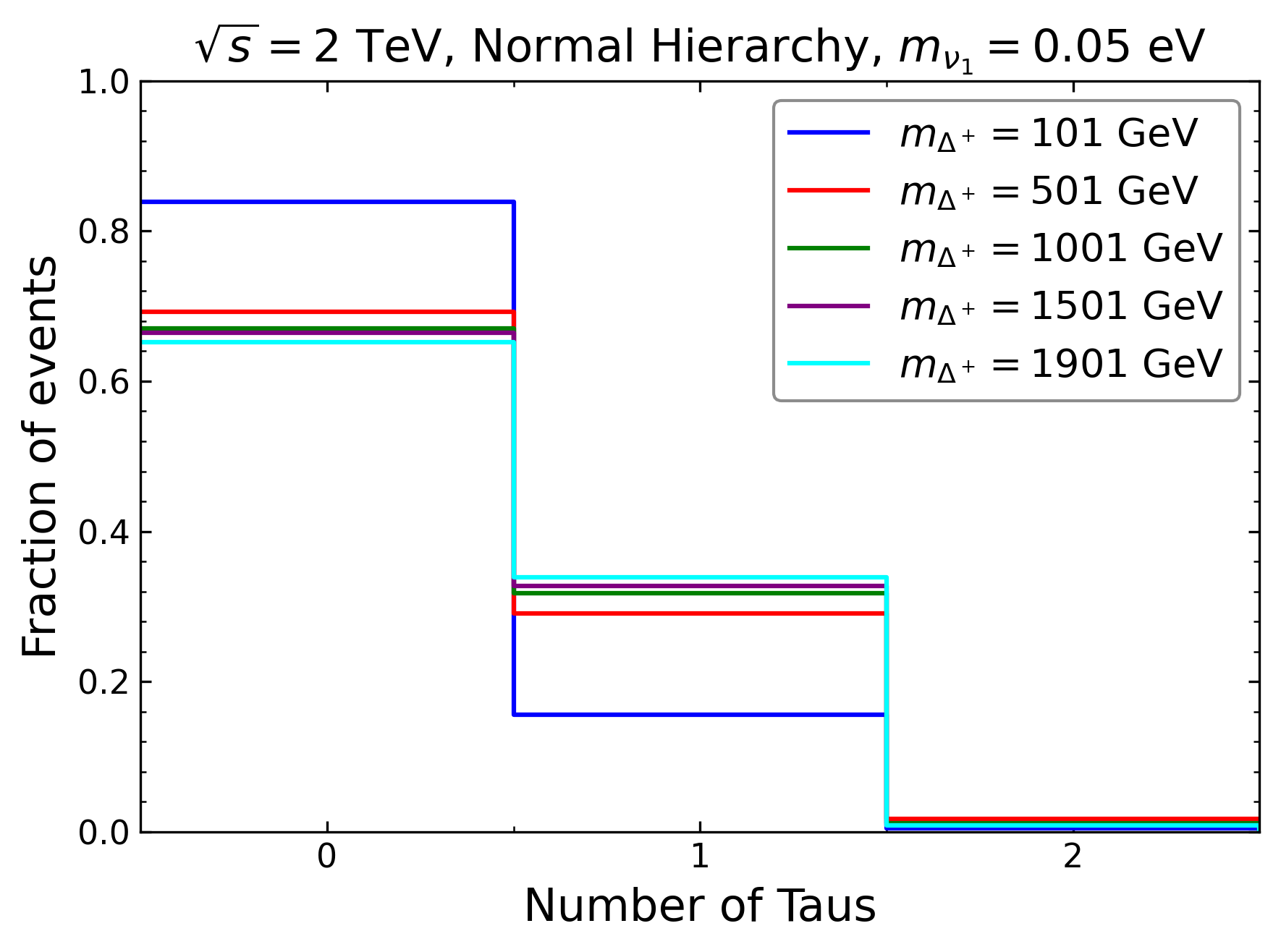}
  \caption{}
  \label{fig:m1normt}
\end{subfigure}
\vspace*{-0.1in}
\caption{Distribution for a) Electrons in the final state and b) Taus in the final state for $m_{\nu_1}=0.05$ eV assuming Normal Hierarchy}
 \label{fig:bm1normet}
\end{figure}

\begin{figure}[ht!]
\centering
\begin{subfigure}[h]{0.45\textwidth}
    \centering
    \includegraphics[width=1\linewidth]{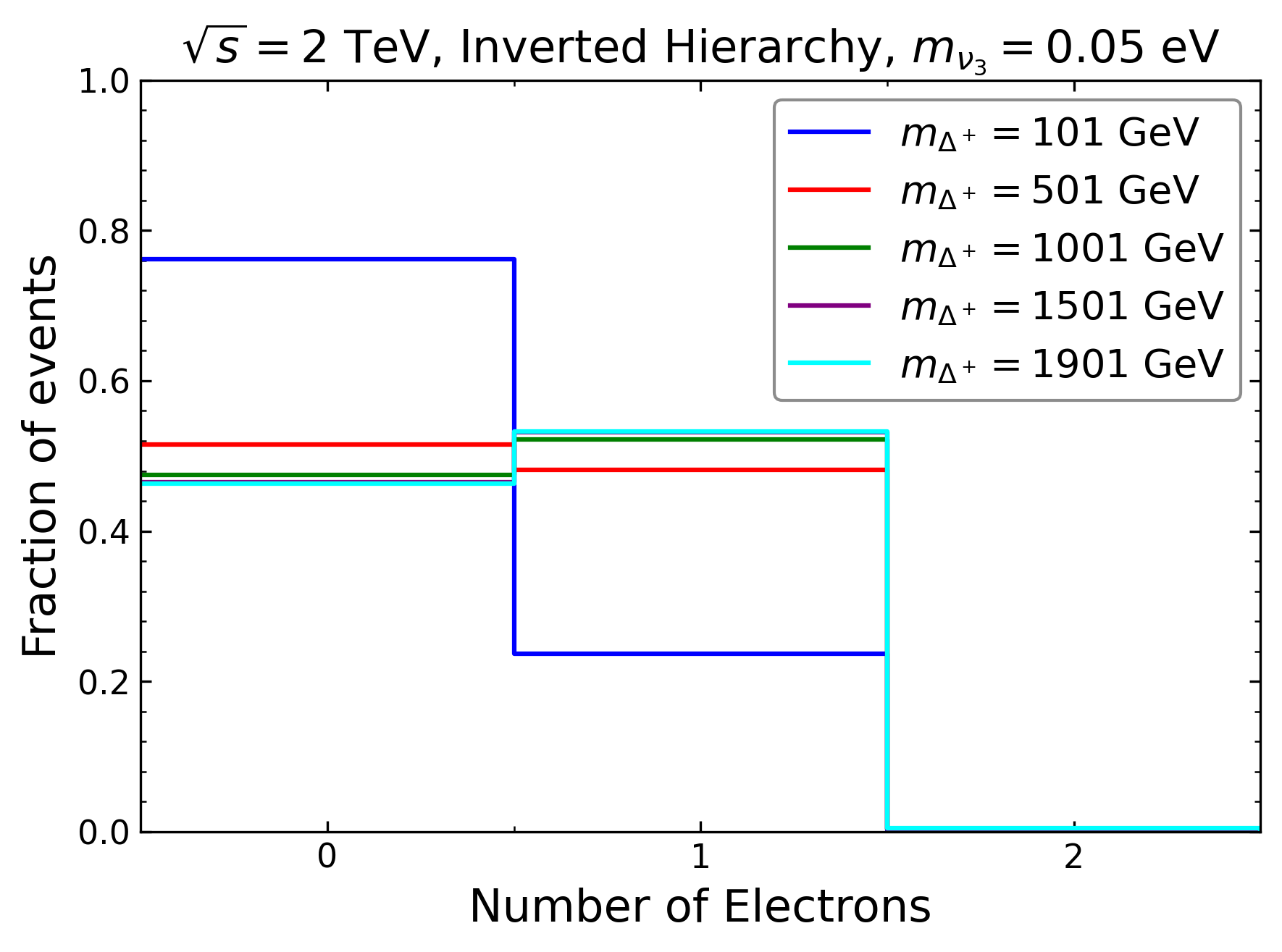}
    \caption{}
    \label{fig:bm1inve}
\end{subfigure}
\begin{subfigure}[h]{0.45\textwidth}
  \centering
  \includegraphics[width=1\linewidth]{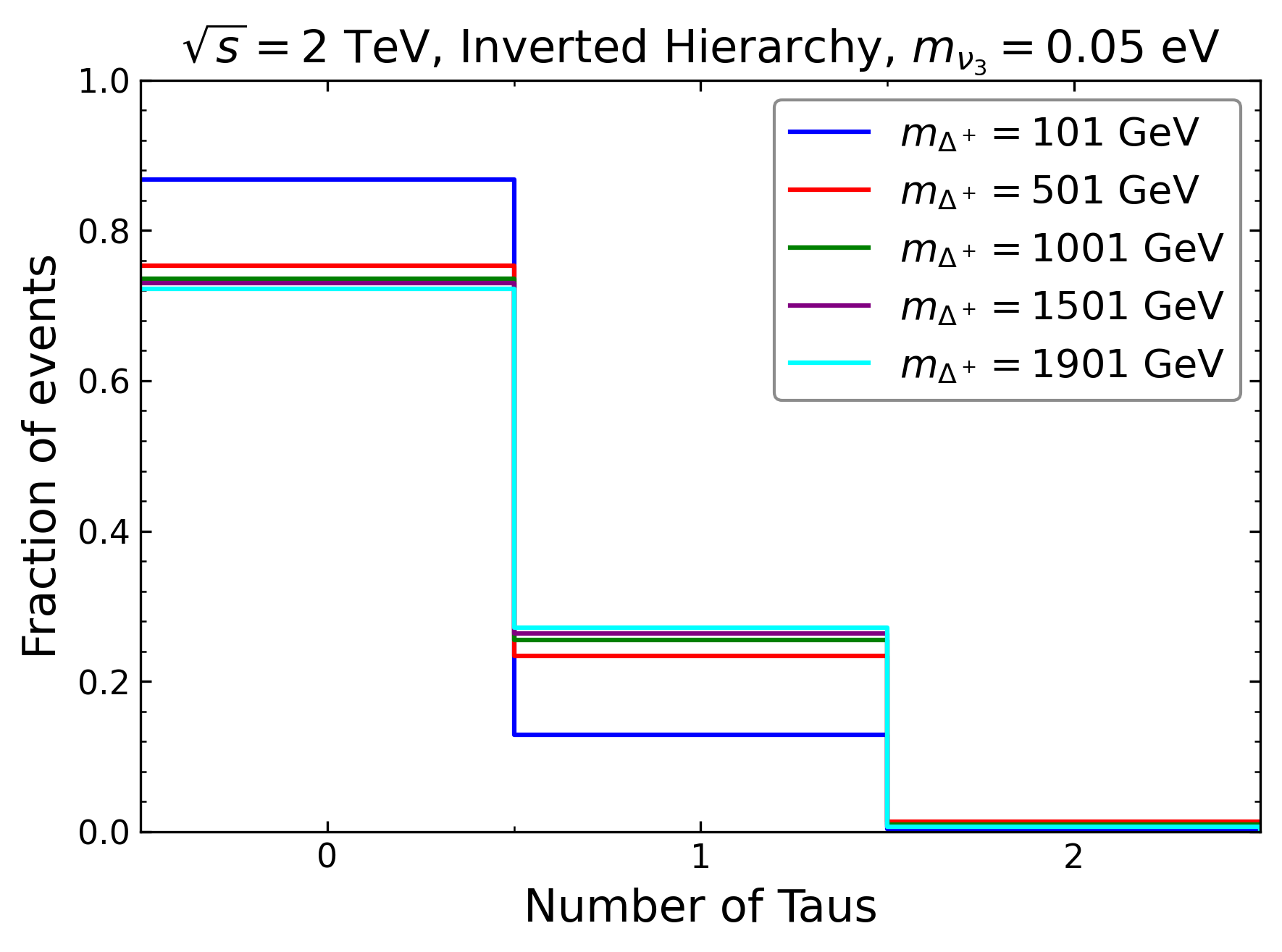}
  \caption{}
  \label{fig:bm1invt}
\end{subfigure}
\vspace*{-0.1in}
\caption{Distribution for a) Electrons in the final state and b) Taus in the final state for $m_{\nu_3}=0.05$ eV assuming Inverted Hierarchy}
\label{fig:bm1invet}   
\end{figure}

\begin{figure}[ht!]
\centering
\begin{subfigure}[h]{0.45\textwidth}
  \centering
  \includegraphics[width=1\linewidth]{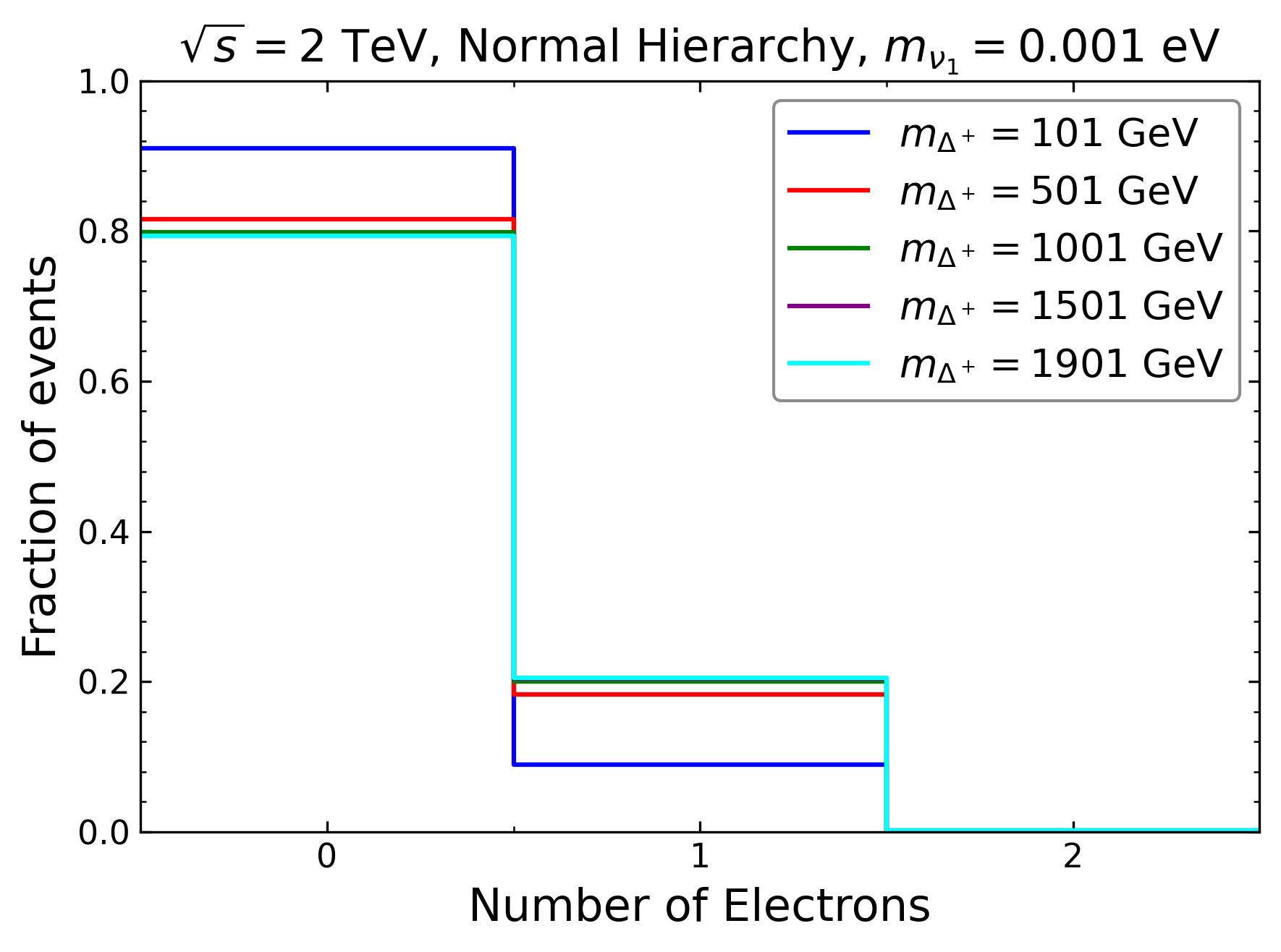}
  \caption{}
  \label{fig:bm2norme}
\end{subfigure}%
\begin{subfigure}[h]{0.45\textwidth}
  \centering
  \includegraphics[width=1\linewidth]{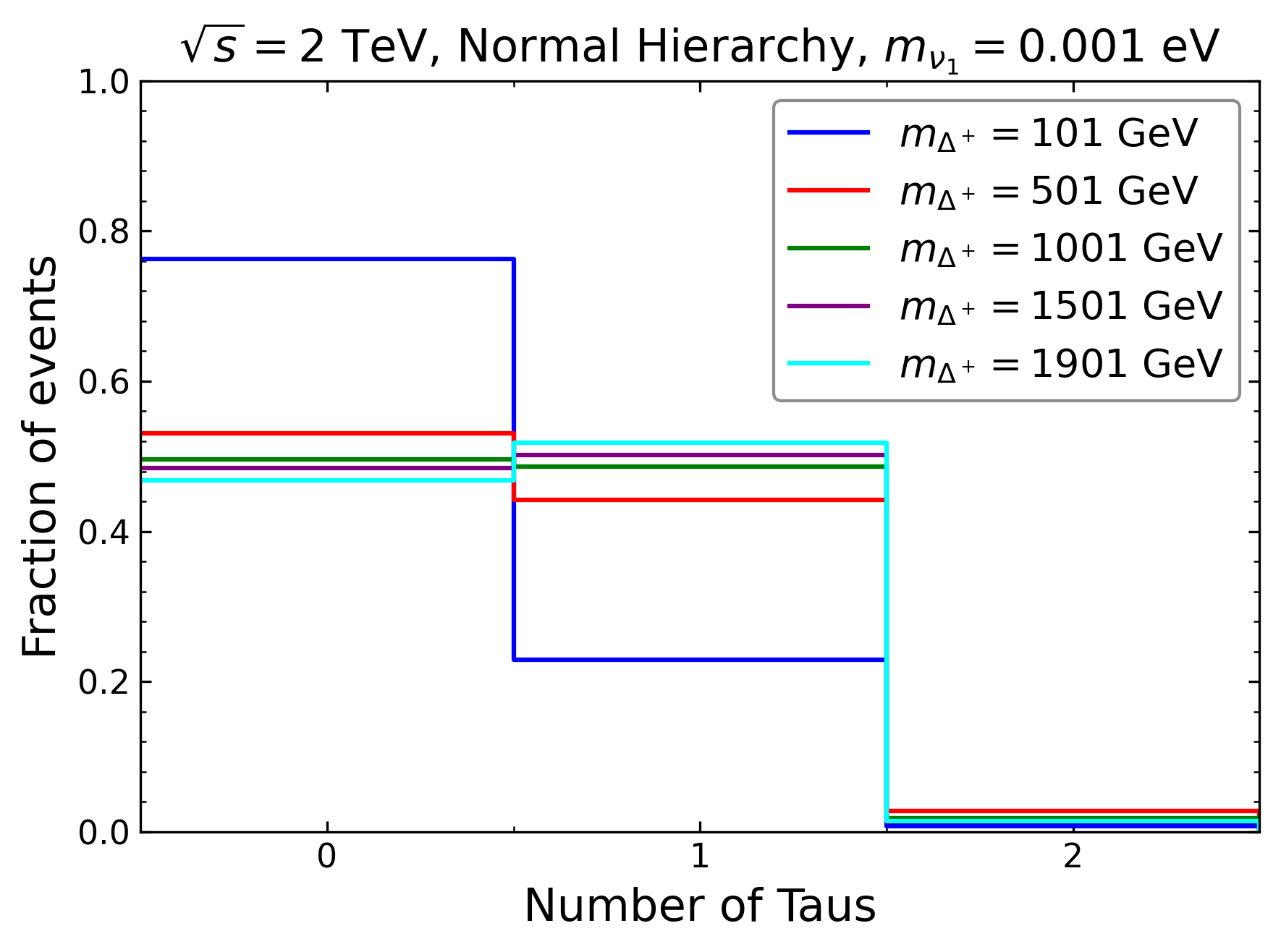}
  \caption{}
  \label{fig:bm2normt}
\end{subfigure}
\vspace*{-0.1in}
\caption{Distribution for a) Electrons in the final state and b) Taus in the final state for $m_{\nu_1}=0.001$ eV assuming Normal Hierarchy}
 \label{fig:bm2normet}
\end{figure}

\begin{figure}[ht!]
\centering
\begin{subfigure}[h]{0.45\textwidth}
    \centering
    \includegraphics[width=1\linewidth]{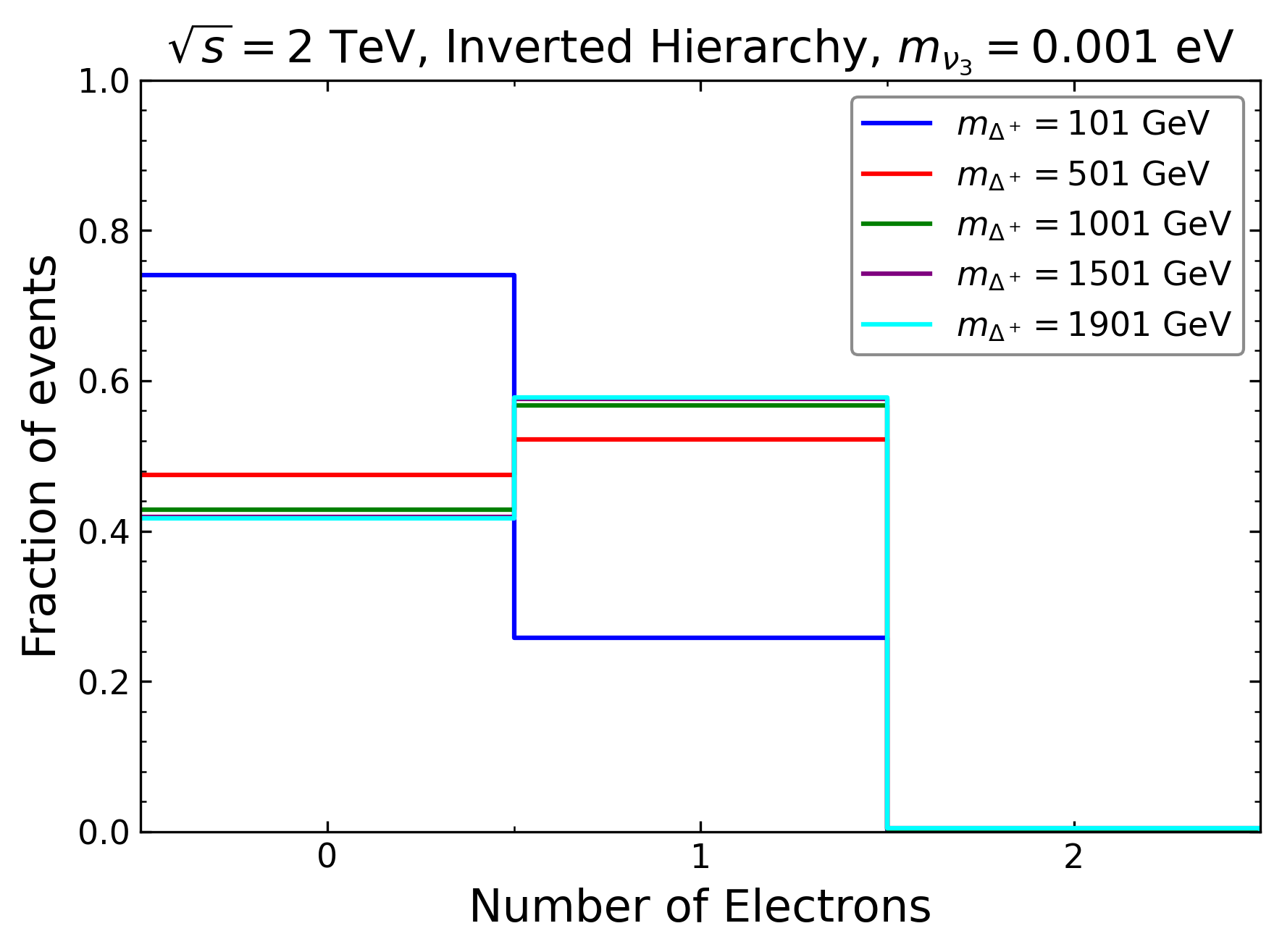}
    \caption{}
    \label{fig:bm2inve}
\end{subfigure}
\begin{subfigure}[h]{0.45\textwidth}
  \centering
  \includegraphics[width=1\linewidth]{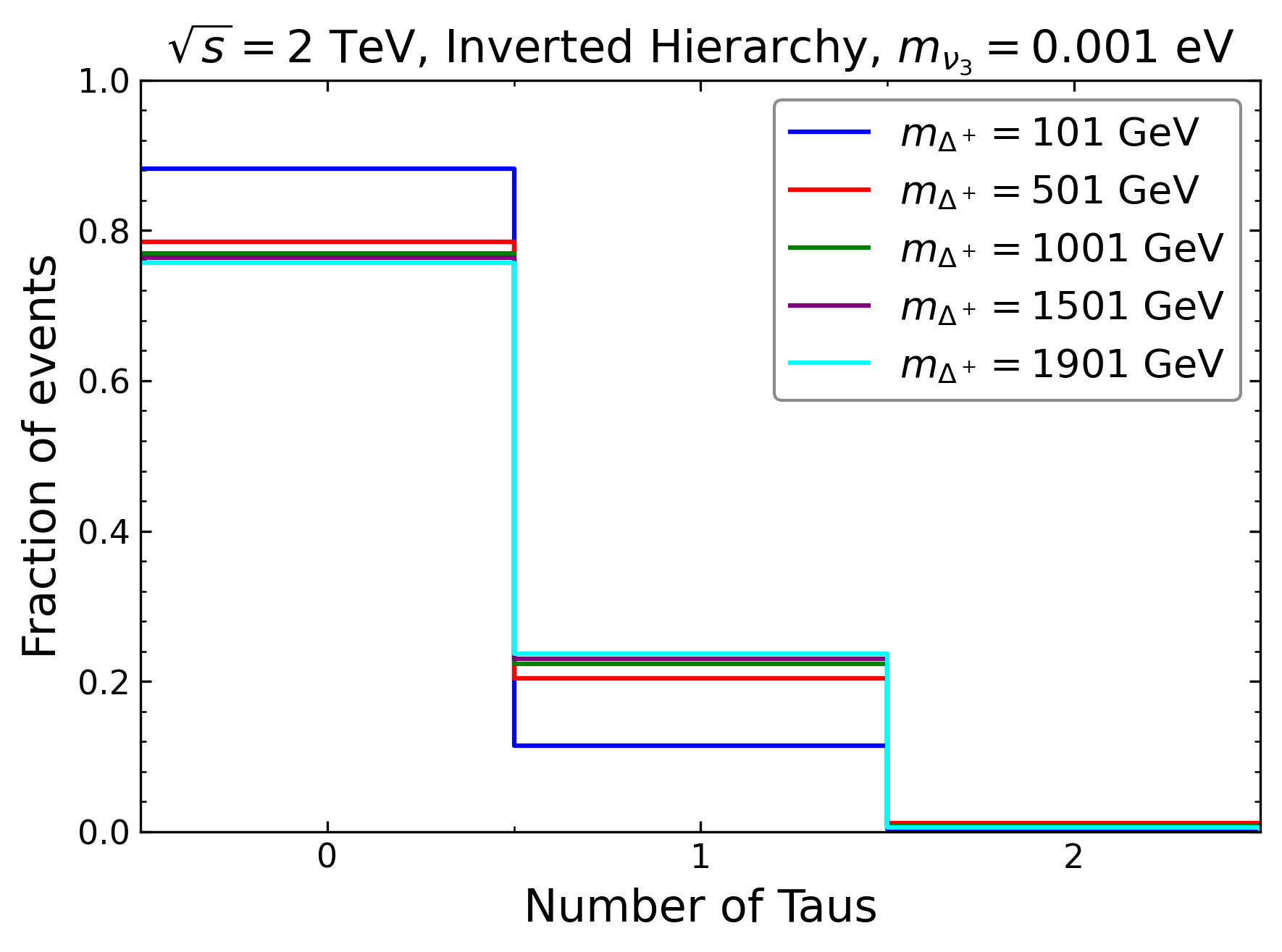}
  \caption{}
  \label{fig:bm2invt}
\end{subfigure}
\vspace*{-0.1in}
\caption{Distribution for a) Electrons in the final state and b) Taus in the final state for $m_{\nu_3}=0.001$ eV assuming Inverted Hierarchy}
\label{fig:bm2invet}   
\end{figure}

%%%%%%%%%%%%%%%%%%%%%%%%%%%%%%%%%%%%%%%%%%%%%%%
As discussed in Sec.~\ref{sec:analysis}, it is possible to obtain a significant amount of cross-section for our signature of interest for both Normal and Inverted Hierarchies in the $\mu^+ \mu^+$ collider, enough to confirm discovery of a singly-charged scalar $\Delta^+$. However, the results obtained in Fig.~\ref{fig:xsecsig} are not sufficient to identify whether the underlying theory respects Normal Hierarchy or Inverted Hierarchy. This important fact can be ensured by the distribution of final state Electrons and Taus originating from the decay of $\Delta^+$. 
\\
The distribution of Electrons and Taus in the final state for Normal and Inverted Hierarchy are shown in Fig.~\ref{fig:bm1normet} and Fig.~\ref{fig:bm1invet}, respectively for benchmark points with the lightest neutrino mass $m_{\nu_{lightest}}=0.05$ eV and $m_{\Delta^+} = $101, 501, 1001, 1501, 1901 GeV. Similar plots for benchmark points with the lightest neutrino mass $m_{\nu_{lightest}}=0.001$ eV and $m_{\Delta^+} = $101, 501, 1001, 1501, 1901 GeV are shown in Fig.~\ref{fig:bm2normet} and Fig.~\ref{fig:bm2invet} for Normal and Inverted Hierarchy, respectively. In this section, the primary goal is to present the nature of the change in the distribution of Electrons and Taus depending on $m_{\Delta^+}$, $m_{\nu_{lightest}}$ and the underlying Hierarchy, albeit some of the chosen benchmark points have been already excluded by the LHC. Therefore, one can easily predict the distribution of Electrons and Taus for Normal and Inverted hierarchies for all of the allowed benchmark points. Although the signature of interest should ideally yield exactly one charged lepton (either $e^+$ or $\tau^+$), there are many events with no $e^+$ or $\tau^+$ in the final state as evident in Fig.~\ref{fig:bm1normet} - Fig.~\ref{fig:bm2invet}. This is because for a huge number of events the final state charged lepton escapes detection for lying outside the range of pseudorapidity that the detector can cover and hence is highly dependent on the detector design. However, this situation improves as $m_{\Delta^+}$ increases and more and more events yield detectable charged lepton in the final state. One might notice there are some events which have 2 charged Electrons/taus in the final state but these events arise due to misidentification of a jet as a lepton. However, one must note that these events are negligibly small in number.

%%%%%%%%%%%%%%%%%%%%%%%%%%%%%%%%%%%%%%%%%%%%
%\textbf{ FOR $m_{\nu_{lightest}}=0.05$ eV}
%%%%%%%%%%%%%%%%%%%%%%%%%%%%%%%%%%%%%%%%%%%%

%%%%%%%%%%%%%%%%%%%%%%%%%%%%%%%%%%%%%%%%%%%%
%\textbf{ FOR $m_{\nu_{lightest}}=0.001$ eV}
%%%%%%%%%%%%%%%%%%%%%%%%%%%%%%%%%%%%%%%%%%%%
%%%%%%%%%%%%%%%%%%%%%%%%%%%%%%%%%%%%%%%%%%%%
%\textbf{ FOR $m_{\nu_{lightest}}=0.05$ eV}
%%%%%%%%%%%%%%%%%%%%%%%%%%%%%%%%%%%%%%%%%%%%
\begin{figure}[ht!]
\centering
\begin{subfigure}[h]{0.45\textwidth}
    \centering
    \includegraphics[width=1\linewidth]{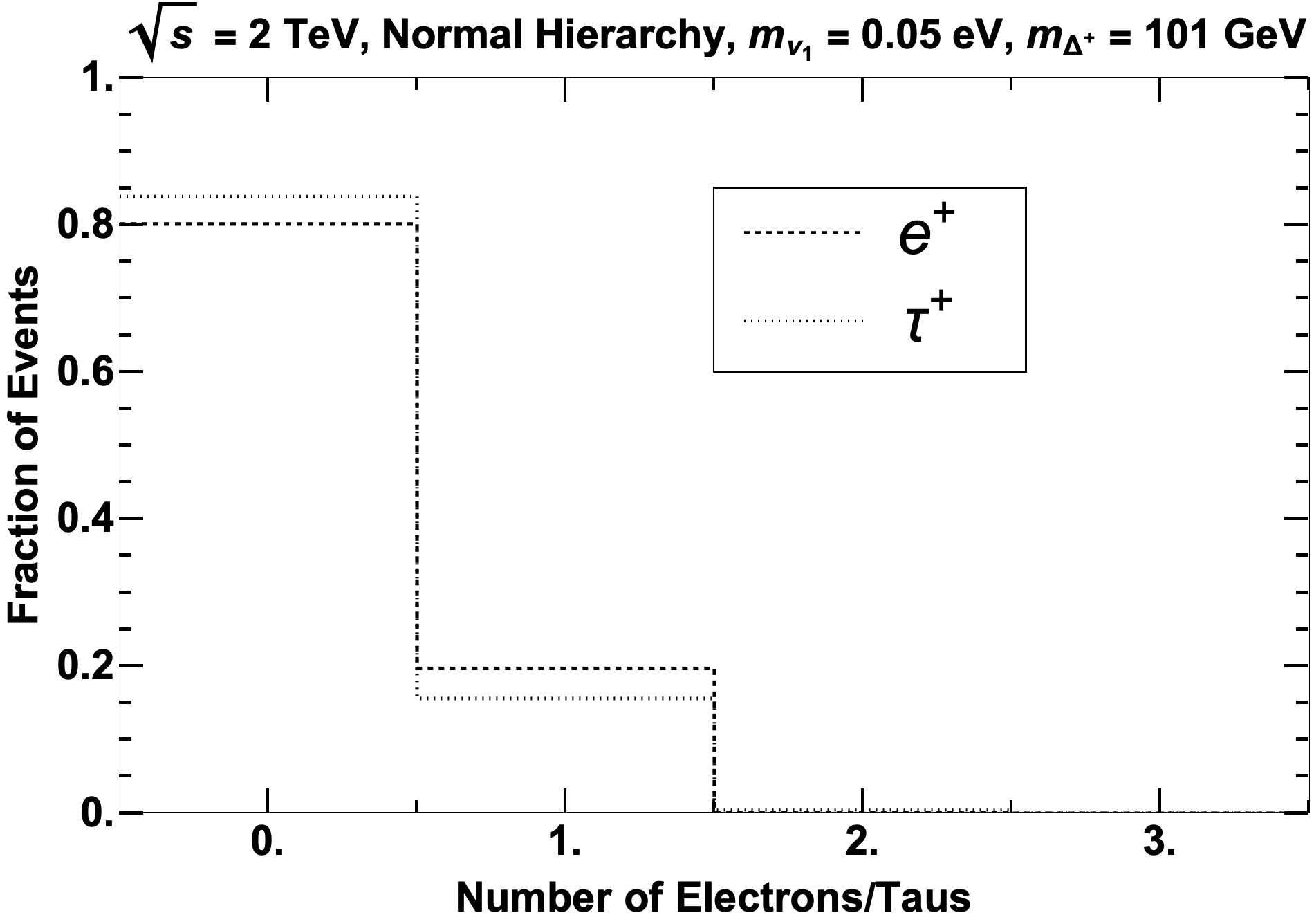}
    \caption{Normal Hierarchy}
    \label{fig:ebm1a}
\end{subfigure}
\begin{subfigure}[h]{0.45\textwidth}
  \centering
  \includegraphics[width=1\linewidth]{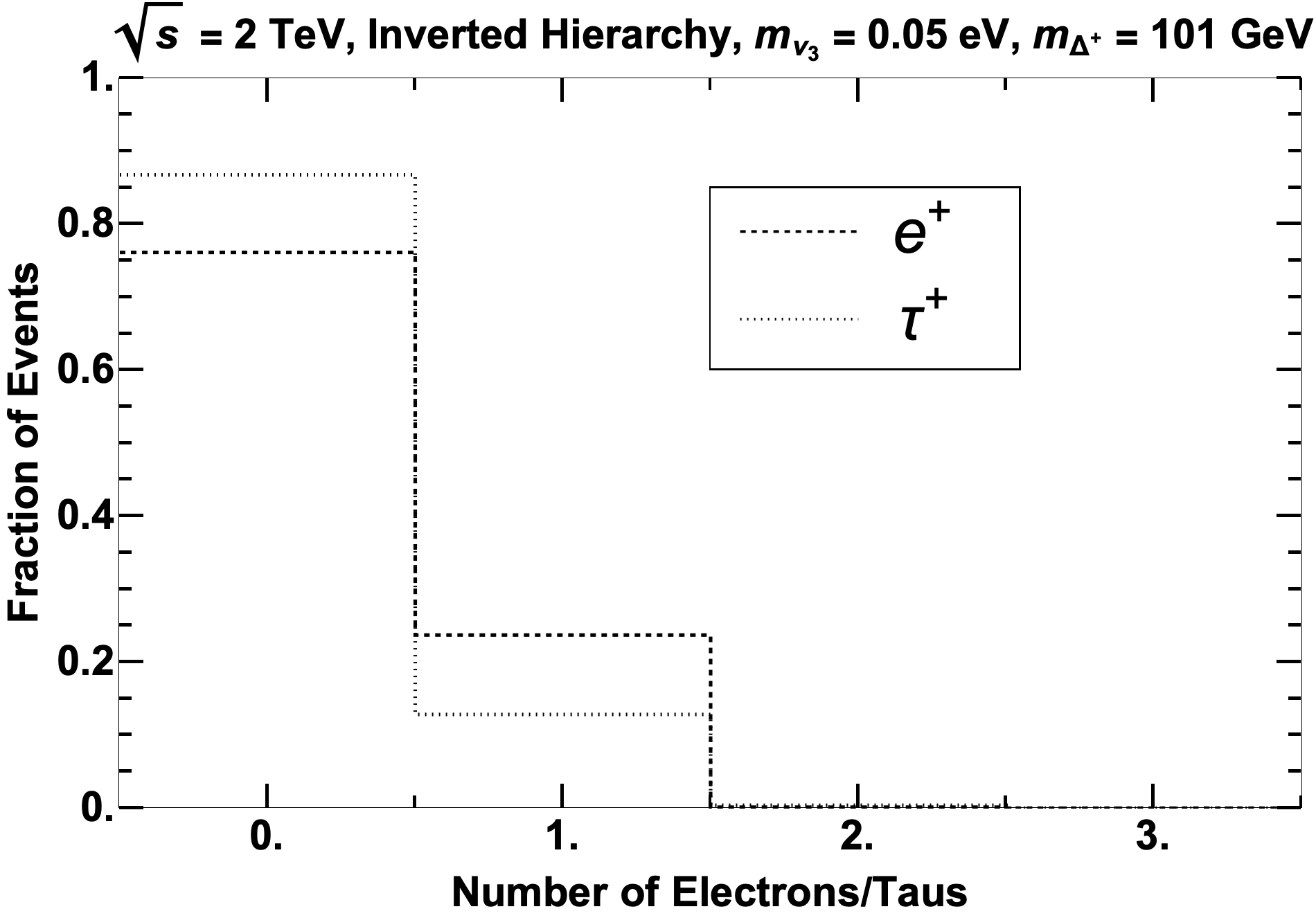}
  \caption{Inverted Hierarchy}
  \label{fig:tbm1a}
\end{subfigure}
\vspace*{-0.1in}
\caption{Distributions of Electrons and Taus for $m_{\nu_{lightest}}=0.05$ eV and $m_{\Delta^+} = 101$ GeV assuming a) Normal Hierarchy and b) Inverted Hierarchy}
\label{fig:etbm1a}   
\end{figure}

\begin{figure}[ht!]
\centering
\begin{subfigure}[h]{0.45\textwidth}
    \centering
    \includegraphics[width=1\linewidth]{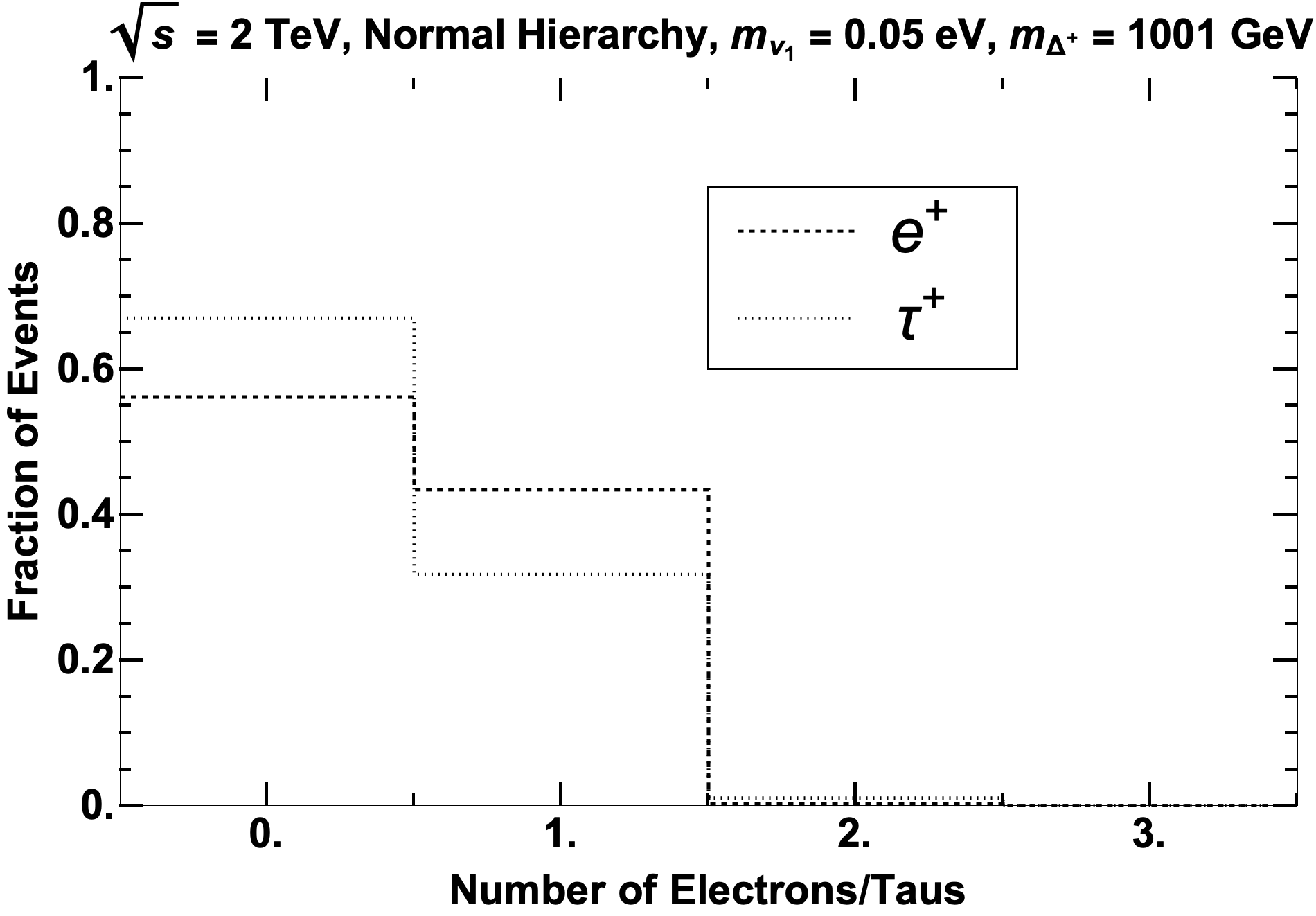}
    \caption{Normal Hierarchy}
    \label{fig:ebm1b}
\end{subfigure}
\begin{subfigure}[h]{0.45\textwidth}
  \centering
  \includegraphics[width=1\linewidth]{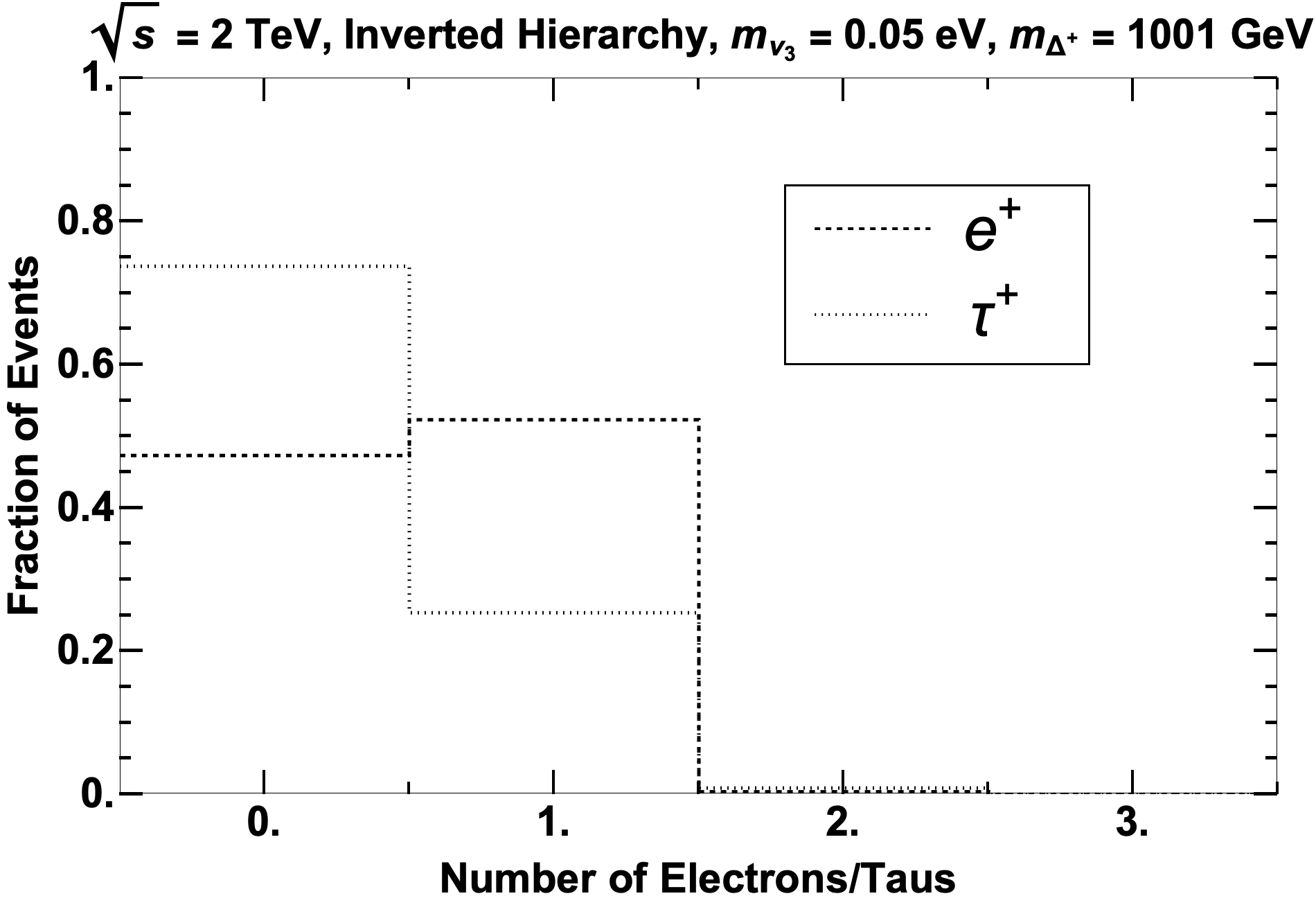}
  \caption{Inverted Hierarchy}
  \label{fig:tbm1b}
\end{subfigure}
\vspace*{-0.1in}
\caption{Distributions of Electrons and Taus for $m_{\nu_{lightest}}=0.05$ eV and $m_{\Delta^+} = 1001$ GeV assuming a) Normal Hierarchy and b) Inverted Hierarchy}
\label{fig:etbm1b}   
\end{figure}

\begin{figure}[ht!]
\centering
\begin{subfigure}[h]{0.45\textwidth}
    \centering
    \includegraphics[width=1\linewidth]{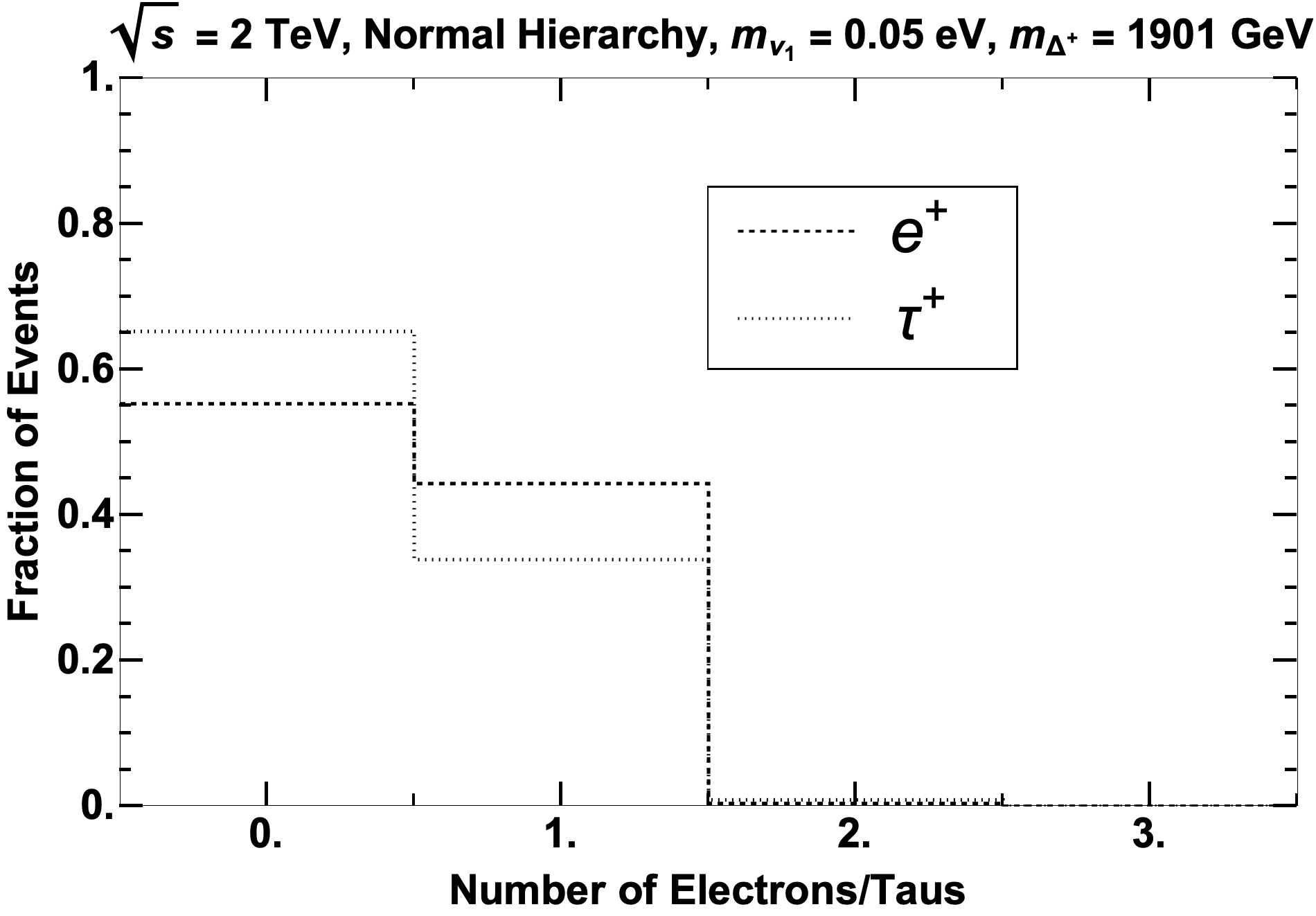}
    \caption{Normal Hierarchy}
    \label{fig:ebm1c}
\end{subfigure}
\begin{subfigure}[h]{0.45\textwidth}
  \centering
  \includegraphics[width=1\linewidth]{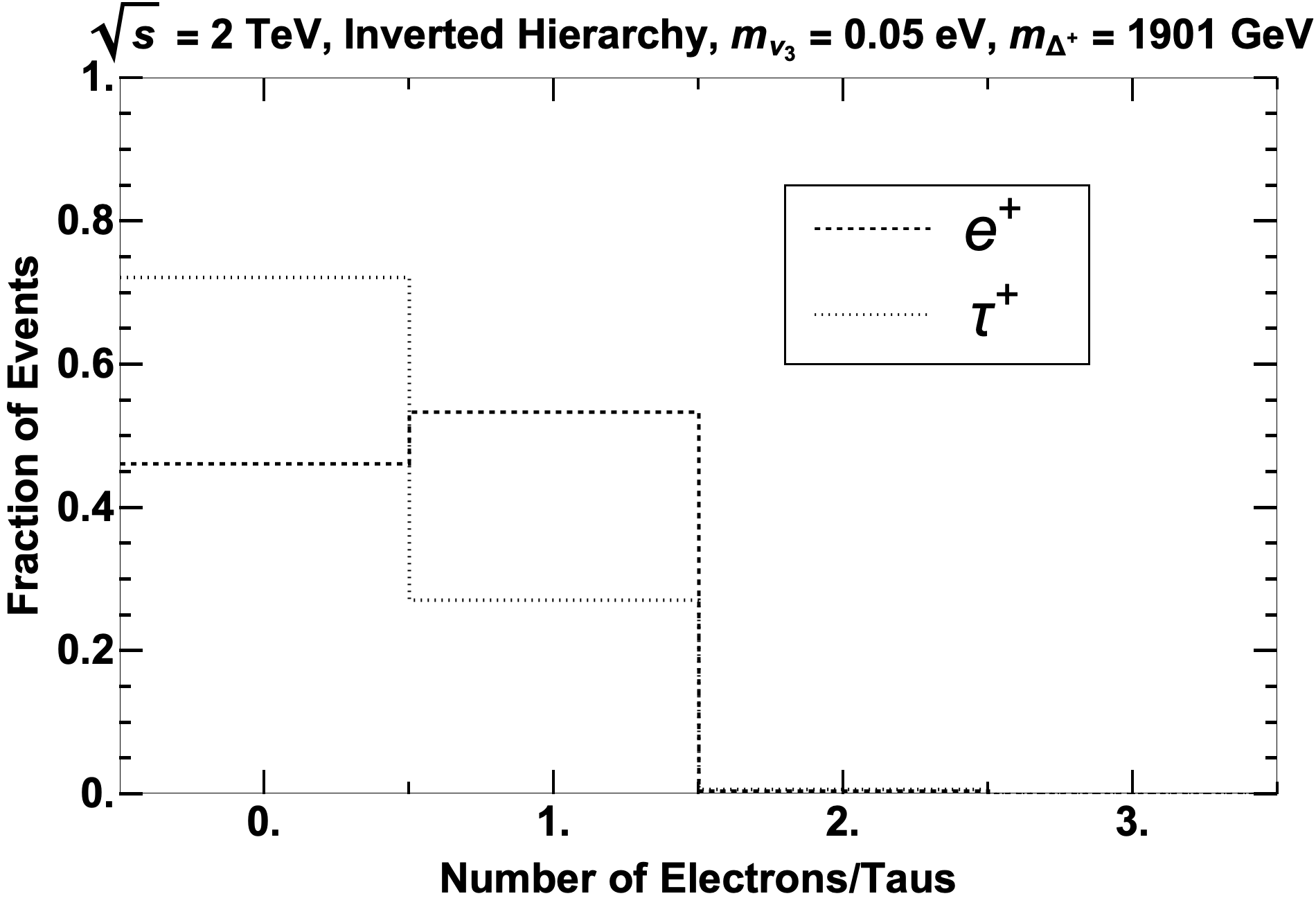}
  \caption{Inverted Hierarchy}
  \label{fig:tbm1c}
\end{subfigure}
\vspace*{-0.1in}
\caption{Distributions of Electrons and Taus for $m_{\nu_{lightest}}=0.05$ eV and $m_{\Delta^+} = 1901$ GeV assuming a) Normal Hierarchy and b) Inverted Hierarchy}
\label{fig:etbm1c}   
\end{figure}
\begin{figure}[ht!]
\centering
\begin{subfigure}[h]{0.45\textwidth}
    \centering
    \includegraphics[width=1\linewidth]{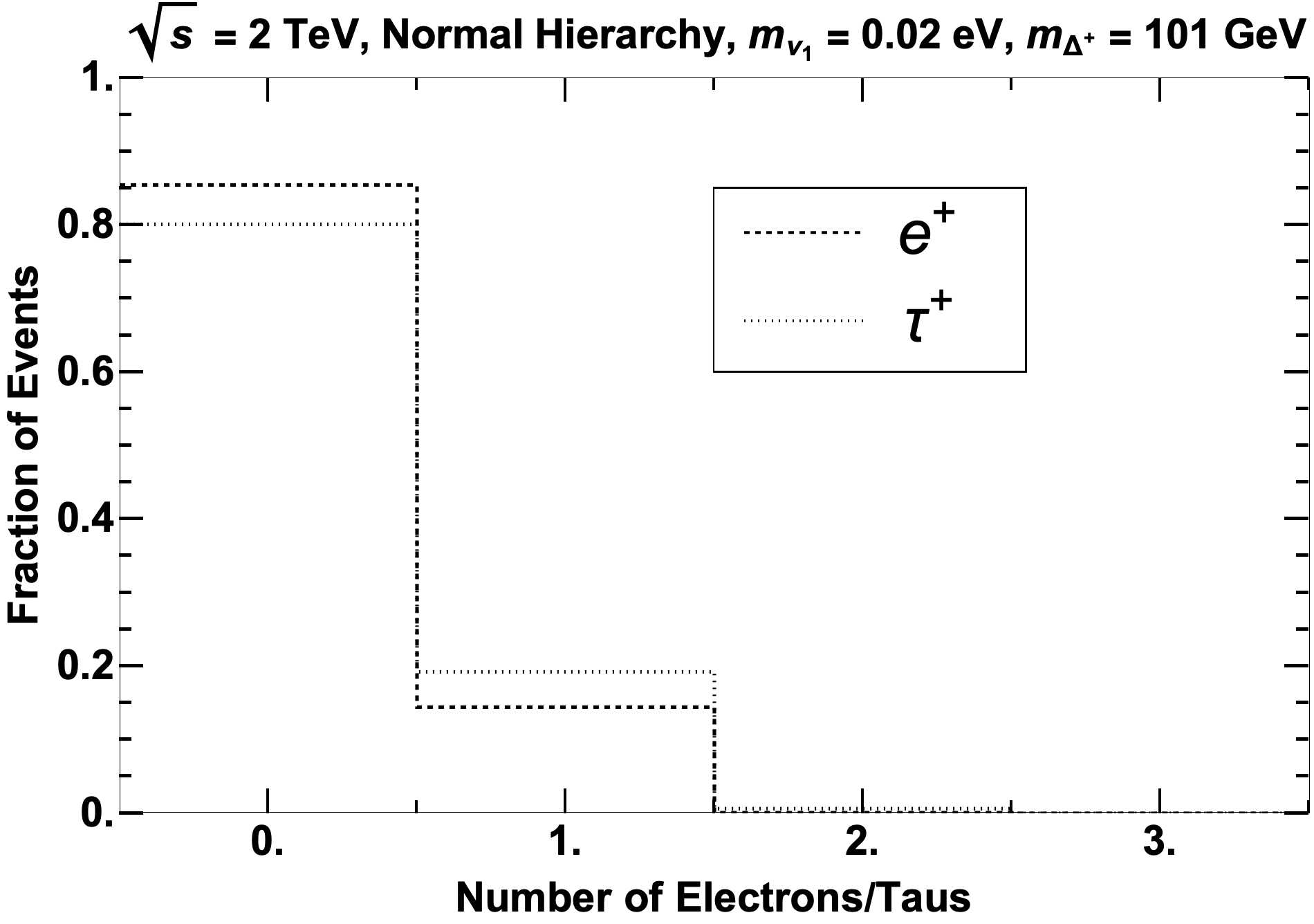}
    \caption{Normal Hierarchy}
    \label{fig:ebm3a}
\end{subfigure}
\begin{subfigure}[h]{0.45\textwidth}
  \centering
  \includegraphics[width=1\linewidth]{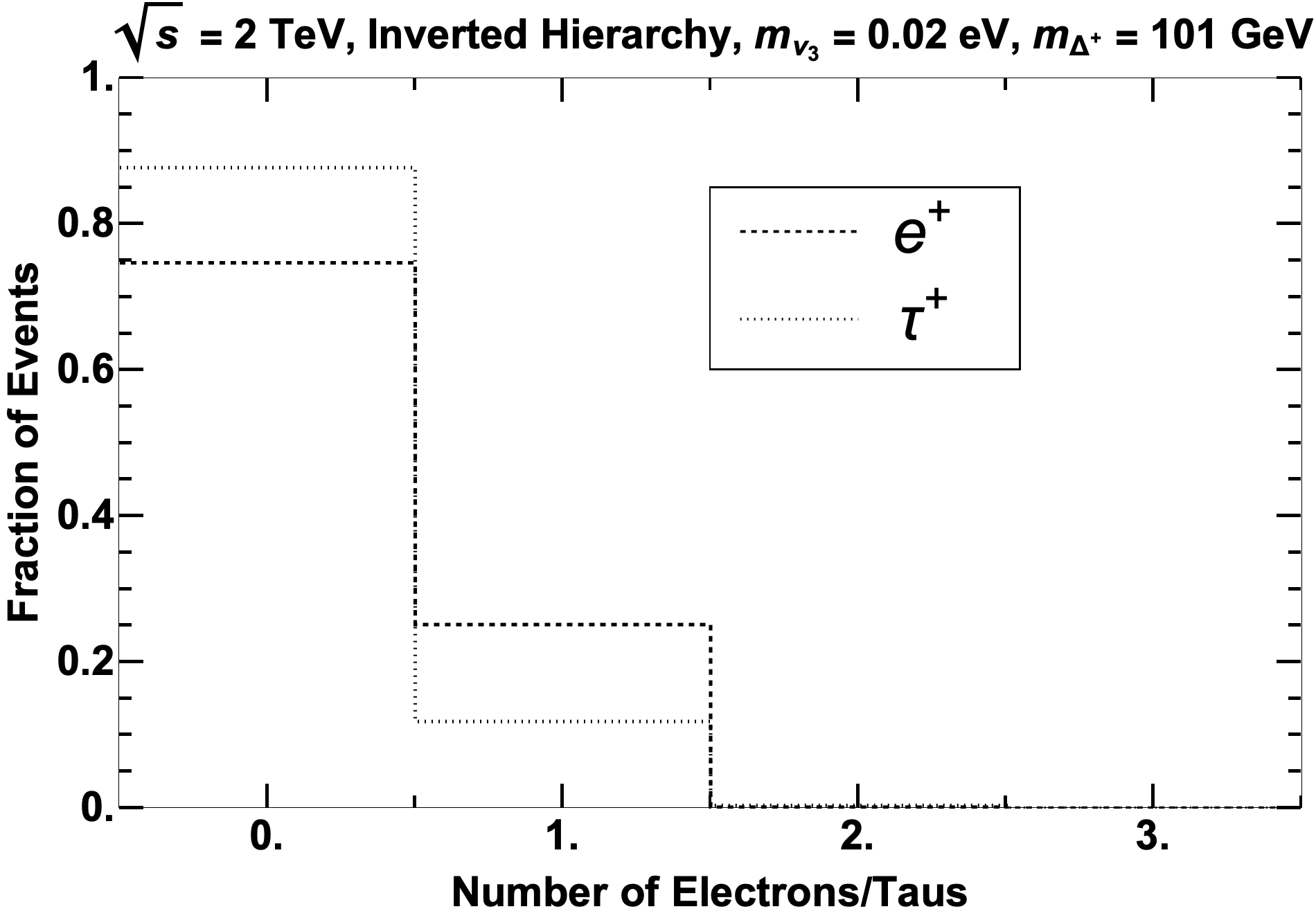}
  \caption{Inverted Hierarchy}
  \label{fig:tbm3a}
\end{subfigure}
\vspace*{-0.1in}
\caption{Distributions of Electrons and Taus for $m_{\nu_{lightest}}=0.02$ eV and $m_{\Delta^+} = 101$ GeV assuming a) Normal Hierarchy and b) Inverted Hierarchy}
\label{fig:etbm3a}   
\end{figure}
\begin{figure}[ht!]
\centering
\begin{subfigure}[h]{0.45\textwidth}
    \centering
    \includegraphics[width=1\linewidth]{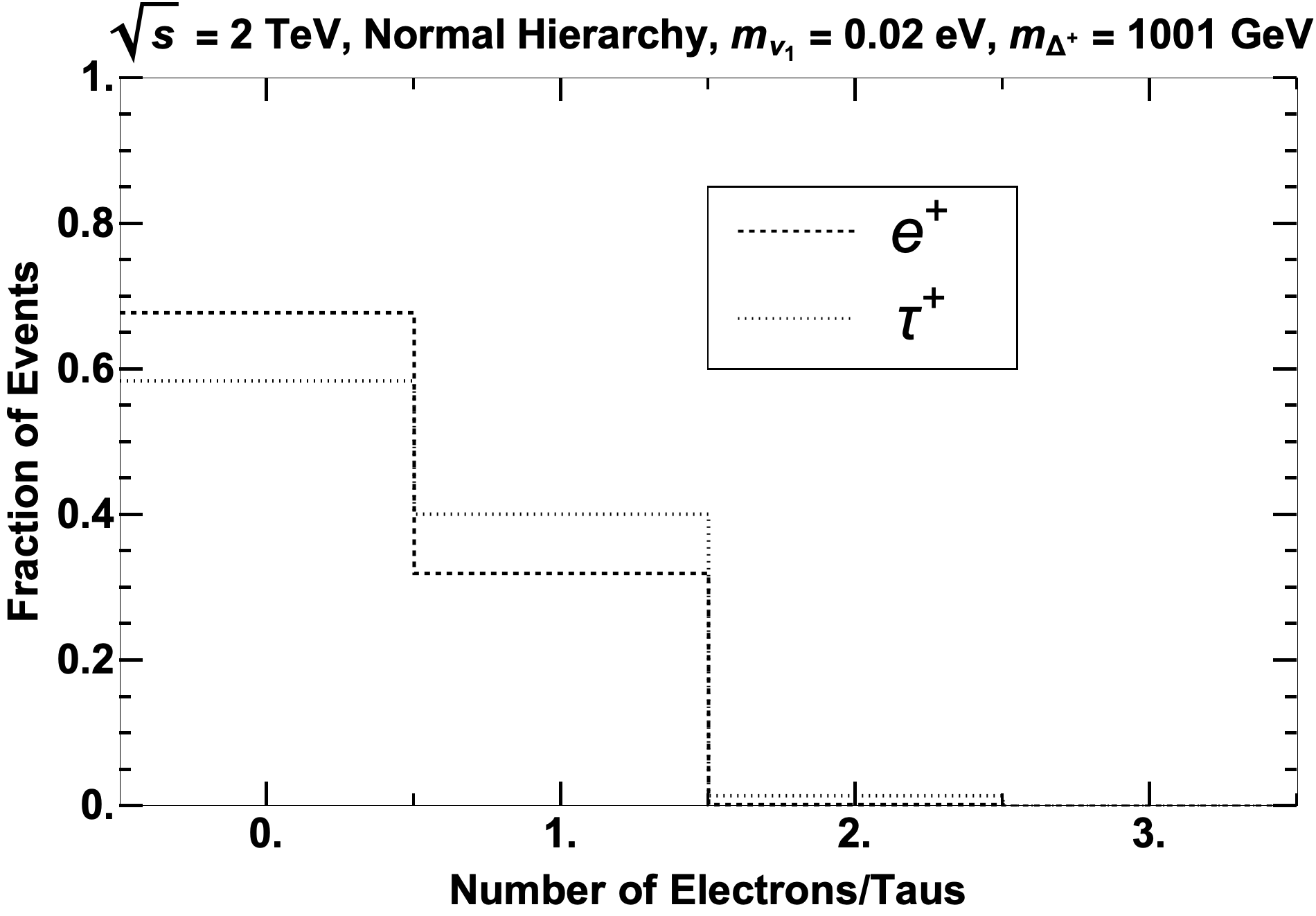}
    \caption{Normal Hierarchy}
    \label{fig:ebm3b}
\end{subfigure}
\begin{subfigure}[h]{0.45\textwidth}
  \centering
  \includegraphics[width=1\linewidth]{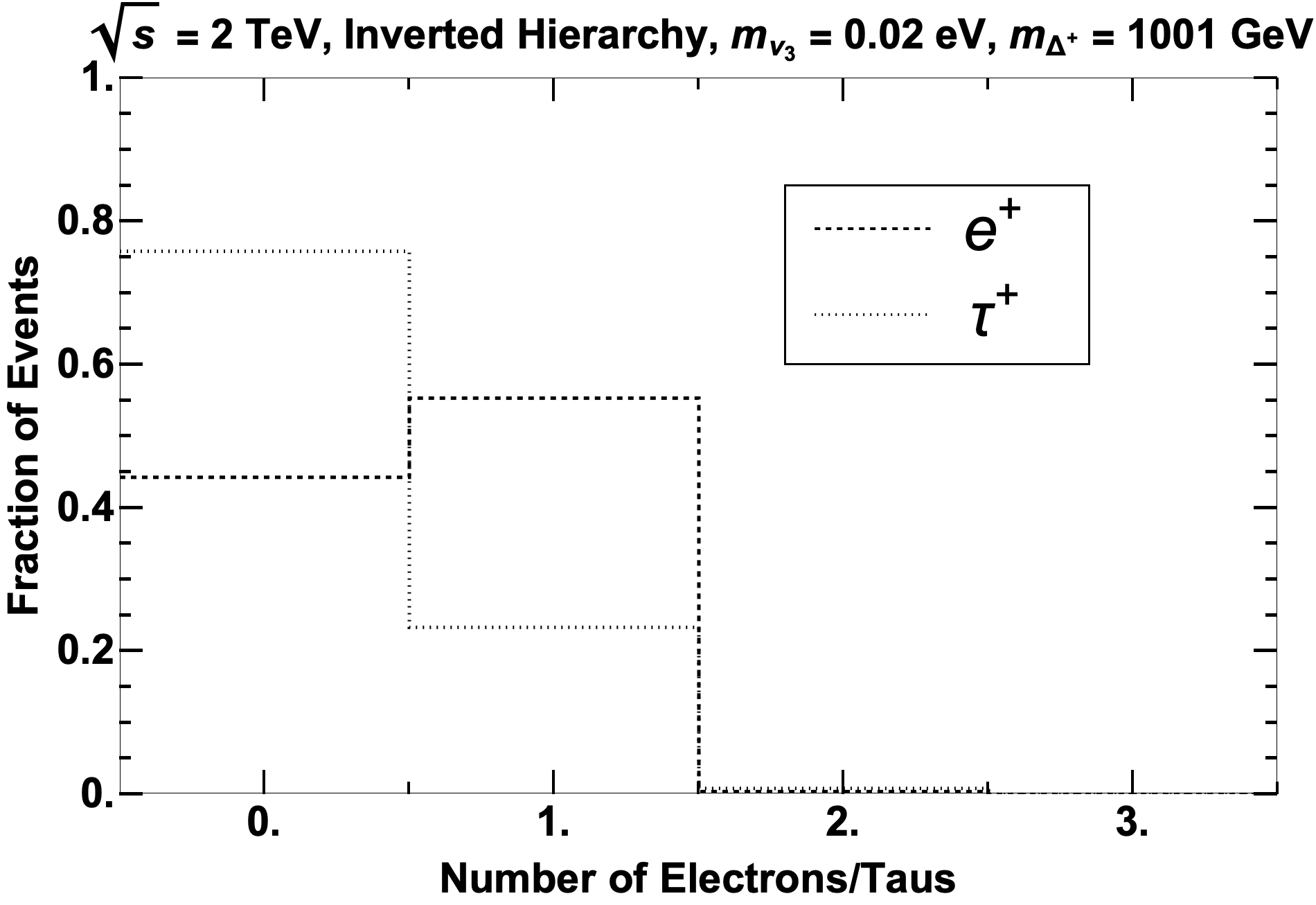}
  \caption{Inverted Hierarchy}
  \label{fig:tbm3b}
\end{subfigure}
\vspace*{-0.1in}
\caption{Distributions of Electrons and Taus for $m_{\nu_{lightest}}=0.02$ eV and $m_{\Delta^+} = 1001$ GeV assuming a) Normal Hierarchy and b) Inverted Hierarchy}
\label{fig:etbm3b}   
\end{figure}
\begin{figure}[ht!]
\centering
\begin{subfigure}[h]{0.45\textwidth}
    \centering
    \includegraphics[width=1\linewidth]{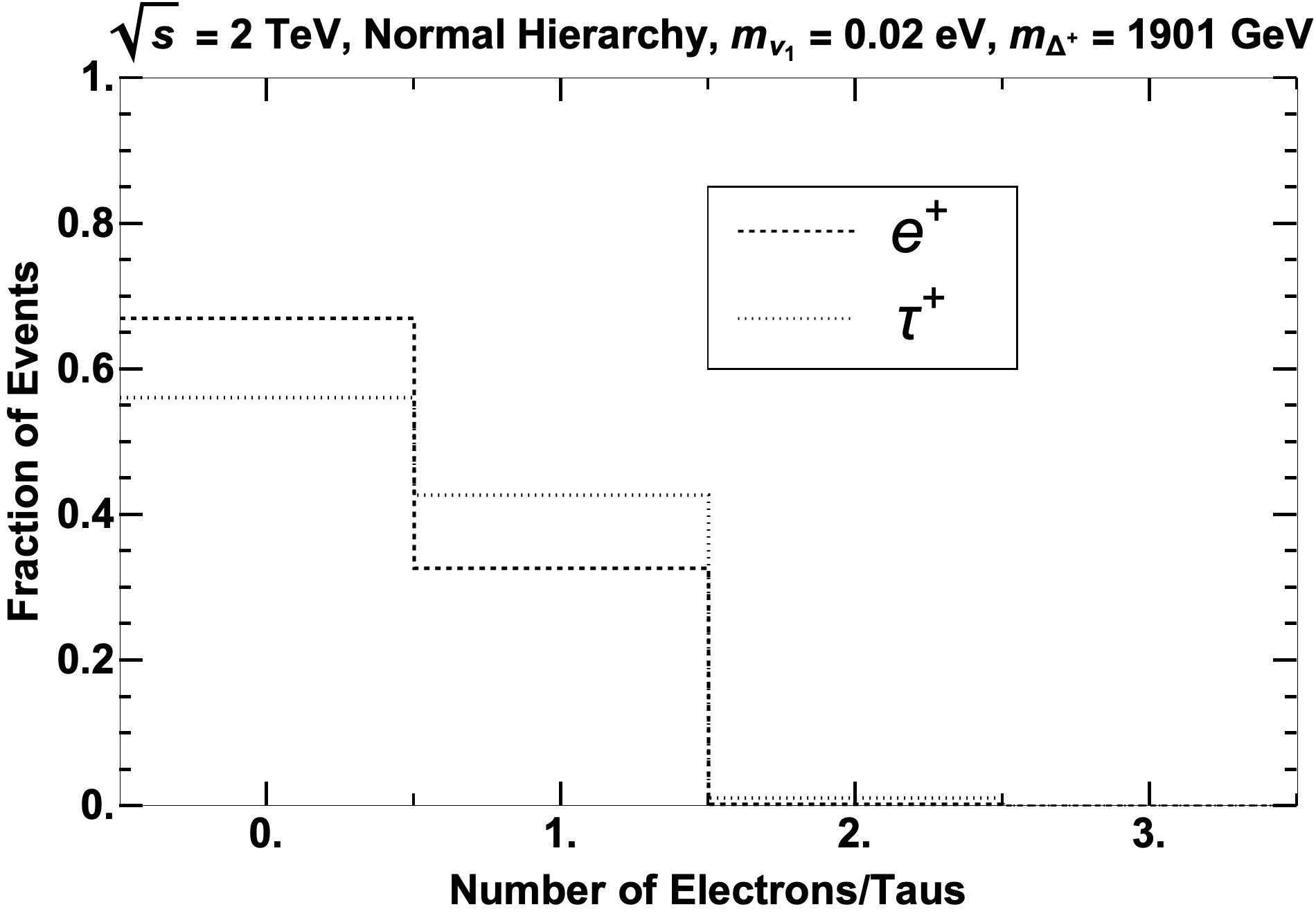}
    \caption{Normal Hierarchy}
    \label{fig:ebm3c}
\end{subfigure}
\begin{subfigure}[h]{0.45\textwidth}
  \centering
  \includegraphics[width=1\linewidth]{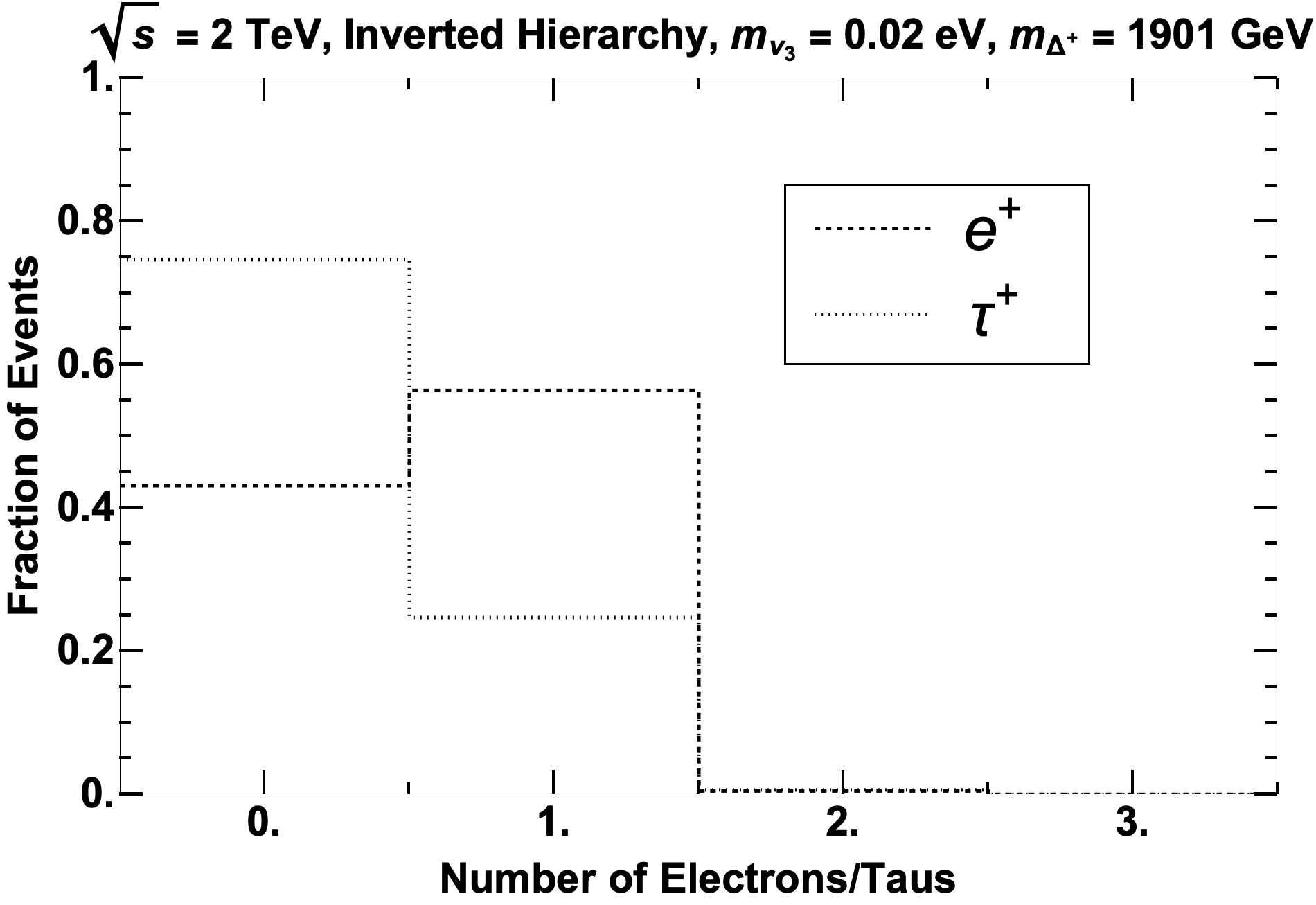}
  \caption{Inverted Hierarchy}
  \label{fig:tbm3c}
\end{subfigure}
\vspace*{-0.1in}
\caption{Distributions of Electrons and Taus for $m_{\nu_{lightest}}=0.02$ eV and $m_{\Delta^+} = 1901$ GeV assuming a) Normal Hierarchy and b) Inverted Hierarchy}
\label{fig:etbm3c}   
\end{figure}
\begin{figure}[ht!]
\centering
\begin{subfigure}[h]{0.45\textwidth}
    \centering
    \includegraphics[width=1\linewidth]{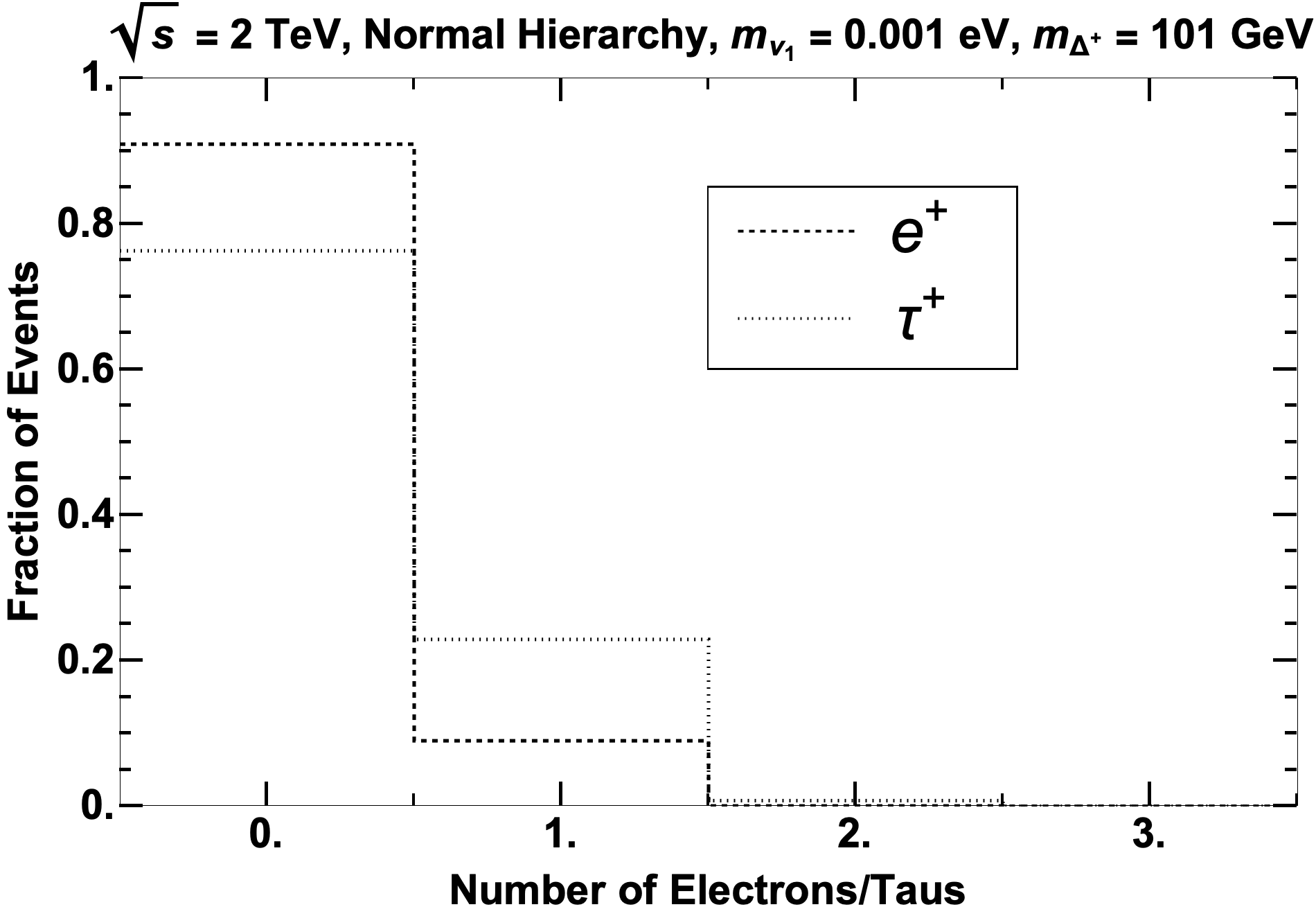}
    \caption{Normal Hierarchy}
    \label{fig:ebm2a}
\end{subfigure}
\begin{subfigure}[h]{0.45\textwidth}
  \centering
  \includegraphics[width=1\linewidth]{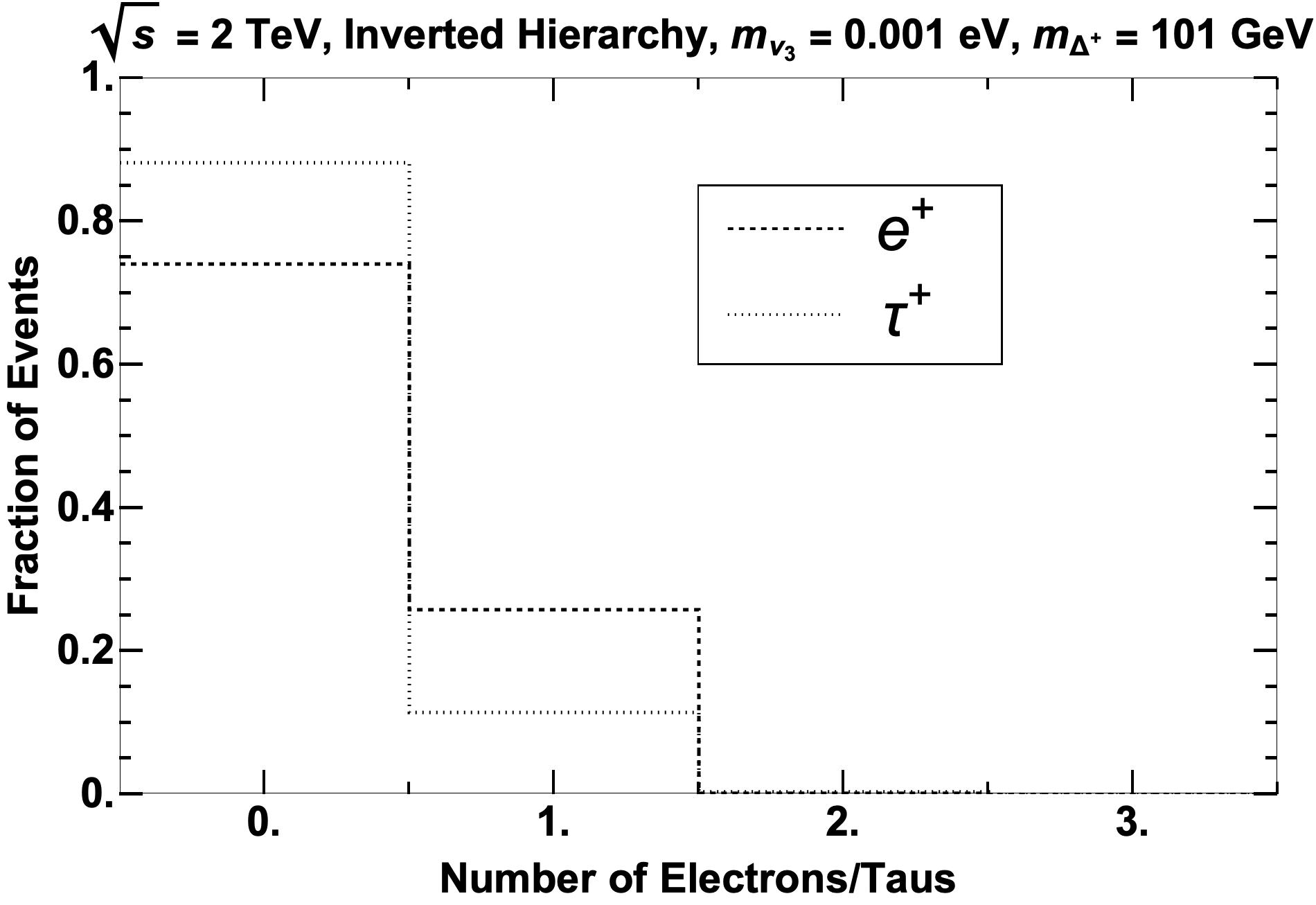}
  \caption{Inverted Hierarchy}
  \label{fig:tbm2a}
\end{subfigure}
\vspace*{-0.1in}
\caption{Distributions of Electrons and Taus for $m_{\nu_{lightest}}=0.001$ eV and $m_{\Delta^+} = 101$ GeV assuming a) Normal Hierarchy and b)Inverted Hierarchy}
\label{fig:etbm2a}   
\end{figure}
\begin{figure}[ht!]
\centering
\begin{subfigure}[h]{0.45\textwidth}
    \centering
    \includegraphics[width=1\linewidth]{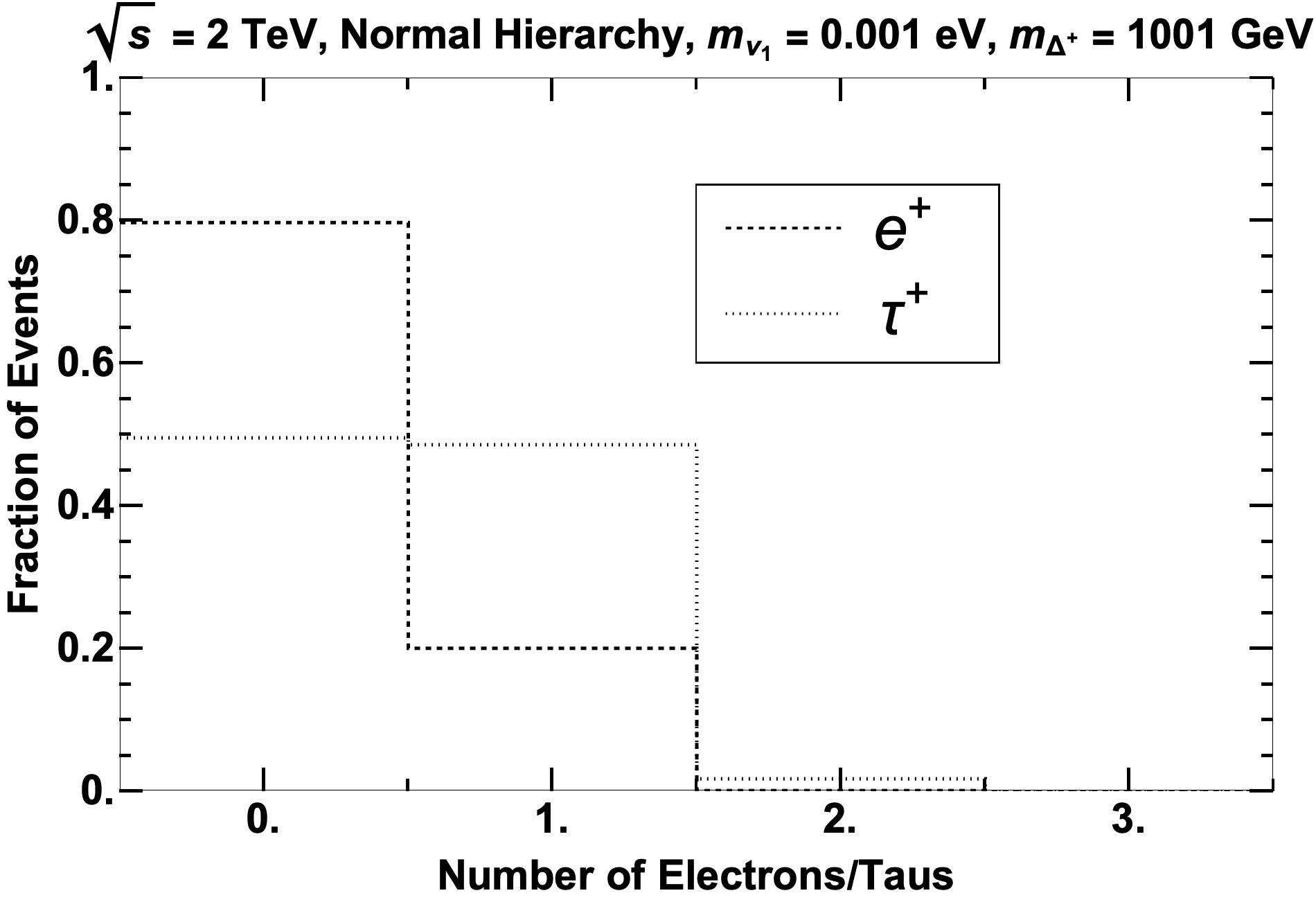}
    \caption{Normal Hierarchy}
    \label{fig:ebm2b}
\end{subfigure}
\begin{subfigure}[h]{0.45\textwidth}
  \centering
  \includegraphics[width=1\linewidth]{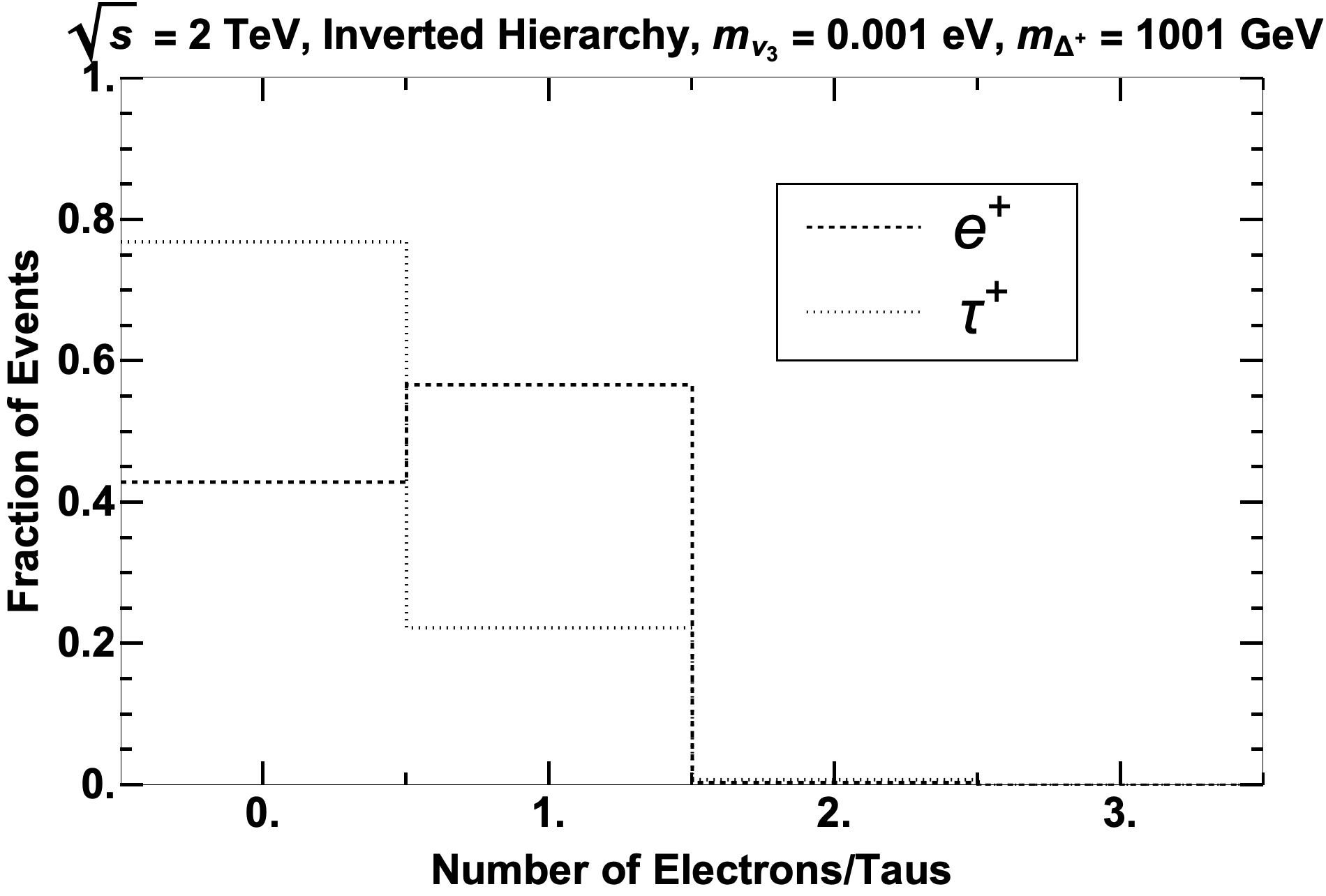}
  \caption{Inverted Hierarchy}
  \label{fig:tbm2b}
\end{subfigure}
\vspace*{-0.1in}
\caption{Distributions of Electrons and Taus for $m_{\nu_{lightest}}=0.001$ eV and $m_{\Delta^+} = 1001$ GeV assuming a) Normal Hierarchy and b)Inverted Hierarchy}
\label{fig:etbm2b}   
\end{figure}
\begin{figure}[ht!]
\centering
\begin{subfigure}[h]{0.45\textwidth}
    \centering
    \includegraphics[width=1\linewidth]{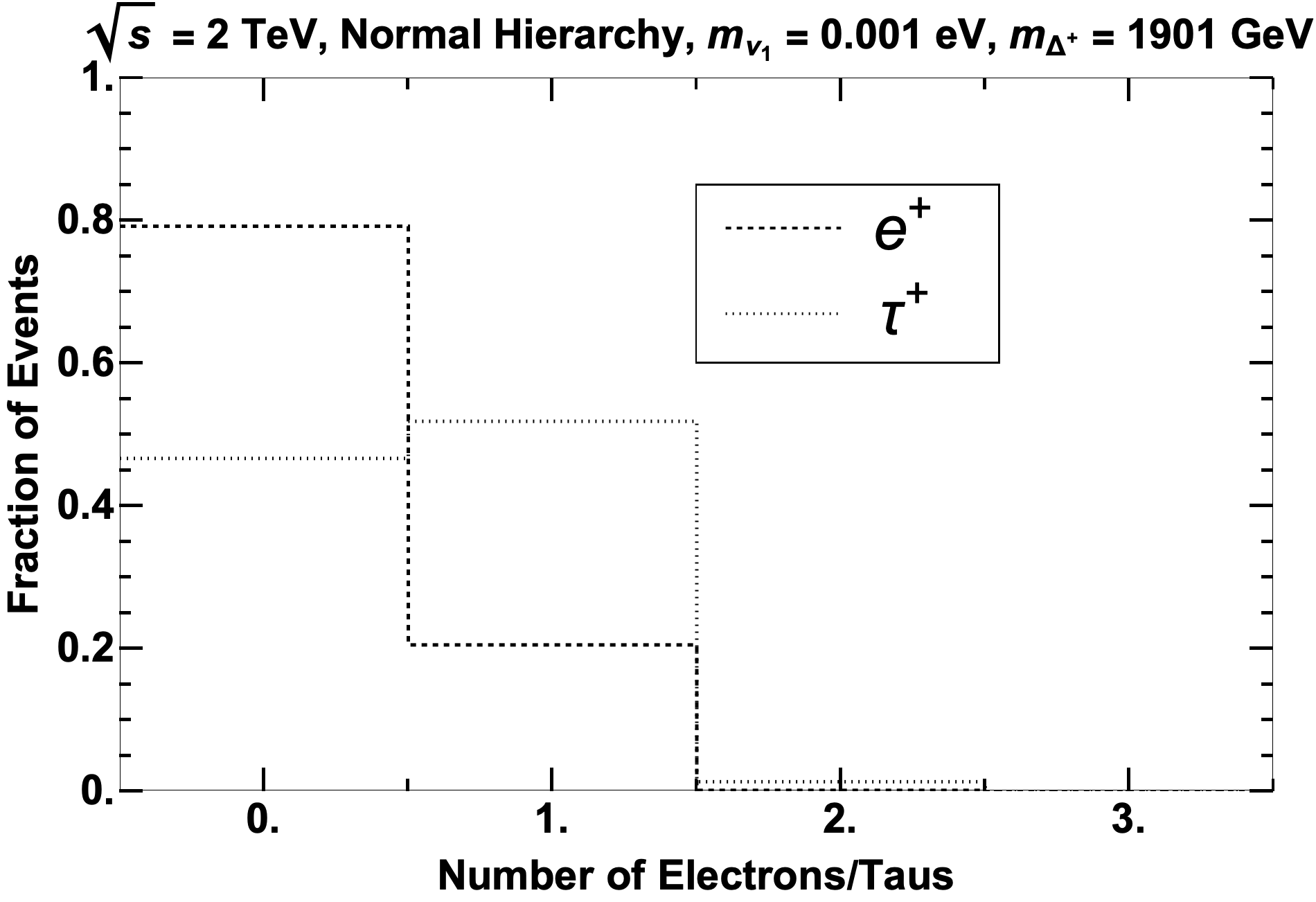}
    \caption{Normal Hierarchy}
    \label{fig:ebm2c}
\end{subfigure}
\begin{subfigure}[h]{0.45\textwidth}
  \centering
  \includegraphics[width=1\linewidth]{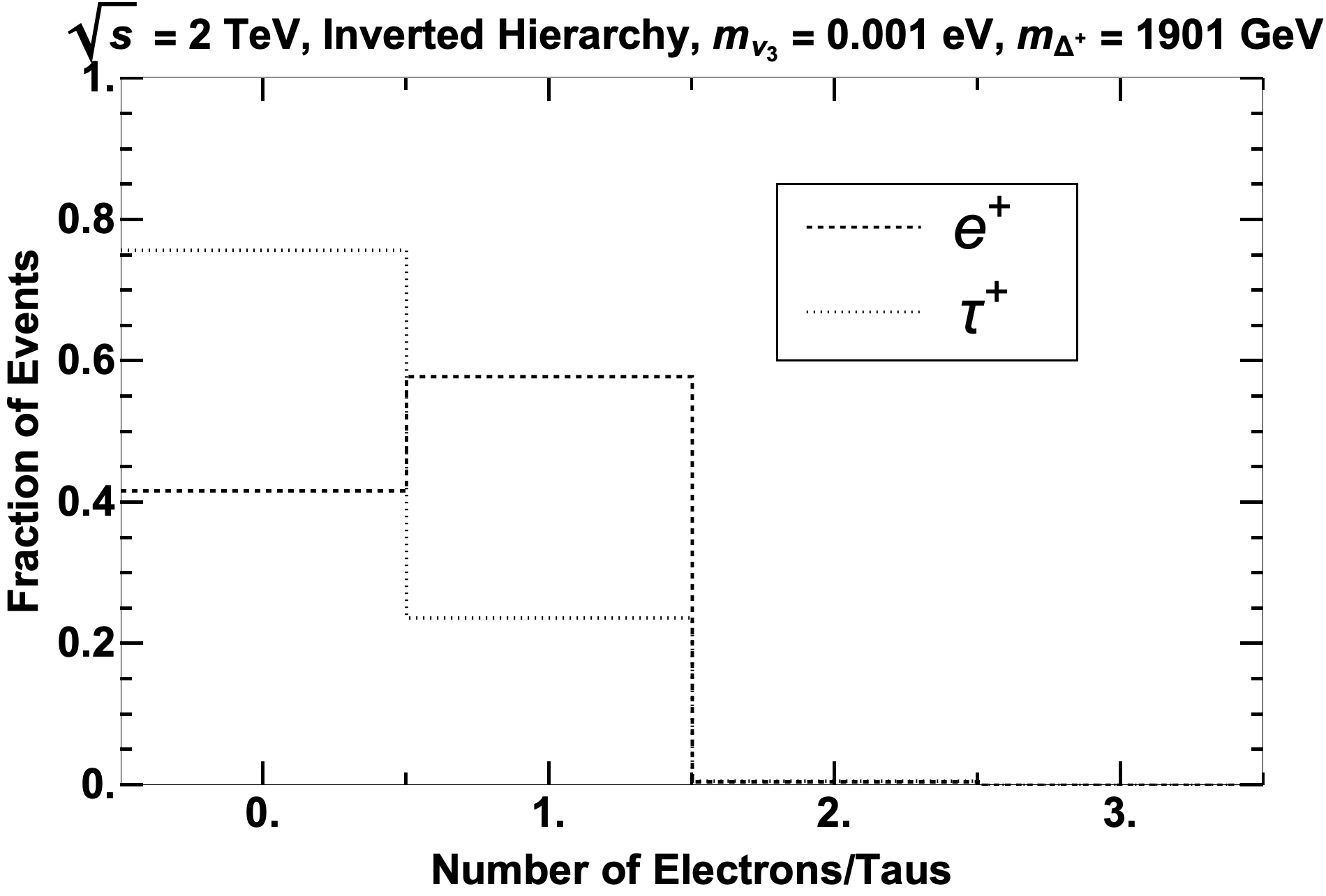}
  \caption{Inverted Hierarchy}
  \label{fig:tbm2c}
\end{subfigure}
\vspace*{-0.1in}
\caption{Distributions of Electrons and Taus for $m_{\nu_{lightest}}=0.001$ eV and $m_{\Delta^+} = 1901$ GeV assuming a) Normal Hierarchy and b)Inverted Hierarchy}
\label{fig:etbm2c}   
\end{figure}

In order to identify whether the theory respects Normal or Inverted hierarchy, one must compare the distribution of final state Electrons and Taus. We compare the distribution of Electrons and Taus in the final state for Normal and Inverted hierarchies in Fig.~\ref{fig:ebm1a} and Fig.~\ref{fig:tbm1a}, respectively, for a particular benchmark point characterized with $m_{\nu_{lightest}}=0.05$ eV and $m_{\Delta^+} = 101$ GeV. Similar plots for two other masses of $\Delta^+$ = 1001 GeV and 1901 GeV have been shown in Fig.~\ref{fig:etbm1b} and Fig.~\ref{fig:etbm1c}, respectively. While one compares the Electron and Tau distribution for a particular benchmark point with a characteristic hierarchy, one must be careful to compare the relevant events i.e., the events which have exactly one charged lepton in the final state. Therefore, comparing such relevant events in Fig.~\ref{fig:etbm1a}, Fig.~\ref{fig:etbm1b} and Fig.~\ref{fig:etbm1c} one would expect that in case of Normal hierarchy there would be more events with Taus in the final state than with final state Electrons while for Inverted hierarchy there would be more events with Electrons in the final state than with final state Taus, as evident from Fig.~\ref{fig:br1}. However, for bm points with lightest neutrino mass = 0.05 eV, there are always more events with Electrons than Taus in the final state irrespective of the underlying hierarchy. The difference between the hierarchies are not observable for bm points with lightest neutrino mass = 0.05 eV because for such high mass of the lightest neutrino the difference in the branching ratios of $\Delta^+$ in different leptonic channels is quite small, as evident from Fig.~\ref{fig:br2}. The difference between the hierarchies for massive lightest neutrino can be improved by modifying the detector structure and taking into consideration several other factors such as uncertainties, response functions etc. However, these factors are not yet known and hence beyond the scope of this work. Therefore, given the detector design as in the default muon collider delphes card,  we find the limit at what mass of the lightest neutrino the difference between the hierarchies could be observed from the flavor of the final state leptons. We find that for lightest neutrino mass $\leq 0.02$ eV, one starts to differentiate between Normal and Inverted hierarchy. We compare the distribution of Electrons and Taus for Normal and Inverted hierarchies for $m_{\nu_{lightest}}=0.02$ eV and $m_{\Delta^+} = 101$, $1001$ and $1901$ GeV, respectively in Fig.~\ref{fig:etbm3a} - Fig.~\ref{fig:etbm3c} where it is evident that indeed if the underlying theory respects Normal hierarchy then there will be more events with Taus in the final state and if the underlying theory respects Inverted hierarchy then there will be more events with Electrons in the final state. The same conclusion holds true for lightest neutrino mass $=0.001$ eV as it is evident from Fig.~\ref{fig:etbm2a} - Fig.~\ref{fig:etbm2c} which compares the distribution of Electrons and Taus for Normal and Inverted hierarchies for $m_{\nu_{lightest}}=0.001$ eV and $m_{\Delta^+} = 101$, $1001$ and $1901$ GeV, respectively.
%\clearpage
%%%%%%%%%%%%%%%%%%%%%%%%%%%%%%%%%%%%%%%%%%%%
%\textbf{ FOR $m_{\nu_{lightest}}=0.02$ eV}
%%%%%%%%%%%%%%%%%%%%%%%%%%%%%%%%%%%%%%%%%%%%

%\clearpage
%%%%%%%%%%%%%%%%%%%%%%%%%%%%%%%%%%%%%%%%%%%%
%\textbf{ FOR $m_{\nu_{lightest}}=0.001$ eV}
%%%%%%%%%%%%%%%%%%%%%%%%%%%%%%%%%%%%%%%%%%%%

%\clearpage
%%%%%%%%%%%%%%%%%%%%%%%%%%%%%%%%%%%%%%%%%%%%%%%
\section{Summary and Outlook}
\label{sec:con}
%%%%%%%%%%%%%%%%%%%%%%%%%%%%%%%%%%%%%%%%%%%%%%%
Although hadron colliders such as the LHC are extremely advantageous in looking for new BSM particles produced via direct production, owing to much higher c.o.m energy as compared to the c.o.m energy in lepton colliders, a lepton collider enjoys a much cleaner environment with reduced SM background and is therefore more advantageous in looking for new BSM particles/processes involving electroweak interactions only. Additionally in lepton colliders the full c.o.m energy is used as a lepton, unlike a hadron, is a fundamental particle and hence lepton colliders offer a higher physics reach then hadron colliders for a given c.o.m energy. Several lepton colliders such as $e^+e^-$ collider, $\mu^+ \mu^-$ collider, $\mu^+ \mu^+$ collider, $\mu^+ e^-$ collider have been proposed to build. However, in this article we focus on the $\mu^+ \mu^+$ mode of the $\mu$TRISTAN~\cite{Hamada:2022mua} that is planned to operate at $\sqrt{s}=2$ TeV. The reasons behind choosing this particular setup among several proposed lepton colliders are two fold: $i)$ Muon colliders lead to smaller loss of energy due to synchrotron radiations than Electron colliders as Muons are heavier than Electrons. $ii)$ Building a $\mu^+ \mu^+$ collider appears to be more convenient than building a $\mu^+ \mu^-$ collider due to the cooling technology developed at J-PARC~\cite{Abe:2019thb} which can apparently yield $\mu^+$ more efficiently than $\mu^-$ such that they are suitable for performing collision at desired c.o.m energy. While several studies have been performed in the $\mu^+ \mu^+$ collider searching for doubly-charged scalar $\Delta^{++}$ and electrically neutral heavy scalar $\Delta^{0}$~\cite{Yang:2023ojm, Fridell:2023gjx, Dev:2023nha, Jia:2024wqi} both of which naturally occur in several BSM scenarios such as ~\cite{Magg:1980ut, Schechter:1980gr, Lazarides:1980nt, Mohapatra:1980yp, Pati:1974yy, Mohapatra:1974hk, Senjanovic:1975rk, Kuchimanchi:1993jg, Babu:2008ep, Babu:2014vba, Basso:2015pka, Zee:1985id, Babu:1988ki, ArkaniHamed:2002qx, Babu:2020hun, Georgi:1985nv, Gunion:1989ci, Babu:2009aq, Bonnet:2009ej, Bhattacharya:2016qsg, Kumericki:2012bh}, there hasn't been any study in search of the singly-charged scalar in the $\mu^+ \mu^+$ collider which also naturally occurs in the above-mentioned BSM scenarios. In this article we look into the discovery prospect of the singly-charged scalar at the $\mu^+ \mu^+$ collider at $\sqrt{s}=2$ TeV in one of the above-mentioned BSM scenarios, namely, the simplest type-II seesaw scenario~\cite{Magg:1980ut, Schechter:1980gr, Mohapatra:1979ia, Lazarides:1980nt} which is an extremely well-motivated BSM scenario as it provides a viable mechanism for neutrino mass generation. We study the associated production of $\Delta^+$ along with a $W$ boson followed by the former decaying leptonically and the latter decaying hadronically.
Additionally, since the type-II seesaw scenario allows for LFV processes which the SM does not, we take into consideration $\Delta^+$ decaying into $e^+$ or $\tau^+$ only as this constraint gets rid of the SM background completely which would otherwise be huge if $\Delta^+$ decaying to $\mu^+$ would also have been considered. On studying the signature of our choice(Fig.~\ref{fig:signal}) for two benchmark points with a mass of the lightest neutrino = 0.05 eV and 0.001 eV, and $m_{\Delta^+}$ varying between 101 and 1901 GeV, we find that the $\mu^+ \mu^+$ collider operating at $\sqrt{s}=2$ TeV is capable of yielding a cross-section with significance above 5$\sigma$ for the entire mass range of $\Delta^+$ for $m_{\nu_{lightest}} = 0.05$ eV assuming Normal or Inverted hierarchy, while for $m_{\nu_{lightest}} = 0.001$ eV, assuming Normal hierarchy $m_{\Delta^+} \lesssim 207~$ GeV and  $1304 \lesssim m_{\Delta^+} \lesssim 1870~$ GeV have significance above 5$\sigma$ and assuming Inverted hierarchy points with $m_{\nu_{lightest}}$ = 0.001 eV and $m_{\Delta^+} \lesssim 140~$ GeV and  $1410 \lesssim m_{\Delta^+} \lesssim 1862~$ GeV have significance above 5$\sigma$. Both the benchmark points with $m_{\nu_{lightest}} = 0.05$ eV and $0.001$ eV have cross-section above 95$\%$ C.L. over the complete mass range of $\Delta^+$ from 101 to 1901 GeV\footnote{Although here we show the cross-section and significance for $101~\text{GeV} \leq m_{\Delta^+} \leq 1901~\text{GeV}$, one should focus on the results obtained for $m_{\Delta^+} \gtrsim 1.0 (1.5)$ TeV, owing to the current(future) limit on $m_{\Delta^{++}}$ obtained from the LHC.}. However, the cross-section or the significance plot is not sufficient to indicate whether the underlying theory respects Normal or Inverted Hierarchy since both the cases yield curves of similar shape as can be seen in Fig.~\ref{fig:xsecsig}. The negligibly small difference between the curves for Normal and Inverted hierarchy, as in Fig.~\ref{fig:xsecsig}, might vanish subject to experimental uncertainties. Therefore, in order to identify the hierarchy one must look into the distribution of the charged leptons in the final state originating from the decay of $\Delta^+$. As discussed in Sec.~\ref{sec:diff}, Normal hierarchy leads $\Delta^+$ to primarily decay to $\tau^+$ while with Inverted hierarchy $\Delta^+$ primarily decays to $e^+$. However, as we see in Sec.~\ref{sec:diff}, with the given detector design the distinction between the underlying hierarchies can be done only if $m_{\nu_{lightest}} \leq 0.02$ eV. With $m_{\nu_{lightest}} > 0.02$ eV, the distinction between the underlying hierarchies could be possible with some modifications in the detector design and considering other factors such as uncertainties, response function etc., which will be referred to in a related future work. Hence to conclude, the $\mu^+ \mu^+$ collider running at $\sqrt{s}=2$ TeV is an excellent tool for discovering a singly-charged scalar which can, in turn, hint towards the existence of the type-II seesaw BSM scenario. This experimental setup can also be used to distinguish between Normal and Inverted hierarchy leading to a highly important deduction regarding the underlying theory.

%%%%%%%%%%%%%%%%%%%%%%%%%%%%%%%%%%%%%%%%%%%%%%%
\section{Acknowledgment}
\label{sec:ack}
%%%%%%%%%%%%%%%%%%%%%%%%%%%%%%%%%%%%%%%%%%%%%%%
We thank Saiyad Ashanujjaman for useful discussions on experimental limits. This work was partly supported by the National Science Centre, Poland, under the OPUS 27 funding scheme, project no.~2024/53/B/ST2/00975. The work of N.O. is supported in part by the United States Department of Energy Grant Nos.~DE-SC0012447, DE-SC0023713, and DE-SC0026347. The work of D.S. was partly supported by the European Research Council (ERC) under the European Union's Horizon 2020 research and innovation programme (Grant agreement No. 949451). SKV is supported by ANRF, DST, Govt. of India Grants, CRG/2021/007170 ``Tiny Effects from Heavy New Physics'' and REDA funds from IISC.

\appendix
\section{Doubly-charged scalar \texorpdfstring{($\Delta^{++}$)}{Delta++} in Muon collider}
\label{sec:dppmu}
Here we present a comparative study for pair production of the doubly-charged scalar in the Muon collider at $\sqrt{s} = 3~\text{and}~10$ TeV.\footnote{Since the LHC predicts a projected limit on $m_{\Delta^{++}} \sim 1.5$ TeV~\cite{Ashanujjaman:2021txz}, therefore we present results for $\sqrt{s} = 10$ TeV as a 3 TeV Muon collider is not sufficient for pair production of the doubly-charged scalar with $m_{\Delta^{++}} \gtrsim 1.5$ TeV.}. The cross-sections for the lightest neutrino mass = 0.05 eV for both Normal and Inverted Hierarchy are presented in Table.~\ref{tab:dppbm1mupmum3} and Table.~\ref{tab:dppbm1mupmum10} at $\sqrt{s} = 3~\text{and}~10$ TeV, respectively. Similar results for lightest neutrino mass = 0.001 eV is presented in Table.~\ref{tab:dppbm2mupmum3} and Table.~\ref{tab:dppbm2mupmum10}.

\begin{table}[ht!]
\centering
\resizebox{\textwidth}{!}{\begin{tabular}{|c|c|c V{5}c |c|c|c V{5}c|c|c|c|c|}
\hline
\multicolumn{3}{|c V{5}}{Masses [GeV]}         & {Normal Hierarchy}    & {Inverted Hierarchy}                                                                                 \\ \hline
  $m_{\Delta^0}$    &    $m_{\Delta^+}$                    &  $m_{\Delta^{++}}$    &        $\sigma$     &   $\sigma$              \\ \hline
    100   & 101   & 103  & 12.3 & 12.36 \\ \hline
%    300 & 301 & 303 & 0.0601 & 0.0603 & 0.0607 &  0.01155 & 0.0742 & 0.0744 & 0.0749 & 0.01162 \\ \hline
    500 & 501 & 503 & 10.25 & 10.32 \\ \hline
    1000 & 1001 & 1003 & 5.0 & 5.022 \\ \hline
\end{tabular}}
\caption{Benchmark Points for Normal and Inverted Hierarchy at $v_{\Delta} = 10^{-9}~$ GeV and lightest neutrino mass = 0.05 eV. The cross-section listed here is for the process $\mu^+ \mu^- > \Delta^{++} \Delta^{--}$ at $\sqrt{s} = 3$~TeV. All the masses and decay widths($\Gamma$) are in units of GeV and the cross-sections $\sigma$ are in units of fb.}
    \label{tab:dppbm1mupmum3}
\end{table}

\begin{table}[ht!]
\centering
\resizebox{\textwidth}{!}{\begin{tabular}{|c|c|c V{5}c |c|c|c V{5}c|c|c|c|c|}
\hline
\multicolumn{3}{|c V{5}}{Masses [GeV]}         & {Normal Hierarchy}    & {Inverted Hierarchy}                                                                                 \\ \hline
  $m_{\Delta^0}$    &    $m_{\Delta^+}$                    &  $m_{\Delta^{++}}$    &        $\sigma$     &   $\sigma$            \\ \hline
    500 & 501 & 503 & 1.1 & 1.1 \\ \hline
    1000 & 1001 & 1003 & 1.04 & 1.05\\ \hline
    1500 & 1501 & 1503 & 0.96 & 0.96 \\ \hline
    1700 & 1701 & 1703 & 0.92 & 0.922 \\ \hline
    1900 & 1901 & 1903 & 0.87 & 0.88 \\ \hline
    2000 & 2001 & 2003 & 0.85 & 0.852 \\ \hline
    3000 & 3001 & 3003 & 0.56 & 0.563 \\ \hline
    4000 & 4001 & 4003 & 0.23 & 0.235 \\ \hline
    4900 & 4901 & 4903 & $8.1 \cdot 10^{-3}$ & $8.2 \cdot 10^{-3}$ \\ \hline
\end{tabular}}
\caption{Benchmark Points for Normal and Inverted Hierarchy at $v_{\Delta} = 10^{-9}~$ GeV and lightest neutrino mass = 0.05 eV. The cross-section listed here is for the process $\mu^+ \mu^- > \Delta^{++} \Delta^{--}$ at $\sqrt{s} = 10$~TeV. All the masses and decay widths($\Gamma$) are in units of GeV and the cross-sections $\sigma$ are in units of fb.}
    \label{tab:dppbm1mupmum10}
\end{table}

\begin{table}[ht!]
\centering
\resizebox{\textwidth}{!}{\begin{tabular}{|c|c|c V{5}c |c|c|c V{5}c|c|c|c|c|}
\hline
\multicolumn{3}{|c V{5}}{Masses [GeV]}         & {Normal Hierarchy}    & {Inverted Hierarchy}                                                                                 \\ \hline
  $m_{\Delta^0}$    &    $m_{\Delta^+}$                    &  $m_{\Delta^{++}}$    &        $\sigma$    &   $\sigma$             \\ \hline
    100 & 101 & 103  & 12.75  & 12.84 \\ \hline
%     300 & 301 & 303 &   $1.538 \cdot 10^{-2}$ &   $1.543 \cdot 10^{-2}$ & $1.553 \cdot 10^{-2}$ & 0.01205 & $2.942 \cdot 10^{-2}$ & $2.952 \cdot 10^{-2}$ & $2.971 \cdot 10^{-2}$ & 0.01214  \\ \hline
    500 & 501 & 503  & 10.71 & 10.78  \\ \hline
    1000 & 1001 & 1003 & 5.246 & 5.289 \\ \hline
\end{tabular}}
\caption{Benchmark Points for Normal and Inverted Hierarchy at $v_{\Delta} = 10^{-9}~$ GeV and lightest neutrino mass = 0.001 eV. The cross-section listed here is for the process $\mu^+ \mu^- > \Delta^{++} \Delta^{--}$ at $\sqrt{s} = 3$~TeV. All the masses and decay widths($\Gamma$) are in units of GeV and the cross-sections $\sigma$ are in units of fb.}
    \label{tab:dppbm2mupmum3}
\end{table}

\begin{table}[ht!]
\centering
\resizebox{\textwidth}{!}{\begin{tabular}{|c|c|c V{5}c |c|c|c V{5}c|c|c|c|c|}
\hline
\multicolumn{3}{|c V{5}}{Masses [GeV]}         & {Normal Hierarchy}    & {Inverted Hierarchy}                                                                                 \\ \hline
  $m_{\Delta^0}$    &    $m_{\Delta^+}$                    &  $m_{\Delta^{++}}$    &        $\sigma$  &   $\sigma$           \\ \hline
    500 & 501 & 503 & 1.14 & 1.144 \\ \hline
    1000 & 1001 & 1003 & 1.085 & 1.092 \\ \hline
    1500 & 1501 & 1503 & 1.0 & 1.01 \\ \hline
    1700 & 1701 & 1703 & 0.96 & 0.965 \\ \hline
    1900 & 1901 & 1903 & 0.911 & 0.92 \\ \hline
    2000 & 2001 & 2003 & 0.87 & 0.9 \\ \hline
    3000 & 3001 & 3003 & 0.59 & 0.6 \\ \hline
    4000 & 4001 & 4003 & 0.246 & 0.25 \\ \hline
    4900 & 4901 & 4903 & $8.6 \cdot 10^{-3}$ &  $8.7 \cdot 10^{-3}$ \\ \hline
\end{tabular}}
\caption{Benchmark Points for Normal and Inverted Hierarchy at $v_{\Delta} = 10^{-9}~$ GeV and lightest neutrino mass = 0.001 eV. The cross-section listed here is for the process $\mu^+ \mu^- > \Delta^{++} \Delta^{--}$ at $\sqrt{s} = 10$~TeV. All the masses and decay widths($\Gamma$) are in units of GeV and the cross-sections $\sigma$ are in units of fb.}
    \label{tab:dppbm2mupmum10}
\end{table}
\clearpage
\section{Doubly-charged scalar \texorpdfstring{($\Delta^{++}$)}{Delta++} in \texorpdfstring{$\mu^+ \mu^+$}{mu+mu+} mode of \texorpdfstring{$\mu$TRISTAN}{muTRISTAN}}
\label{sec:dpp}
This section shows another comparative study for production of the doubly-charged scalar in the $\mu$TRISTAN at $\sqrt{s} = 2$~TeV. The cross-sections for the lightest neutrino mass = 0.05 eV for both Normal and Inverted Hierarchy are presented in Table.~\ref{tab:dppbm1mupmup}. Similar results for lightest neutrino mass = 0.001 eV is presented in Table.~\ref{tab:dppbm2mupmup}
\begin{table}[ht!]
\centering
\resizebox{\textwidth}{!}{\begin{tabular}{|c|c|c V{5}c |c|c|c V{5}c|c|c|c|c|}
\hline
\multicolumn{3}{|c V{5}}{Masses [GeV]}         & \multicolumn{4}{c V{5}}{Normal Hierarchy}    & \multicolumn{4}{c|}{Inverted Hierarchy}                                                                                 \\ \hline
  $m_{\Delta^0}$    &    $m_{\Delta^+}$                    &  $m_{\Delta^{++}}$    &      $\Gamma_{\Delta^0}$      &   $\Gamma_{\Delta^+}$   &  $\Gamma_{\Delta^{++}}$   &  $\sigma$  &   $\Gamma_{\Delta^0}$      &   $\Gamma_{\Delta^+}$   &  $\Gamma_{\Delta^{++}}$       &     $\sigma$                  \\ \hline
    100   & 101   & 103  &  0.02  & 0.0202   & 0.0206 & $1.419 \cdot 10^{-2}$ & 0.0247 & 0.0249 & 0.0254 & $1.146 \cdot 10^{-2}$ \\ \hline
   % 200   & 201 &   203   & 0.0401 & 0.0403 &   0.0407 & 0.0494 & 0.0497 & 0.05 \\ \hline
    300 & 301 & 303 & 0.0601 & 0.0603 & 0.0607 & $1.478 \cdot 10^{-2}$ & 0.0742 & 0.0744 & 0.0749 & $1.194 \cdot 10^{-2}$ \\ \hline
   % 400 & 401 & 403 & 0.0802 & 0.0804 & 0.0808  & 0.0989 & 0.0991 & 0.0996  \\ \hline
    500 & 501 & 503 & 0.1002 & 0.1004 & 0.1008 & $1.608 \cdot 10^{-2}$ & 0.1236 & 0.1239 & 0.1243 & $1.299 \cdot 10^{-2}$ \\ \hline
   % 600 & 601 & 603 & 0.1202 & 0.1204 & 0.1208 & 0.1483 & 0.1486 & 0.1491\\ \hline
   % 700 & 701 & 703 & 0.1403 & 0.1405 & 0.1409 & 0.173 & 0.1733 & 0.174 \\ \hline
   % 800 & 801 & 803 & 0.1603 & 0.1605 & 0.1609 & 0.1978 & 0.198 & 0.1985\\ \hline
   % 900 & 901 & 903 & 0.1804 & 0.1806 & 0.181 & 0.2225 & 0.2227 & 0.2232\\ \hline
    1000 & 1001 & 1003 & 0.2004 & 0.2006 & 0.201 & $2.519 \cdot 10^{-2}$ & 0.2472 & 0.2475 & 0.248 & $2.035 \cdot 10^{-2}$ \\ \hline
   % 1100 & 1101 & 1103 & 0.2205 & 0.2207 & 0.2211 & 0.2719 & 0.2722 & 0.2727 \\ \hline
   % 1200 & 1201 & 1203 & 0.2405 & 0.2407 & 0.2411 & 0.2967 & 0.2969 & 0.2974 \\ \hline
   % 1300 & 1301 & 1303 & 0.2605 & 0.2607 & 0.2611 & 0.3214 & 0.3216 & 0.3221 \\ \hline
   % 1400 & 1401 & 1403 & 0.2806 & 0.2808 & 0.2812 & 0.3461 & 0.3463 & 0.3468 \\ \hline
    1500 & 1501 & 1503 & 0.3006 & 0.3008 & 0.3012 & $7.448 \cdot 10^{-2}$ & 0.3708 & 0.3711 & 0.3716 & $6.018 \cdot 10^{-2}$ \\ \hline
   % 1600 & 1601 & 1603 & 0.3207 & 0.3209 & 0.3213 & 0.3955 & 0.3958 & 0.3963 \\ \hline
   % 1700 & 1701 & 1703 & 0.3407 & 0.3409 & 0.3413 & 0.4203 & 0.4205 & 0.421 \\ \hline
   % 1800 & 1801 & 1803 & 0.3607 & 0.3609 & 0.3613 & 0.445 & 0.4452 & 0.4457 \\ \hline
    1900 & 1901 & 1903 & 0.3808 & 0.381 & 0.3814 & 1.575 & 0.4697 & 0.47 & 0.4704 & 1.273 \\ \hline
\end{tabular}}
\caption{Benchmark Points for Normal and Inverted Hierarchy at $v_{\Delta} = 10^{-9}~$ GeV and lightest neutrino mass = 0.05 eV. The cross-section listed here is for the process $\mu^+ \mu^+ > \Delta^{++} > \mu^+ \mu^+$ at $\sqrt{s} = 2$~TeV. All the masses and decay widths($\Gamma$) are in units of GeV and the cross-sections $\sigma$ are in units of fb.}
    \label{tab:dppbm1mupmup}
\end{table}

\begin{table}[ht!]
\centering
\resizebox{\textwidth}{!}{\begin{tabular}{|c|c|c V{5}c |c|c|c V{5}c|c|c|c|c|}
\hline
\multicolumn{3}{|c V{5}}{Masses [GeV]}         & \multicolumn{4}{c V{5}}{Normal Hierarchy}    & \multicolumn{4}{c|}{Inverted Hierarchy}                                                                                 \\ \hline
  $m_{\Delta^0}$    &    $m_{\Delta^+}$                    &  $m_{\Delta^{++}}$    &      $\Gamma_{\Delta^0}$   &   $\Gamma_{\Delta^+}$    &  $\Gamma_{\Delta^{++}}$ & $\sigma$   &      $\Gamma_{\Delta^0}$   &   $\Gamma_{\Delta^+}$  &  $\Gamma_{\Delta^{++}}$ & $\sigma$                \\ \hline
    100 & 101 & 103  &  $5.13 \cdot 10^{-3}$  &  $5.18 \cdot 10^{-3}$  & $5.28 \cdot 10^{-3}$ & $8.531 \cdot 10^{-4}$ & $9.806 \cdot 10^{-3}$ & $9.902 \cdot 10^{-3}$ & $1.01 \cdot 10^{-2}$ & $2.195 \cdot 10^{-4}$\\ \hline
    % 200 & 201 & 203  &  $1.025 \cdot 10^{-2}$  &  $1.03 \cdot 10^{-2}$  & $1.041 \cdot 10^{-2}$ & $1.961 \cdot 10^{-2}$ & $1.971 \cdot 10^{-2}$ & $1.99 \cdot 10^{-2}$\\ \hline
     300 & 301 & 303 &   $1.538 \cdot 10^{-2}$ &   $1.543 \cdot 10^{-2}$ & $1.553 \cdot 10^{-2}$ & $8.889 \cdot 10^{-4}$ & $2.942 \cdot 10^{-2}$ & $2.952 \cdot 10^{-2}$ & $2.971 \cdot 10^{-2}$ & $2.287 \cdot 10^{-4}$ \\ \hline
    % 400 & 401 & 403 &  $2.051 \cdot 10^{-2}$  &  $2.056 \cdot 10^{-2}$  & $2.066 \cdot 10^{-2}$ &  $3.922 \cdot 10^{-2}$  & $3.932 \cdot 10^{-2}$ & $3.952 \cdot 10^{-2}$ \\ \hline
    500 & 501 & 503 & $2.563 \cdot 10^{-2}$   &  $2.568 \cdot 10^{-2}$  & $2.58 \cdot 10^{-2}$ & $9.67 \cdot 10^{-4}$ & $4.903 \cdot 10^{-2}$ & $4.913 \cdot 10^{-2}$ & $4.932 \cdot 10^{-2}$ & $2.488 \cdot 10^{-4}$\\ \hline
    % 600 & 601 & 603 & $3.076 \cdot 10^{-2}$   & $3.081 \cdot 10^{-2}$   & $3.091 \cdot 10^{-2}$ & $5.884 \cdot 10^{-2}$  & $5.893 \cdot 10^{-2}$ & $5.913 \cdot 10^{-2}$\\ \hline
    % 700 & 701 & 703 &  $3.589 \cdot 10^{-2}$  &  $3.594 \cdot 10^{-2}$  & $3.604 \cdot 10^{-2}$ & $6.864 \cdot 10^{-2}$  & $6.874 \cdot 10^{-2}$ & $6.894 \cdot 10^{-2}$\\ \hline
    % 800 & 801 & 803 &  $4.101 \cdot 10^{-2}$  &  $4.106 \cdot 10^{-2}$  & $4.117 \cdot 10^{-2}$ & $7.845 \cdot 10^{-2}$  & $7.855 \cdot 10^{-2}$ & $7.874 \cdot 10^{-2}$\\ \hline
    % 900 & 901 & 903 &  $4.614 \cdot 10^{-2}$  &   $4.619 \cdot 10^{-2}$ & $4.629 \cdot 10^{-2}$ & $8.825 \cdot 10^{-2}$  & $8.835 \cdot 10^{-2}$ & $8.855 \cdot 10^{-2}$\\ \hline
    1000 & 1001 & 1003 &  $5.126 \cdot 10^{-2}$  & $5.132 \cdot 10^{-2}$   & $5.142 \cdot 10^{-2}$ & $1.515 \cdot 10^{-3}$ & $9.806 \cdot 10^{-2}$ & $9.816 \cdot 10^{-2}$ & $9.836 \cdot 10^{-2}$ & $3.897 \cdot 10^{-4}$ \\ \hline
    % 1100 & 1101 & 1103 & $5.639 \cdot 10^{-2}$   &  $5.644 \cdot 10^{-2}$  & $5.654 \cdot 10^{-2}$ & $1.079 \cdot 10^{-1}$ & $1.08 \cdot 10^{-1}$ & $1.082 \cdot 10^{-1}$\\ \hline
    % 1200 & 1201 & 1203 &  $6.152 \cdot 10^{-2}$  &  $6.157 \cdot 10^{-2}$  & $6.167 \cdot 10^{-2}$ & $1.177 \cdot 10^{-1}$ & $1.178 \cdot 10^{-1}$ & $1.18 \cdot 10^{-1}$\\ \hline
    % 1300 & 1301 & 1303 &  $6.664 \cdot 10^{-2}$  & $6.67 \cdot 10^{-2}$   & $6.68 \cdot 10^{-2}$  & $1.275 \cdot 10^{-1}$ & $1.276 \cdot 10^{-1}$ & $1.278 \cdot 10^{-1}$\\ \hline
    % 1400 & 1401 & 1403 &  $7.177 \cdot 10^{-2}$  &  $7.182 \cdot 10^{-2}$  & $7.193 \cdot 10^{-2}$ & $1.373 \cdot 10^{-1}$ & $1.374 \cdot 10^{-1}$ & $1.376 \cdot 10^{-1}$ \\ \hline
    1500 & 1501 & 1503 &  $7.69 \cdot 10^{-2}$  & $7.695 \cdot 10^{-2}$   & $7.705 \cdot 10^{-2}$ & $4.479 \cdot 10^{-3}$ & $1.471 \cdot 10^{-1}$ & $1.472 \cdot 10^{-1}$ & $1.474 \cdot 10^{-1}$ & $1.153 \cdot 10^{-3}$  \\ \hline
    % 1600 & 1601 & 1603 &  $8.203 \cdot 10^{-2}$  &  $8.208 \cdot 10^{-2}$  & $8.218 \cdot 10^{-2}$ & $1.569 \cdot 10^{-1}$ & $1.57 \cdot 10^{-1}$ & $1.572 \cdot 10^{-1}$ \\ \hline
    % 1700 & 1701 & 1703 & $8.715 \cdot 10^{-2}$   &  $8.72 \cdot 10^{-2}$  & $8.731 \cdot 10^{-2}$ & $1.667 \cdot 10^{-1}$ & $1.668 \cdot 10^{-1}$ & $1.67 \cdot 10^{-1}$\\ \hline
    % 1800 & 1801 & 1803 &  $9.228 \cdot 10^{-2}$  & $9.233 \cdot 10^{-2}$   & $9.243 \cdot 10^{-2}$ & $1.765 \cdot 10^{-1}$ & $1.766 \cdot 10^{-1}$ & $1.768 \cdot 10^{-1}$ \\ \hline
    1900 & 1901 & 1903 &  $9.741 \cdot 10^{-2}$  &  $9.746 \cdot 10^{-2}$  & $9.756 \cdot 10^{-2}$ & $9.472 \cdot 10^{-2}$ & $1.863 \cdot 10^{-1}$ & $1.864 \cdot 10^{-1}$ & $1.866 \cdot 10^{-1}$ & $2.437 \cdot 10^{-2}$ \\ \hline
\end{tabular}}
\caption{Benchmark Points for Normal and Inverted Hierarchy at $v_{\Delta} = 10^{-9}~$ GeV and lightest neutrino mass = 0.001 eV. The cross-section listed here is for the process $\mu^+ \mu^+ > \Delta^{++} > \mu^+ \mu^+$ at $\sqrt{s} = 2$~TeV. All the masses and decay widths($\Gamma$) are in units of GeV and the cross-sections $\sigma$ are in units of fb.}
    \label{tab:dppbm2mupmup}
\end{table}
\newpage
\section{Analytical formula for the signal process}
\label{sec:analytical}

This section explains the fall and rise nature of the curve obtained in Fig.~\ref{fig:prodxsec} for cross-section of our signature of interest. For the processes (a) and (b) shown in Fig.~\ref{fig:signal},
the corresponding helicity amplitudes ${\cal M}_{(a)}(s_1,s_2,s_3)$ and ${\cal M}_{(b)}(s_1,s_2,s_3)$ are given as follows.
Here, $s_1=\pm$, $s_2=\pm$ are the helicities of the initial anti-Muons, and $s_3=\pm,0$ is the helicity of the final state $W$ boson.
\begin{eqnarray}
{\cal M}_{(a)}(-,-,0)/Y_{\mu \mu}^* &=& - {\cal M}_{(a)}(+,+,0) /Y_{\mu \mu}^* \nonumber \\
&=&
-\frac{\sqrt{2} \sqrt{s}\,
\sqrt{
(s - m_\Delta^2 - m_W^2)^2 - 4\,m_\Delta^2 m_W^2
}}
{m_W\,(s - m_\Delta^2)}.
\end{eqnarray}
\begin{eqnarray}
{\cal M}_{(b)}(-,-,\mp)/Y_{\mu \mu}^*&=& - {\cal M}_{(b)}(+,+,\pm)/Y_{\mu \mu}^*  \nonumber \\
&=&
\frac{\sin\theta}{2}\,
\frac{s - m_W^2 + m_\Delta^2 \pm \sqrt{(s - m_\Delta^2 - m_W^2)^2 - 4 m_\Delta^2 m_W^2}}
     {s- m_W^2 - m_\Delta^2 - \cos\theta\,\sqrt{(s - m_\Delta^2 - m_W^2)^2 - 4 m_\Delta^2 m_W^2}}
\nonumber \\
&-&
\;\frac{\sin\theta}{2}\,
\frac{s- m_W^2 + m_\Delta^2  \pm \sqrt{(s - m_\Delta^2 - m_W^2)^2 - 4 m_\Delta^2 m_W^2}}
     {s - m_W^2 - m_\Delta^2 + \cos\theta\,\sqrt{(s - m_\Delta^2 - m_W^2)^2 - 4 m_\Delta^2 m_W^2}} .
\end{eqnarray}
\begin{eqnarray}
{\cal M}_{(b)}(-,-,0)/Y_{\mu \mu}^* &=& - {\cal M}_{(b)}(+,+,0)/Y_{\mu \mu}^* \nonumber \\
&=&
\frac{
\sqrt{s}\;\left(
\sqrt{(s + m_W^2 - m_\Delta^2)^2 - 4 \,m_\Delta^2 m_W^2}
\;-\; \cos\theta\,\bigl(s - m_\Delta^2 - m_W^2\bigr)
\right)
}{
\sqrt{2}\,m_W\,
\left(
s-m_\Delta^2 - m_W^2
\;-\; \cos\theta\,
\sqrt{(s - m_\Delta^2 - m_W^2)^2 - 4\,m_\Delta^2 m_W^2}
\right)
}  \nonumber \\
&+&
\frac{
\sqrt{s} \; \left(
\sqrt{(s + m_W^2 - m_\Delta^2)^2 - 4\,m_\Delta^2 m_W^2}
\;+\; \cos\theta\,\bigl(s - m_\Delta^2 - m_W^2\bigr)
\right)
}{
\sqrt{2}\,m_W\,
\left(
s-m_\Delta^2 - m_W^2 
\;+\; \cos\theta\,
\sqrt{(s - m_\Delta^2 - m_W^2)^2 - 4\,m_\Delta^2 m_W^2}
\right)
}.
\end{eqnarray}
Other helicity combinations are all zero. 
For the process (b), we have neglected the $\nu_\ell$ masses.
The cross-term between ${\cal M}_{(a)}(\pm,\pm,0)$ and ${\cal M}_{(b)}(\pm,\pm,0)$ yields destructive interference that first decreases and then increases with increasing $m_{\Delta}$. This fall and rise behaviour of the destructive interference term explains the typical fall and rise nature of the curves with increasing $m_{\Delta}$ as seen in Fig.~\ref{fig:prodxsec} and Fig.~\ref{fig:xsecsig}.

%%%%%%%%%%%%%%%%%%%%%%%%%%%%%%%%%%%%%%%%%%%%%
%%%%%%%%%%%%%%%%%%%%%%%%%%%%%%%%%%%%%%%%%%%%%
\bibliographystyle{utphys}

\bibliography{reference}

\providecommand{\href}[2]{#2}\begingroup\raggedright\begin{thebibliography}{10}

\bibitem{Aad:2012tfa}
{\bfseries ATLAS} Collaboration, G.~Aad {\em et~al.}, ``{Observation of a new particle in the search for the Standard model Higgs boson with the ATLAS detector at the LHC},'' \href{http://dx.doi.org/10.1016/j.physletb.2012.08.020}{{\em Phys. Lett. B} {\bfseries 716} (2012) 1--29}, \href{http://arxiv.org/abs/1207.7214}{{\ttfamily arXiv:1207.7214 [hep-ex]}}.

\bibitem{Chatrchyan:2012ufa}
{\bfseries CMS} Collaboration, S.~Chatrchyan {\em et~al.}, ``{Observation of a New Boson at a Mass of 125 GeV with the CMS Experiment at the LHC},'' \href{http://dx.doi.org/10.1016/j.physletb.2012.08.021}{{\em Phys. Lett. B} {\bfseries 716} (2012) 30--61}, \href{http://arxiv.org/abs/1207.7235}{{\ttfamily arXiv:1207.7235 [hep-ex]}}.

\bibitem{Mohapatra:2006gs}
R.~N. Mohapatra and A.~Y. Smirnov, ``{Neutrino Mass and New Physics},'' \href{http://dx.doi.org/10.1146/annurev.nucl.56.080805.140534}{{\em Ann. Rev. Nucl. Part. Sci.} {\bfseries 56} (2006) 569--628}, \href{http://arxiv.org/abs/hep-ph/0603118}{{\ttfamily arXiv:hep-ph/0603118}}.

\bibitem{deGouvea:2016qpx}
A.~de~Gouv{\^e}a, ``{Neutrino Mass Models},'' \href{http://dx.doi.org/10.1146/annurev-nucl-102115-044600}{{\em Ann. Rev. Nucl. Part. Sci.} {\bfseries 66} (2016) 197--217}.

\bibitem{Branchina:2013jra}
V.~Branchina and E.~Messina, ``{Stability, Higgs Boson Mass and New Physics},'' \href{http://dx.doi.org/10.1103/PhysRevLett.111.241801}{{\em Phys. Rev. Lett.} {\bfseries 111} (2013) 241801}, \href{http://arxiv.org/abs/1307.5193}{{\ttfamily arXiv:1307.5193 [hep-ph]}}.

\bibitem{Susskind:1978ms}
L.~Susskind, ``{Dynamics of Spontaneous Symmetry Breaking in the Weinberg-Salam Theory},'' \href{http://dx.doi.org/10.1103/PhysRevD.20.2619}{{\em Phys. Rev. D} {\bfseries 20} (1979) 2619--2625}.

\bibitem{Veltman:1980mj}
M.~J.~G. Veltman, ``{The Infrared - Ultraviolet Connection},'' {\em Acta Phys. Polon. B} {\bfseries 12} (1981) 437.

\bibitem{Sakharov:1967dj}
A.~D. Sakharov, ``{Violation of CP Invariance, C asymmetry, and baryon asymmetry of the universe},'' \href{http://dx.doi.org/10.1070/PU1991v034n05ABEH002497}{{\em Pisma Zh. Eksp. Teor. Fiz.} {\bfseries 5} (1967) 32--35}.

\bibitem{Peccei:1977np}
R.~D. Peccei and H.~R. Quinn, ``{Some Aspects of Instantons},'' \href{http://dx.doi.org/10.1007/BF02730110}{{\em Nuovo Cim. A} {\bfseries 41} (1977) 309}.

\bibitem{Peccei:1977hh}
R.~D. Peccei and H.~R. Quinn, ``{CP Conservation in the Presence of Instantons},'' \href{http://dx.doi.org/10.1103/PhysRevLett.38.1440}{{\em Phys. Rev. Lett.} {\bfseries 38} (1977) 1440--1443}.

\bibitem{Peccei:1977ur}
R.~D. Peccei and H.~R. Quinn, ``{Constraints Imposed by CP Conservation in the Presence of Instantons},'' \href{http://dx.doi.org/10.1103/PhysRevD.16.1791}{{\em Phys. Rev. D} {\bfseries 16} (1977) 1791--1797}.

\bibitem{Magg:1980ut}
M.~Magg and C.~Wetterich, ``{Neutrino Mass Problem and Gauge Hierarchy},'' \href{http://dx.doi.org/10.1016/0370-2693(80)90825-4}{{\em Phys. Lett. B} {\bfseries 94} (1980) 61--64}.

\bibitem{Schechter:1980gr}
J.~Schechter and J.~W.~F. Valle, ``{Neutrino Masses in SU(2) x U(1) Theories},'' \href{http://dx.doi.org/10.1103/PhysRevD.22.2227}{{\em Phys. Rev. D} {\bfseries 22} (1980) 2227}.

\bibitem{Mohapatra:1979ia}
R.~N. Mohapatra and G.~Senjanovic, ``{Neutrino Mass and Spontaneous Parity Nonconservation},'' \href{http://dx.doi.org/10.1103/PhysRevLett.44.912}{{\em Phys. Rev. Lett.} {\bfseries 44} (1980) 912}.

\bibitem{Lazarides:1980nt}
G.~Lazarides, Q.~Shafi, and C.~Wetterich, ``{Proton Lifetime and Fermion Masses in an SO(10) Model},'' \href{http://dx.doi.org/10.1016/0550-3213(81)90354-0}{{\em Nucl. Phys. B} {\bfseries 181} (1981) 287--300}.

\bibitem{Deppisch:2015qwa}
F.~F. Deppisch, P.~S. Bhupal~Dev, and A.~Pilaftsis, ``{Neutrinos and Collider Physics},'' \href{http://dx.doi.org/10.1088/1367-2630/17/7/075019}{{\em New J. Phys.} {\bfseries 17} no.~7, (2015) 075019}, \href{http://arxiv.org/abs/1502.06541}{{\ttfamily arXiv:1502.06541 [hep-ph]}}.

\bibitem{Cai:2017mow}
Y.~Cai, T.~Han, T.~Li, and R.~Ruiz, ``{Lepton Number Violation: Seesaw Models and Their Collider Tests},'' \href{http://dx.doi.org/10.3389/fphy.2018.00040}{{\em Front. in Phys.} {\bfseries 6} (2018) 40}, \href{http://arxiv.org/abs/1711.02180}{{\ttfamily arXiv:1711.02180 [hep-ph]}}.

\bibitem{Calibbi:2017uvl}
L.~Calibbi and G.~Signorelli, ``{Charged Lepton Flavour Violation: An Experimental and Theoretical Introduction},'' \href{http://dx.doi.org/10.1393/ncr/i2018-10144-0}{{\em Riv. Nuovo Cim.} {\bfseries 41} no.~2, (2018) 71--174}, \href{http://arxiv.org/abs/1709.00294}{{\ttfamily arXiv:1709.00294 [hep-ph]}}.

\bibitem{Davidson:2022jai}
S.~Davidson, B.~Echenard, R.~H. Bernstein, J.~Heeck, and D.~G. Hitlin, ``{Charged Lepton Flavor Violation},'' \href{http://arxiv.org/abs/2209.00142}{{\ttfamily arXiv:2209.00142 [hep-ex]}}.

\bibitem{Delahaye:2019omf}
J.~P. Delahaye, M.~Diemoz, K.~Long, B.~Mansouli{\'e}, N.~Pastrone, L.~Rivkin, D.~Schulte, A.~Skrinsky, and A.~Wulzer, ``{Muon Colliders},'' \href{http://arxiv.org/abs/1901.06150}{{\ttfamily arXiv:1901.06150 [physics.acc-ph]}}.

\bibitem{Long:2020wfp}
K.~Long, D.~Lucchesi, M.~Palmer, N.~Pastrone, D.~Schulte, and V.~Shiltsev, ``{Muon colliders to expand frontiers of particle physics},'' \href{http://dx.doi.org/10.1038/s41567-020-01130-x}{{\em Nature Phys.} {\bfseries 17} no.~3, (2021) 289--292}, \href{http://arxiv.org/abs/2007.15684}{{\ttfamily arXiv:2007.15684 [physics.acc-ph]}}.

\bibitem{AlAli:2021let}
H.~Al~Ali {\em et~al.}, ``{The muon Smasher{\textquoteright}s guide},'' \href{http://dx.doi.org/10.1088/1361-6633/ac6678}{{\em Rept. Prog. Phys.} {\bfseries 85} no.~8, (2022) 084201}, \href{http://arxiv.org/abs/2103.14043}{{\ttfamily arXiv:2103.14043 [hep-ph]}}.

\bibitem{MuonCollider:2022xlm}
{\bfseries Muon Collider} Collaboration, J.~de~Blas {\em et~al.}, ``{The physics case of a 3 TeV muon collider stage},'' \href{http://arxiv.org/abs/2203.07261}{{\ttfamily arXiv:2203.07261 [hep-ph]}}.

\bibitem{Accettura:2023ked}
C.~Accettura {\em et~al.}, ``{Towards a muon collider},'' \href{http://dx.doi.org/10.1140/epjc/s10052-023-11889-x}{{\em Eur. Phys. J. C} {\bfseries 83} no.~9, (2023) 864}, \href{http://arxiv.org/abs/2303.08533}{{\ttfamily arXiv:2303.08533 [physics.acc-ph]}}. [Erratum: Eur.Phys.J.C 84, 36 (2024)].

\bibitem{Aime:2022flm}
C.~Aime {\em et~al.}, ``{Muon Collider Physics Summary},'' \href{http://arxiv.org/abs/2203.07256}{{\ttfamily arXiv:2203.07256 [hep-ph]}}.

\bibitem{Hamada:2022mua}
Y.~Hamada, R.~Kitano, R.~Matsudo, H.~Takaura, and M.~Yoshida, ``{$\mu$TRISTAN},'' \href{http://dx.doi.org/10.1093/ptep/ptac059}{{\em PTEP} {\bfseries 2022} no.~5, (2022) 053B02}, \href{http://arxiv.org/abs/2201.06664}{{\ttfamily arXiv:2201.06664 [hep-ph]}}.

\bibitem{Heusch:1995yw}
C.~A. Heusch and F.~Cuypers, ``{Physics with like-sign muon beams in a TeV muon collider},'' \href{http://dx.doi.org/10.1063/1.49345}{{\em AIP Conf. Proc.} {\bfseries 352} (1996) 219--231}, \href{http://arxiv.org/abs/hep-ph/9508230}{{\ttfamily arXiv:hep-ph/9508230}}.

\bibitem{Abe:2019thb}
M.~Abe {\em et~al.}, ``{A New Approach for Measuring the Muon Anomalous Magnetic Moment and Electric Dipole Moment},'' \href{http://dx.doi.org/10.1093/ptep/ptz030}{{\em PTEP} {\bfseries 2019} no.~5, (2019) 053C02}, \href{http://arxiv.org/abs/1901.03047}{{\ttfamily arXiv:1901.03047 [physics.ins-det]}}.

\bibitem{Bossi:2020yne}
F.~Bossi and P.~Ciafaloni, ``{Lepton Flavor Violation at muon-electron colliders},'' \href{http://dx.doi.org/10.1007/JHEP10(2020)033}{{\em JHEP} {\bfseries 10} (2020) 033}, \href{http://arxiv.org/abs/2003.03997}{{\ttfamily arXiv:2003.03997 [hep-ph]}}.

\bibitem{Lu:2020dkx}
M.~Lu, A.~M. Levin, C.~Li, A.~Agapitos, Q.~Li, F.~Meng, S.~Qian, J.~Xiao, and T.~Yang, ``{The physics case for an electron-muon collider},'' \href{http://dx.doi.org/10.1155/2021/6693618}{{\em Adv. High Energy Phys.} {\bfseries 2021} (2021) 6693618}, \href{http://arxiv.org/abs/2010.15144}{{\ttfamily arXiv:2010.15144 [hep-ph]}}.

\bibitem{Lichtenstein:2023iut}
G.~Lichtenstein, M.~A. Schmidt, G.~Valencia, and R.~R. Volkas, ``{Complementarity of $\mu$TRISTAN and Belle II in searches for charged-lepton flavour violation},'' \href{http://dx.doi.org/10.1016/j.physletb.2023.138144}{{\em Phys. Lett. B} {\bfseries 845} (2023) 138144}, \href{http://arxiv.org/abs/2307.11369}{{\ttfamily arXiv:2307.11369 [hep-ph]}}.

\bibitem{Das:2022mmh}
A.~Das, T.~Nomura, and T.~Shimomura, ``{Multi muon/anti-muon signals via productions of gauge and scalar bosons in a $U(1)_{L_\mu - L_\tau }$ model at muonic colliders},'' \href{http://dx.doi.org/10.1140/epjc/s10052-023-11955-4}{{\em Eur. Phys. J. C} {\bfseries 83} no.~9, (2023) 786}, \href{http://arxiv.org/abs/2212.11674}{{\ttfamily arXiv:2212.11674 [hep-ph]}}.

\bibitem{Yang:2023ojm}
J.-L. Yang, C.-H. Chang, and T.-F. Feng, ``{Leptonic di-flavor and di-number violation processes at high energy colliders*},'' \href{http://dx.doi.org/10.1088/1674-1137/ad17b0}{{\em Chin. Phys. C} {\bfseries 48} no.~4, (2024) 043101}, \href{http://arxiv.org/abs/2302.13247}{{\ttfamily arXiv:2302.13247 [hep-ph]}}.

\bibitem{Fridell:2023gjx}
K.~Fridell, R.~Kitano, and R.~Takai, ``{Lepton flavor physics at {\ensuremath{\mu}}$^{+}${\ensuremath{\mu}}$^{+}$ colliders},'' \href{http://dx.doi.org/10.1007/JHEP06(2023)086}{{\em JHEP} {\bfseries 06} (2023) 086}, \href{http://arxiv.org/abs/2304.14020}{{\ttfamily arXiv:2304.14020 [hep-ph]}}.

\bibitem{Dev:2023nha}
P.~S.~B. Dev, J.~Heeck, and A.~Thapa, ``{Neutrino mass models at $\mu $TRISTAN},'' \href{http://dx.doi.org/10.1140/epjc/s10052-024-12496-0}{{\em Eur. Phys. J. C} {\bfseries 84} no.~2, (2024) 148}, \href{http://arxiv.org/abs/2309.06463}{{\ttfamily arXiv:2309.06463 [hep-ph]}}.

\bibitem{Jia:2024wqi}
J.-C. Jia, Z.-L. Han, F.~Huang, Y.~Jin, and H.~Li, ``{Production of single doubly charged Higgs bosons at muon colliders},'' \href{http://dx.doi.org/10.1103/PhysRevD.111.015009}{{\em Phys. Rev. D} {\bfseries 111} no.~1, (2025) 015009}, \href{http://arxiv.org/abs/2409.16582}{{\ttfamily arXiv:2409.16582 [hep-ph]}}.

\bibitem{Kitano:2025xaj}
R.~Kitano, I.~Low, R.~Matsudo, S.~Okawa, and S.~Roy, ``{Heavy Neutral Lepton at Same-Sign Muon Collider},'' \href{http://arxiv.org/abs/2510.18390}{{\ttfamily arXiv:2510.18390 [hep-ph]}}.

\bibitem{Das:2024gfg}
A.~Das and Y.~Orikasa, ``{Z' induced forward dominant processes in $\mu$TRISTAN experiment},'' \href{http://dx.doi.org/10.1016/j.physletb.2024.138577}{{\em Phys. Lett. B} {\bfseries 851} (2024) 138577}, \href{http://arxiv.org/abs/2401.00696}{{\ttfamily arXiv:2401.00696 [hep-ph]}}.

\bibitem{Das:2024kyk}
A.~Das, J.~Li, S.~Mandal, T.~Nomura, and R.~Zhang, ``{Testing tree level TeV scale seesaw scenarios in {\ensuremath{\mu}}TRISTAN},'' \href{http://dx.doi.org/10.1103/df3g-32t9}{{\em Phys. Rev. D} {\bfseries 112} no.~3, (2025) 035008}, \href{http://arxiv.org/abs/2410.21956}{{\ttfamily arXiv:2410.21956 [hep-ph]}}.

\bibitem{Zyla:2020zbs}
{\bfseries Particle Data Group} Collaboration, P.~A. Zyla {\em et~al.}, ``{Review of Particle Physics},'' \href{http://dx.doi.org/10.1093/ptep/ptaa104}{{\em PTEP} {\bfseries 2020} no.~8, (2020) 083C01}.

\bibitem{Ghosh:2025gdx}
N.~Ghosh, S.~K. Rai, T.~Samui, and A.~Sarkar, ``{Investigating the leptonic couplings of doubly charged scalars at the muon collider},'' \href{http://arxiv.org/abs/2506.00966}{{\ttfamily arXiv:2506.00966 [hep-ph]}}.

\bibitem{ATLAS:2012hi}
{\bfseries ATLAS} Collaboration, G.~Aad {\em et~al.}, ``{Search for doubly-charged Higgs bosons in like-sign dilepton final states at $\sqrt{s}=7$ TeV with the ATLAS detector},'' \href{http://dx.doi.org/10.1140/epjc/s10052-012-2244-2}{{\em Eur. Phys. J. C} {\bfseries 72} (2012) 2244}, \href{http://arxiv.org/abs/1210.5070}{{\ttfamily arXiv:1210.5070 [hep-ex]}}.

\bibitem{ATLAS:2014kca}
{\bfseries ATLAS} Collaboration, G.~Aad {\em et~al.}, ``{Search for anomalous production of prompt same-sign lepton pairs and pair-produced doubly charged Higgs bosons with $ \sqrt{s}=8 $ TeV $pp$ collisions using the ATLAS detector},'' \href{http://dx.doi.org/10.1007/JHEP03(2015)041}{{\em JHEP} {\bfseries 03} (2015) 041}, \href{http://arxiv.org/abs/1412.0237}{{\ttfamily arXiv:1412.0237 [hep-ex]}}.

\bibitem{ATLAS:2017xqs}
{\bfseries ATLAS} Collaboration, M.~Aaboud {\em et~al.}, ``{Search for doubly charged Higgs boson production in multi-lepton final states with the ATLAS detector using proton{\textendash}proton collisions at $\sqrt{s}=13\,\text {TeV}$},'' \href{http://dx.doi.org/10.1140/epjc/s10052-018-5661-z}{{\em Eur. Phys. J. C} {\bfseries 78} no.~3, (2018) 199}, \href{http://arxiv.org/abs/1710.09748}{{\ttfamily arXiv:1710.09748 [hep-ex]}}.

\bibitem{ATLAS:2018ceg}
{\bfseries ATLAS} Collaboration, M.~Aaboud {\em et~al.}, ``{Search for doubly charged scalar bosons decaying into same-sign $W$ boson pairs with the ATLAS detector},'' \href{http://dx.doi.org/10.1140/epjc/s10052-018-6500-y}{{\em Eur. Phys. J. C} {\bfseries 79} no.~1, (2019) 58}, \href{http://arxiv.org/abs/1808.01899}{{\ttfamily arXiv:1808.01899 [hep-ex]}}.

\bibitem{ATLAS:2021jol}
{\bfseries ATLAS} Collaboration, G.~Aad {\em et~al.}, ``{Search for doubly and singly charged Higgs bosons decaying into vector bosons in multi-lepton final states with the ATLAS detector using proton-proton collisions at $ \sqrt{\mathrm{s}} $ = 13 TeV},'' \href{http://dx.doi.org/10.1007/JHEP06(2021)146}{{\em JHEP} {\bfseries 06} (2021) 146}, \href{http://arxiv.org/abs/2101.11961}{{\ttfamily arXiv:2101.11961 [hep-ex]}}.

\bibitem{CMS:2012dun}
{\bfseries CMS} Collaboration, S.~Chatrchyan {\em et~al.}, ``{A Search for a Doubly-Charged Higgs Boson in $pp$ Collisions at $\sqrt{s}=7$ TeV},'' \href{http://dx.doi.org/10.1140/epjc/s10052-012-2189-5}{{\em Eur. Phys. J. C} {\bfseries 72} (2012) 2189}, \href{http://arxiv.org/abs/1207.2666}{{\ttfamily arXiv:1207.2666 [hep-ex]}}.

\bibitem{CMS:2014mra}
{\bfseries CMS} Collaboration, V.~Khachatryan {\em et~al.}, ``{Study of vector boson scattering and search for new physics in events with two same-sign leptons and two jets},'' \href{http://dx.doi.org/10.1103/PhysRevLett.114.051801}{{\em Phys. Rev. Lett.} {\bfseries 114} no.~5, (2015) 051801}, \href{http://arxiv.org/abs/1410.6315}{{\ttfamily arXiv:1410.6315 [hep-ex]}}.

\bibitem{CMS:2017fhs}
{\bfseries CMS} Collaboration, A.~M. Sirunyan {\em et~al.}, ``{Observation of electroweak production of same-sign W boson pairs in the two jet and two same-sign lepton final state in proton-proton collisions at $\sqrt{s} = $ 13 TeV},'' \href{http://dx.doi.org/10.1103/PhysRevLett.120.081801}{{\em Phys. Rev. Lett.} {\bfseries 120} no.~8, (2018) 081801}, \href{http://arxiv.org/abs/1709.05822}{{\ttfamily arXiv:1709.05822 [hep-ex]}}.

\bibitem{ATLAS:2022pbd}
{\bfseries ATLAS} Collaboration, G.~Aad {\em et~al.}, ``{Search for doubly charged Higgs boson production in multi-lepton final states using 139~fb$^{-1}$ of proton{\textendash}proton collisions at $\sqrt{s}$ = 13~TeV with the ATLAS detector},'' \href{http://dx.doi.org/10.1140/epjc/s10052-023-11578-9}{{\em Eur. Phys. J. C} {\bfseries 83} no.~7, (2023) 605}, \href{http://arxiv.org/abs/2211.07505}{{\ttfamily arXiv:2211.07505 [hep-ex]}}.

\bibitem{Ashanujjaman:2021txz}
S.~Ashanujjaman and K.~Ghosh, ``{Revisiting type-II see-saw: present limits and future prospects at LHC},'' \href{http://dx.doi.org/10.1007/JHEP03(2022)195}{{\em JHEP} {\bfseries 03} (2022) 195}, \href{http://arxiv.org/abs/2108.10952}{{\ttfamily arXiv:2108.10952 [hep-ph]}}.

\bibitem{Fuks:2019clu}
B.~Fuks, M.~Nemev{\v{s}}ek, and R.~Ruiz, ``{Doubly Charged Higgs Boson Production at Hadron Colliders},'' \href{http://dx.doi.org/10.1103/PhysRevD.101.075022}{{\em Phys. Rev. D} {\bfseries 101} no.~7, (2020) 075022}, \href{http://arxiv.org/abs/1912.08975}{{\ttfamily arXiv:1912.08975 [hep-ph]}}.

\bibitem{Alwall:2011uj}
J.~Alwall, M.~Herquet, F.~Maltoni, O.~Mattelaer, and T.~Stelzer, ``{MadGraph 5 : Going Beyond},'' \href{http://dx.doi.org/10.1007/JHEP06(2011)128}{{\em JHEP} {\bfseries 06} (2011) 128}, \href{http://arxiv.org/abs/1106.0522}{{\ttfamily arXiv:1106.0522 [hep-ph]}}.

\bibitem{Sjostrand:2014zea}
T.~Sj\"ostrand, S.~Ask, J.~R. Christiansen, R.~Corke, N.~Desai, P.~Ilten, S.~Mrenna, S.~Prestel, C.~O. Rasmussen, and P.~Z. Skands, ``{An introduction to PYTHIA 8.2},'' \href{http://dx.doi.org/10.1016/j.cpc.2015.01.024}{{\em Comput. Phys. Commun.} {\bfseries 191} (2015) 159--177}, \href{http://arxiv.org/abs/1410.3012}{{\ttfamily arXiv:1410.3012 [hep-ph]}}.

\bibitem{deFavereau:2013fsa}
{\bfseries DELPHES 3} Collaboration, J.~de~Favereau, C.~Delaere, P.~Demin, A.~Giammanco, V.~Lema\^\i{}tre, A.~Mertens, and M.~Selvaggi, ``{DELPHES 3, A modular framework for fast simulation of a generic collider experiment},'' \href{http://dx.doi.org/10.1007/JHEP02(2014)057}{{\em JHEP} {\bfseries 02} (2014) 057}, \href{http://arxiv.org/abs/1307.6346}{{\ttfamily arXiv:1307.6346 [hep-ex]}}.

\bibitem{Selvaggi:2717698}
M.~Selvaggi, ``{A Delphes parameterisation of the FCC-hh detector},'' tech. rep., CERN, Geneva, 2020.
\newblock \url{https://cds.cern.ch/record/2717698}.

\bibitem{Leogrande:2019qbe}
E.~Leogrande, P.~Roloff, U.~Schnoor, and M.~Weber, ``{A DELPHES card for the CLIC detector},'' \href{http://arxiv.org/abs/1909.12728}{{\ttfamily arXiv:1909.12728 [hep-ex]}}.

\bibitem{Boronat:2014hva}
M.~Boronat, J.~Fuster, I.~Garcia, E.~Ros, and M.~Vos, ``{A robust jet reconstruction algorithm for high-energy lepton colliders},'' \href{http://dx.doi.org/10.1016/j.physletb.2015.08.055}{{\em Phys. Lett. B} {\bfseries 750} (2015) 95--99}, \href{http://arxiv.org/abs/1404.4294}{{\ttfamily arXiv:1404.4294 [hep-ex]}}.

\bibitem{AlipourTehrani:1742993}
N.~Alipour~Tehrani and P.~Roloff, ``{Optimisation Studies for the CLIC Vertex-Detector Geometry},''. \url{https://cds.cern.ch/record/1742993}.

\bibitem{Mohapatra:1980yp}
R.~N. Mohapatra and G.~Senjanovic, ``{Neutrino Masses and Mixings in Gauge Models with Spontaneous Parity Violation},'' \href{http://dx.doi.org/10.1103/PhysRevD.23.165}{{\em Phys. Rev. D} {\bfseries 23} (1981) 165}.

\bibitem{Pati:1974yy}
J.~C. Pati and A.~Salam, ``{Lepton Number as the Fourth Color},'' \href{http://dx.doi.org/10.1103/PhysRevD.10.275}{{\em Phys. Rev. D} {\bfseries 10} (1974) 275--289}. [Erratum: Phys.Rev.D 11, 703--703 (1975)].

\bibitem{Mohapatra:1974hk}
R.~N. Mohapatra and J.~C. Pati, ``{Left-Right Gauge Symmetry and an Isoconjugate Model of CP Violation},'' \href{http://dx.doi.org/10.1103/PhysRevD.11.566}{{\em Phys. Rev. D} {\bfseries 11} (1975) 566--571}.

\bibitem{Senjanovic:1975rk}
G.~Senjanovic and R.~N. Mohapatra, ``{Exact Left-Right Symmetry and Spontaneous Violation of Parity},'' \href{http://dx.doi.org/10.1103/PhysRevD.12.1502}{{\em Phys. Rev. D} {\bfseries 12} (1975) 1502}.

\bibitem{Kuchimanchi:1993jg}
R.~Kuchimanchi and R.~N. Mohapatra, ``{No parity violation without R-parity violation},'' \href{http://dx.doi.org/10.1103/PhysRevD.48.4352}{{\em Phys. Rev. D} {\bfseries 48} (1993) 4352--4360}, \href{http://arxiv.org/abs/hep-ph/9306290}{{\ttfamily arXiv:hep-ph/9306290}}.

\bibitem{Babu:2008ep}
K.~S. Babu and R.~N. Mohapatra, ``{Minimal Supersymmetric Left-Right Model},'' \href{http://dx.doi.org/10.1016/j.physletb.2008.09.018}{{\em Phys. Lett. B} {\bfseries 668} (2008) 404--409}, \href{http://arxiv.org/abs/0807.0481}{{\ttfamily arXiv:0807.0481 [hep-ph]}}.

\bibitem{Babu:2014vba}
K.~S. Babu and A.~Patra, ``{Higgs Boson Spectra in Supersymmetric Left-Right Models},'' \href{http://dx.doi.org/10.1103/PhysRevD.93.055030}{{\em Phys. Rev. D} {\bfseries 93} no.~5, (2016) 055030}, \href{http://arxiv.org/abs/1412.8714}{{\ttfamily arXiv:1412.8714 [hep-ph]}}.

\bibitem{Basso:2015pka}
L.~Basso, B.~Fuks, M.~E. Krauss, and W.~Porod, ``{Doubly-charged Higgs and vacuum stability in left-right supersymmetry},'' \href{http://dx.doi.org/10.1007/JHEP07(2015)147}{{\em JHEP} {\bfseries 07} (2015) 147}, \href{http://arxiv.org/abs/1503.08211}{{\ttfamily arXiv:1503.08211 [hep-ph]}}.

\bibitem{Zee:1985id}
A.~Zee, ``{Quantum Numbers of Majorana Neutrino Masses},'' \href{http://dx.doi.org/10.1016/0550-3213(86)90475-X}{{\em Nucl. Phys. B} {\bfseries 264} (1986) 99--110}.

\bibitem{Babu:1988ki}
K.~S. Babu, ``{Model of 'Calculable' Majorana Neutrino Masses},'' \href{http://dx.doi.org/10.1016/0370-2693(88)91584-5}{{\em Phys. Lett. B} {\bfseries 203} (1988) 132--136}.

\bibitem{ArkaniHamed:2002qx}
N.~Arkani-Hamed, A.~G. Cohen, E.~Katz, A.~E. Nelson, T.~Gregoire, and J.~G. Wacker, ``{The Minimal moose for a little Higgs},'' \href{http://dx.doi.org/10.1088/1126-6708/2002/08/021}{{\em JHEP} {\bfseries 08} (2002) 021}, \href{http://arxiv.org/abs/hep-ph/0206020}{{\ttfamily arXiv:hep-ph/0206020}}.

\bibitem{Babu:2020hun}
K.~S. Babu, P.~S.~B. Dev, S.~Jana, and A.~Thapa, ``{Unified framework for $B$-anomalies, muon $g - 2$ and neutrino masses},'' \href{http://dx.doi.org/10.1007/JHEP03(2021)179}{{\em JHEP} {\bfseries 03} (2021) 179}, \href{http://arxiv.org/abs/2009.01771}{{\ttfamily arXiv:2009.01771 [hep-ph]}}.

\bibitem{Georgi:1985nv}
H.~Georgi and M.~Machacek, ``{DOUBLY CHARGED HIGGS BOSONS},'' \href{http://dx.doi.org/10.1016/0550-3213(85)90325-6}{{\em Nucl. Phys. B} {\bfseries 262} (1985) 463--477}.

\bibitem{Gunion:1989ci}
J.~F. Gunion, R.~Vega, and J.~Wudka, ``{Higgs triplets in the standard model},'' \href{http://dx.doi.org/10.1103/PhysRevD.42.1673}{{\em Phys. Rev. D} {\bfseries 42} (1990) 1673--1691}.

\bibitem{Babu:2009aq}
K.~S. Babu, S.~Nandi, and Z.~Tavartkiladze, ``{New Mechanism for Neutrino Mass Generation and Triply Charged Higgs Bosons at the LHC},'' \href{http://dx.doi.org/10.1103/PhysRevD.80.071702}{{\em Phys. Rev. D} {\bfseries 80} (2009) 071702}, \href{http://arxiv.org/abs/0905.2710}{{\ttfamily arXiv:0905.2710 [hep-ph]}}.

\bibitem{Bonnet:2009ej}
F.~Bonnet, D.~Hernandez, T.~Ota, and W.~Winter, ``{Neutrino masses from higher than d=5 effective operators},'' \href{http://dx.doi.org/10.1088/1126-6708/2009/10/076}{{\em JHEP} {\bfseries 10} (2009) 076}, \href{http://arxiv.org/abs/0907.3143}{{\ttfamily arXiv:0907.3143 [hep-ph]}}.

\bibitem{Bhattacharya:2016qsg}
S.~Bhattacharya, S.~Jana, and S.~Nandi, ``{Neutrino Masses and Scalar Singlet Dark Matter},'' \href{http://dx.doi.org/10.1103/PhysRevD.95.055003}{{\em Phys. Rev. D} {\bfseries 95} no.~5, (2017) 055003}, \href{http://arxiv.org/abs/1609.03274}{{\ttfamily arXiv:1609.03274 [hep-ph]}}.

\bibitem{Kumericki:2012bh}
K.~Kumericki, I.~Picek, and B.~Radovcic, ``{TeV-scale Seesaw with Quintuplet Fermions},'' \href{http://dx.doi.org/10.1103/PhysRevD.86.013006}{{\em Phys. Rev. D} {\bfseries 86} (2012) 013006}, \href{http://arxiv.org/abs/1204.6599}{{\ttfamily arXiv:1204.6599 [hep-ph]}}.

\end{thebibliography}\endgroup

\end{document}